\documentclass[fleqn,usenatbib]{mnras}

\usepackage{newtxtext,newtxmath}

\usepackage[T1]{fontenc}
\usepackage{ae,aecompl}
\DeclareRobustCommand{\VAN}[3]{#2}
\let\VANthebibliography\thebibliography
\def\thebibliography{\DeclareRobustCommand{\VAN}[3]{##3}\VANthebibliography}
\usepackage{hyperref} 

\usepackage{graphicx}   
\usepackage{amsmath}	
\usepackage{longtable} 
\def\aj{AJ}
\def\actaa{Acta Astron.}
\def\araa{ARA\&A}
\def\apj{ApJ}
\def\apss{Ap\&SS}
\def\mnras{MNRAS}
\def\na{New A}
\def\pasp{PASP}
\def\pasj{PASJ}
\def\rmxaa{Rev. Mexicana Astron. Astrofis.}
\def\solphys{Sol.~Phys.}

\title[Investigating kinematics and dynamics of three open clusters towards Galactic anti-center]
  	{Investigating kinematics and dynamics of three open clusters towards Galactic anti-center}

\author[Rangwal et al.]
   {Geeta Rangwal$^{1}$\thanks{E-mail: geetarangwal91@gmail.com},
	R. K. S. Yadav$^{2}$, D. Bisht$^{3}$, Alok Durgapal$^{4}$,
	Devesh P. Sariya$^{5}$\\
	\\
	$^{1}$ Indian Institute of Astrophysics, Bangalore, India \\
	$^{2}$ Aryabhatta Research Institute of Observational Sciences,
       	Manora Peak, Nainital 263129, India\\
	$^{3}$ Indian Centre for Space Physics, 466 Barakhola, Singabari road, Netai Nagar, Kolkata, 700099, India\\
	$^{4}$Center of Advanced Study, Department of Physics,
       	D. S. B. Campus, Kumaun University Nainital 263002, India\\
	$^{5}$ Department of Physics and Institute of Astronomy, National Tsing-Hua University, Hsin-Chu, Taiwan \\
	}

\date{Accepted XXX. Received YYY; in original form ZZZ}

\pubyear{2023}

\begin{document}
\label{firstpage}
\pagerange{\pageref{firstpage}--\pageref{lastpage}}
\maketitle


\begin{abstract}
    We present the intra-cluster kinematics and dynamics of three open clusters: 
    NGC 1193, NGC 2355, and King 12 by incorporating kinematical and photometric data from
    Gaia DR3, as well as a ground-based telescope. After selecting cluster members based on proper motion data, clusters' fundamental and 
    structural parameters are investigated. We found the clusters at distances of 4.45, 1.97, and 3.34 kpc from the Sun in the direction of the Galactic 
    anticenter. The luminosity function of the cluster NGC 1193 is flat, whereas it advances towards the fainter ends of the other two clusters. 
    We observed a dip in the luminosity function of King 12. The mass function slopes for all 
    three clusters differ from the solar neighbourhood reported by Salpeter,
    with NGC 1193 and NGC 2355 being flatter and King 12 having a higher 
    value than the Salpeter value. The intra-cluster kinematics depict that stars in King 12 are moving outwards due to tidal forces from the 
    Galactic disc, which we confirmed by plotting the cluster's orbit in the Galaxy. Stars in NGC 2355 are moving with smaller relative 
    velocities and have zero mean relative motion, which signifies that the cluster is neither contracting nor evaporating. The Galactic 
    orbits of NGC 1193 suggest that it is orbiting farther from the Galactic disc, and so is less impacted by the Galactic tidal forces.
\end{abstract}




\section{Introduction} \label{sec:intro}

Open clusters (OCs) of our Galaxy have an extensive range of ages that can cover the Galactic disc's entire lifetime. The older OCs are tracing the thick disc,
while the younger ones are found in spiral arms in the thin disc where the
perturbation is numerous. Motion and spatial distribution of these objects in the
Galaxy can assist in comprehending the gravitational potential and perturbations which
shape the configuration and dynamics of the Galaxy \citep{2018A&A...619A.155S}.
In addition to the evolution of an isolated system, dynamical evolution of the
stars in open clusters are influenced by the disruptive effects of the tidal forces
arising from the Galactic disc and the molecular clouds present in the Galactic disc \citep{1995ARA&A..33..381F}.
The theoretical models suggest that generally, in the time scale of $10^{8}$ to $10^{9}$ years,
all the open clusters will disrupt, and the disruption time scales depend on the
mass as well as the core radius of the clusters \citep{1958ApJ...127...17S, 1958ApJ...127..544S}. Most of the OCs in the Milky Way survive
due to their higher mass and central concentration and the path they follow to revolve around the Galactic centre, which avoids the influence of large
disruptive forces \citep{1995ARA&A..33..381F}. So old open clusters are worth
exploring to test these theories. Also, we can predict the survival of the 
younger systems according to these theories.

\begin{figure*}
	\centering
	\includegraphics[width=17cm, height=7cm]{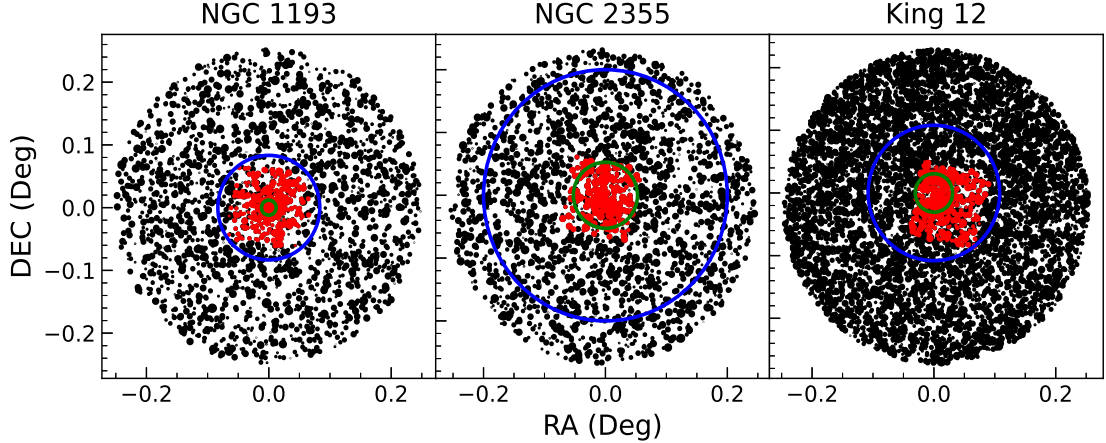}
	\caption{ Identification charts for the clusters
   	NGC 1193, NGC 2355 and King 12. The different sizes of filled circles indicate
    the magnitudes of the stars in $G$ filter as the shortest size denotes stars of $G$ $\sim 19$ mag. The black stars symbolise stars taken from Gaia DR3 in a field of 15
    arcmin radius and red stars are observed from 104-cm Sampurnanand Telescope having a field of view of 6.8 $\times$ 6.5 $arcmin^2$. The blue
    circle designates the dimensions of the cluster region, while the green circle represents the core region of the clusters.
    }
	\label{id}
  \end{figure*}

\begin{table*}
   \centering
   \caption{ Basic information of the clusters investigated in the present article, as given in the WEBDA database.}
   \begin{tabular}{ccccccccc}
   \hline\hline
  Cluster &  $\alpha_{2000}$ & $\delta_{2000}$  & $l$ (deg)  & $b$ (deg) & $d_{\odot}$ (pc) & $E(B-V)$ & $log(t)$ & $[Fe/H]$ (dex) \\
  \hline
   NGC 1193 & 03:05:32 &  +44:23:00 & 146.749 & -12.199 & 4300 & 0.12 & 9.90 & -0.29   \\
   NGC 2355 & 07:16:59 &  +13:45:00 & 203.39 & 11.803 & 2200 & 0.12 & 8.85 & -0.07   \\
   King 12 & 23:53:00 &  +61:58:00  & 116.124 & -0.130 & 2378 & 0.59 & 7.04 & -	\\
  \hline
  \end{tabular}
  \label{bpara}
  \end{table*}

For this study, we selected three open clusters, NGC 1193, NGC 2355
and King 12, found towards the Galactic anti-centre. These clusters are
at different dynamical stages and have distinct spatial locations
in the Galactic disc. NGC 1193 is one of the Galaxy’s oldest and
most well-populated open clusters of faint stars.
\citet{1988AcA....38..339K} studied this cluster using CCD data and determined
age, metallicity, and distance modulus as 8 Gyr, 0.001 dex, and 13.8 mag. 
They also identified five blue straggler stars in this cluster.
\citet{2005AN....326...19T} studied this cluster extensively using USNO-B1.0
catalogue. They found age, metallicity, $E(B-V)$, and distance as 8.0 Gyr, 0.008,
0.10 $\pm$ 0.06 mag and 5250 $\pm$ 240 pc, respectively. \citet{2008JKAS...41..147K}
conducted an $UBVRI$ photometric study and estimated the fundamental parameters
for this cluster as $E(B-V)$, $[Fe/H]$, $(m-M)_{0}$ and age as 0.19 $\pm$ 0.04,
-0.45 $\pm$ 0.12, 13.3 $\pm$ 0.15 and, 5 Gyr respectively.

NGC 2355 is an intermediate open cluster at a distance of $\sim$ 2 kpc from the Sun.
\citet{1991AcA....41..279K} studied the central part of this cluster using CCD photometric
data in $UBV$ filters and calculated the fundamental parameters $E(B-V)$, $[Fe/H]$ and distance as 0.12 mag, 0.13 dex, and 2.2 kpc respectively.
\citet{2000A&A...357..484S} studied it using photometric and spectroscopic data
and found that the cluster is at a distance of 1.65 kpc from Sun, has reddening, age, and $[Fe/H]$
equal to 0.16 mag, 1 Gyr, and -0.07 $\pm$ 0.11 dex, respectively. They concluded 
that the turn-off stars of this cluster are fast rotators. They also found a blue straggler and one
unusual straggler star above the main sequence. \citet{2022AJ....164...40W} conducted a time-series study 
for NGC 2355 and found 88 variable stars in this cluster.

King 12 is a very young open cluster having a heliocentric distance of $\sim$ 3 kpc. This cluster was 
first studied by \citet{1984Ap&SS.105..315M} using photoelectric data, and they identified 30 stars in the cluster 
field. They calculated the distance of the
cluster as 2.49 kpc. \citet{2013BASI...41..209T} calculated the cluster's age, reddening, and distance as 10 $\pm$ 0.1 Myr, 0.63 $\pm$ 0.05 mag and 2.5 $\pm$ 0.1 kpc respectively
in their CCD photometric study of this cluster. 
\citet{2012ApJ...761..155D} studied
this cluster using CFHT MegaCam data and determined its age to be 20 Myr. They used data of brighter stars from  \citet{1984Ap&SS.105..315M} due to photometric saturation.
They observed breaks in luminosity function in this cluster
and argued that this might be the onset of pre-main sequence stars. \citet{2013MNRAS.429.1102G}
conducted a CCD photometric study for this cluster and determined age, distance,
and reddening for the cluster as 70 Myr, $2490^{+180}_{-170}$ pc and 0.51 $\pm$
0.05 mag respectively. They also found a gap in the luminosity function of the
cluster, but the gap is for a different magnitude bin compared to the gap
reported by \citet{2012ApJ...761..155D}. They inferred this gap as a discontinuity
in the star formation process inside the cluster. \citet{2014NewA...26...77L} also
studied these clusters using photometric data in $UBVRI$ filters and determined the 
radius as 4.0 arcmin, reddening as 0.58 mag, and age between 10 to 500 Myr. They also found a mass function 
slope comparable to the \citet{Salpeter1955ApJ...121..161S} value.

Despite numerous investigations on these clusters in the literature, the values of the cluster's properties remain inconsistent. The study of the dynamics of these clusters has not yet been conducted. The general information
on the clusters taken from the WEBDA database is listed in Table \ref{bpara}.
The kinematical data from Gaia \citep{2016A&A...595A...1G, 2016A&A...595A...2G, 2018A&A...616A...9L, 2018A&A...618A..93C} is revolutionizing our understanding of the
dynamics of clusters in the Galaxy. Its third data release  \citep{2016A&A...595A...1G,
2016A&A...595A...2G, 2018A&A...616A..11G, 2021A&A...649A...1G, 2021A&A...649A...6G}
is providing precise kinematic and photometric data with
minimum uncertainty. For the present analysis, we used kinematic and photometric data from Gaia DR3 and ground-
based photometric data to study the dynamics of the three clusters
NGC 1193, NGC 2355, and King 12. The description of data is provided in sections \ref{sec:obs} and \ref{sec:kin}.
In section \ref{sec:vpds}, the process of selecting cluster members is discussed, and
the fundamental properties of the clusters are determined in section \ref{sec:ana}.
The dynamical properties of the clusters are discussed in section \ref{sec:luminosity}. In section
\ref{dynamics}, the internal and external dynamics of the clusters are discussed.
Finally, we concluded the article in section \ref{con}.

\section{Observational data and reduction} \label{sec:obs}

We have observed the three clusters, namely NGC 1193, NGC 2355, and King 12, using the PyLoN CCD
 mounted on 104-cm Sampurnanad
telescope located at ARIES, Nainital India, in the standard Johnson $B$, $V$, and
$I$-band filters. The size of CCD is 1340 $\times$ 1300 pixels with a pixel size of 20 microns and covers an area of 6.8 $\times$ 6.5
arcmin$^2$ in the sky. The observations were carried out on 14 November 2017 and
4 October 2018. Besides science images, many biases,
flat images, and two standard fields, PG 2331 and PG 2336, were observed
during the observing nights. A log of observational data is provided
in Table \ref{log}.
For the pre-reduction process of CCD images, such as bias subtraction, flat-fielding, and 
cosmic ray correction, we used the $imred$ package available in $IRAF$ software. 
The reduction of images, including
determining the accurate positions and magnitude of the stars, was carried out with the help of PSF
photometry available in DAOPHOT II software originally developed by \citet{1987PASP...99..191S}. In this analysis, we used the Fortran version of this software, updated by Stetson in 2004. The reduction 
process started with finding the stars in every image taken from the CCD using an automatic algorithm
of DAOPHOT II. Since the field is crowded, we adopted the profile-fitting photometry routine available in 
DAOPHOT II. This constructs a point spread function using bright stars in the field and
then applies it to all the stars. The output of this process is a catalogue of
positions and instrumental magnitudes of the stars in the image.

\begin{table}
   \centering
   \caption{Log of the observations carried out with 104-Sampurnanand Telescope
    for the clusters NGC 1193, NGC 2355 and King 12, as well as the standard fields PG 2331 and PG 2336.
    }
   \begin{tabular}{ccc}
   \hline\hline
  Filters & Exposure Time & Date   \\
    	& (in seconds)	&   \\
  \hline
  &  {\bf NGC 1193}  &	\\
   $B$   & 420 $\times$ 3  & 4 Oct 2018	\\
   	& 180 $\times$ 3 &  "   \\
   $V$   & 420 $\times$ 3  &  "   \\
   	& 120 $\times$ 3  &  "   \\
   $I$ & 240 $\times$ 3 &  "	\\
   	&  60 $\times$ 3 &  " \\
 	& {\bf NGC 2355}   &   \\
   $B$   & 420 $\times$ 3  & 14 Nov 2017	\\
   	& 180 $\times$ 3 &  "   \\
   	& 180 $\times$ 2 & 4 Oct 2018  \\
   $V$   & 420 $\times$ 3  & 14 Nov 2017   \\
   	& 120 $\times$ 3  &  "   \\
   	& 120 $\times$ 2 &  4 Oct 2018 \\
   $I$ & 240 $\times$ 3 & 14 Nov 2017	\\
   	&  60 $\times$ 3 &  " \\
   	&  60 $\times$ 2 & 4 Oct 2018 \\
   & {\bf King 12}  &  \\
  $B$	& 420 $\times$ 3  &  14 Nov 2017	\\
   	& 180 $\times$ 3 & 	"  	\\
   	& 180 $\times$ 2 & 4 Oct 2018  \\
  $V$	& 240 $\times$ 2 & 14 Nov 2017	\\
     & 120 $\times$ 3  &  "   \\
   	& 120 $\times$ 2 & 4 Oct 2018 \\
   $I$ & 240 $\times$ 3 & 14 Nov 2017	\\
   	&  60 $\times$ 3 &  " \\
   	&  60 $\times$ 2 &  4 Oct 2018 \\
   	& {\bf PG 2331}  & \\
   $B$ & 90 $\times$ 14 & 4 Oct 2018  \\
   $V$ & 60 $\times$ 14 & "  \\
   $I$ & 30 $\times$ 14 & " \\
   	& {\bf PG 2336}  & \\
   $B$ & 90 $\times$ 2  & " \\
   $ V$ & 60 $\times$ 2 &  " \\
   $I$ & 30 $\times$ 2 & " \\
  \hline
  \end{tabular}
  \label{log}
  \end{table}

 \begin{table*}
 \centering
 \caption{ The colour coefficients ($C_{X}$) and zero-points
       	($Z_{X}$) for respective filters
      	used for the calibration equations. $X$ represents different filter systems.
      	}
   \begin{tabular}{ccccccccc}
  \hline\hline
   	$C_{B}$ & $C_{V}$ & $C_{I}$ & $Z_{B}$  & $Z_{V}$ & $Z_{I}$\\
  \hline
  	$-0.047 \pm 0.016$ &  $ 0.078 \pm 0.009 $ & $-0.044 \pm 0.006$ & $2.739 \pm 0.012$ & $2.657 \pm 0.007$ & $2.760 \pm 0.007$	\\
 \hline
 \end{tabular}
  \label{zp}
 \end{table*}

  \begin{figure}
	\centering
	\includegraphics[width=8.8cm]{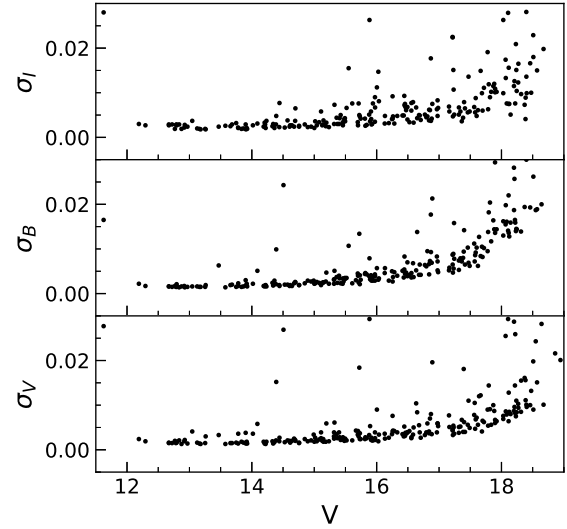}
	\caption{The photometric errors in three filters namely $B$, $V$ and
 	$I$ are plotted against the calibrated $V$ magnitude for the NGC 2355.}
	\label{er}
  \end{figure}

The instrumental magnitudes of the stars were calibrated into the standard Johnson
and Kron-Cousin system using the standard fields observed with the science frames.
This calibration was also done using the different calibration routines
in the DAOPHOT II package.
The transformation equations used to calibrate the instrumental magnitudes are as follows: \\

\begin{equation}
b = B + C_{B} \times (B - V) + Z_{B} + k_{B} X
\end{equation}

\begin{equation}
v = V + C_{V} \times (B - V) + Z_{V} + k_{V} X 
\end{equation}

\begin{equation}
i = I + C_{I} \times (V - I) + Z_{I} + k_{I} X\\
\end{equation}

where $B$, $V$, $I$ and $b$, $v$, $i$ are the standards and instrumental magnitudes
while $C_{B}$, $C_{V}$, and $C_{I}$ are the colour coefficients
for $B$, $V$, and $I$ filters respectively. $Z_{B}$, $Z_{V}$
and $Z_{I}$ are the zero-points for respective filters and $X$ is the airmass.
The colour coefficients and zero points obtained during the transformation
for each cluster are listed in Table \ref{zp}. The typical error in our photometry in different filters
as a function of $V$ mag is plotted in Fig. \ref{er}. It is perceptible from this figure
that the error is less than 0.05 mag up to $V=19^{th}$ for $B$, $V$ and $I$ filters. We converted $X$ and $Y$
pixel position of stars into the right ascension (RA) and declination (DEC) of 
$J2000$ using the $ccmap$ and $cctran$ tasks in the IRAF. 

\begin{figure}
	\centering
	\includegraphics[width=8.5cm, height=6cm]{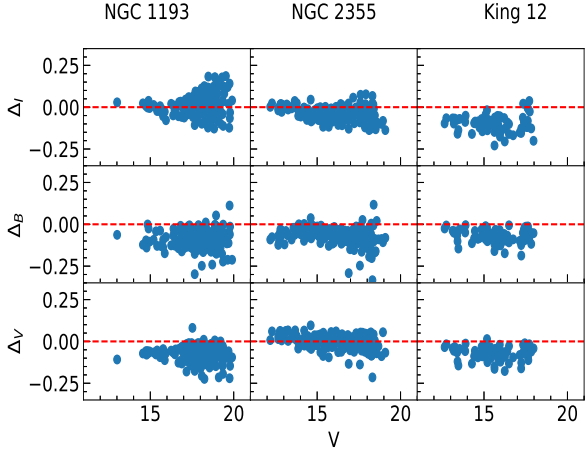}
	\caption{The difference of magnitudes from the current study and literature in three filters, namely $B$, $V$ and
 	$I$ are plotted against the $V$ magnitude for the three clusters under study. The dotted red line denotes the zero difference.}
	\label{comparision}
  \end{figure}

We also compared our photometry with the literature values shown in Fig. \ref{comparision}, the red dotted line denotes the zero difference. For this comparison, we used data from \citet{2008JKAS...41..147K}, \citet{1999JKAS...32....7A} and  \citet{2014NewA...26...77L} for the clusters NGC 1193, NGC 2355 and King 12 respectively. 
Our photometry looks analogous to the literature values in each filter. The mean differences are -0.06 $\pm$ 0.04, -0.001 $\pm$ 0.04, 
-0.07 $\pm$ 0.04 mag in $V$ filter, -0.09 $\pm$ 0.05, -0.06 $\pm$ 0.050, -0.07 $\pm$ 0.04 mag in $B$ filter and 0.008 $\pm$ 0.0, -0.03 $\pm$ 
0.04, -0.09 $\pm$ 0.05 mag in $I$ filter for the clusters NGC 1193, NGC 2355 and King 12, respectively.

\section{Photometric and Kinematical data from Gaia DR3} \label{sec:kin}

Gaia, a space observatory of the European Space Agency, unleashed its third data release
(Gaia DR3) on 13 June 2022. Gaia DR3 provides full astrometric solutions
for 1.46 billion sources, including positions,
proper motion, and parallax for stars having magnitude $3 \leq G \leq 21$.
Gaia DR3 also includes photometric data in $G$, $G_{BP}$ and $G_{RP}$
filters for 1.806 billion, 1.54 billion, and 1.55 billion stars, respectively, 
and radial velocity data for 33 million stars \citep{2021A&A...649A...1G, 2021A&A...649A...6G}.
Since the field of view of the observed clusters using the 104-cm Sampurnanand telescope
is small, we took advantage of Gaia DR3 data with more extensive spatial coverage. We used photometric data
in three Gaia bands which are $G$, $G_{BP}$, and $G_{RP}$, for a field having
30 arcmin diameter to cover the potential extent of the clusters. In addition, kinematical data for these stars (proper motion, radial
velocity, and parallax) is taken from the Gaia DR3 catalogue. In this analysis,
we only incorporated the stars having a maximum proper motion error of 1 mas $yr^{-1}$.


\section{Selection of cluster members and mean proper motion} \label{sec:vpds}

\begin{figure*}
	\centering
    \includegraphics[width=5.8cm]{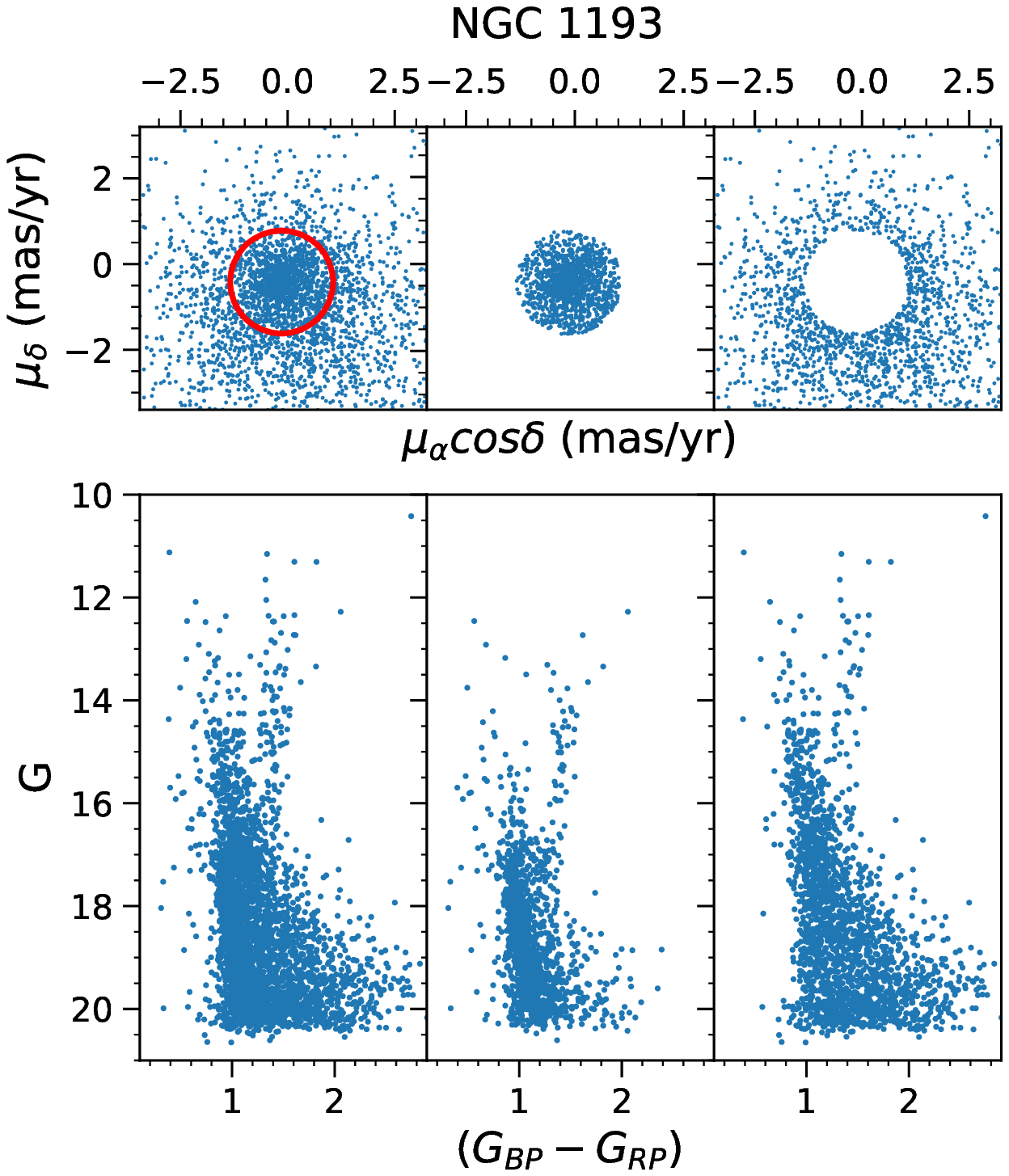}
    \includegraphics[width=5.8cm]{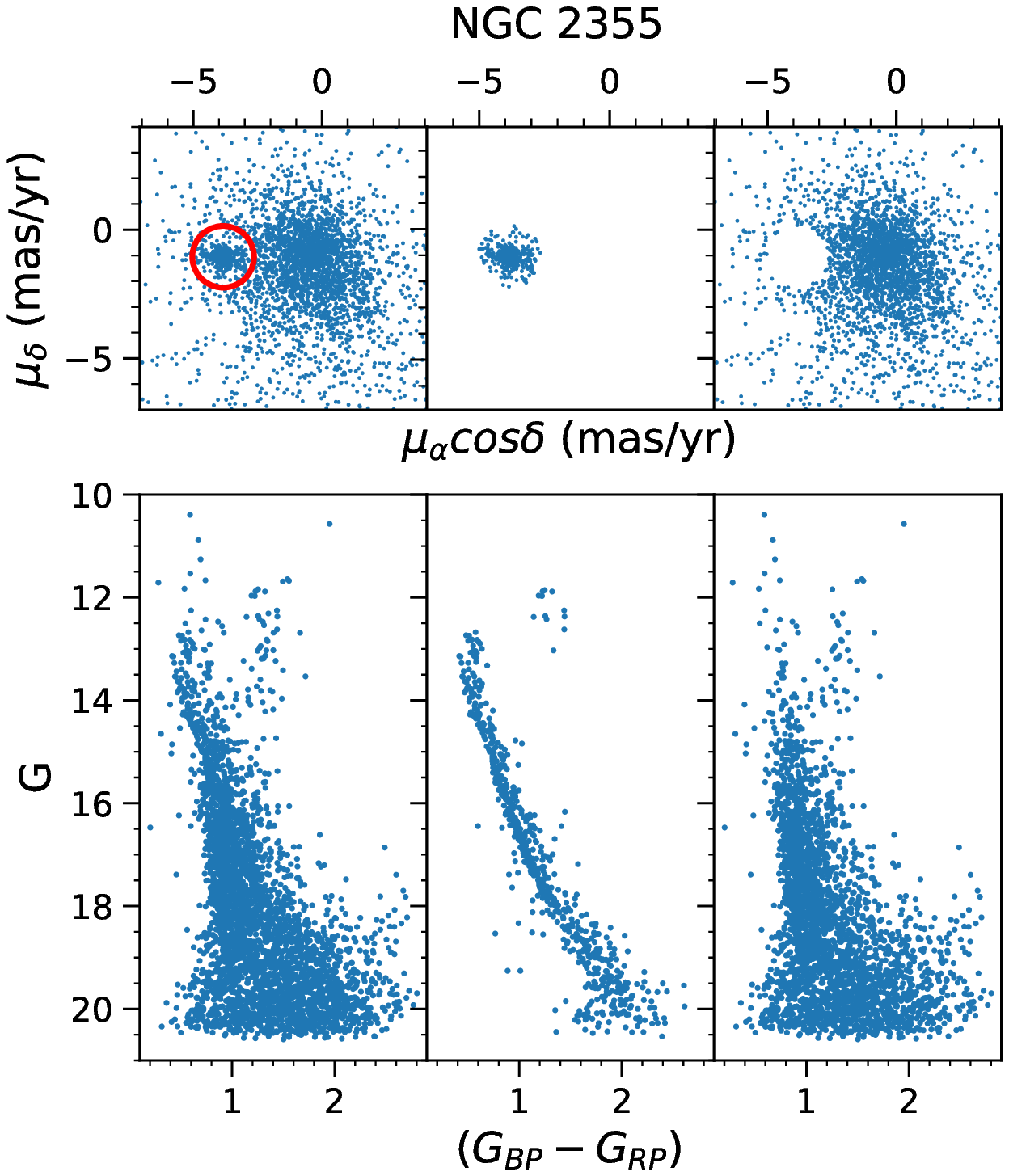}
    \includegraphics[width=5.8cm]{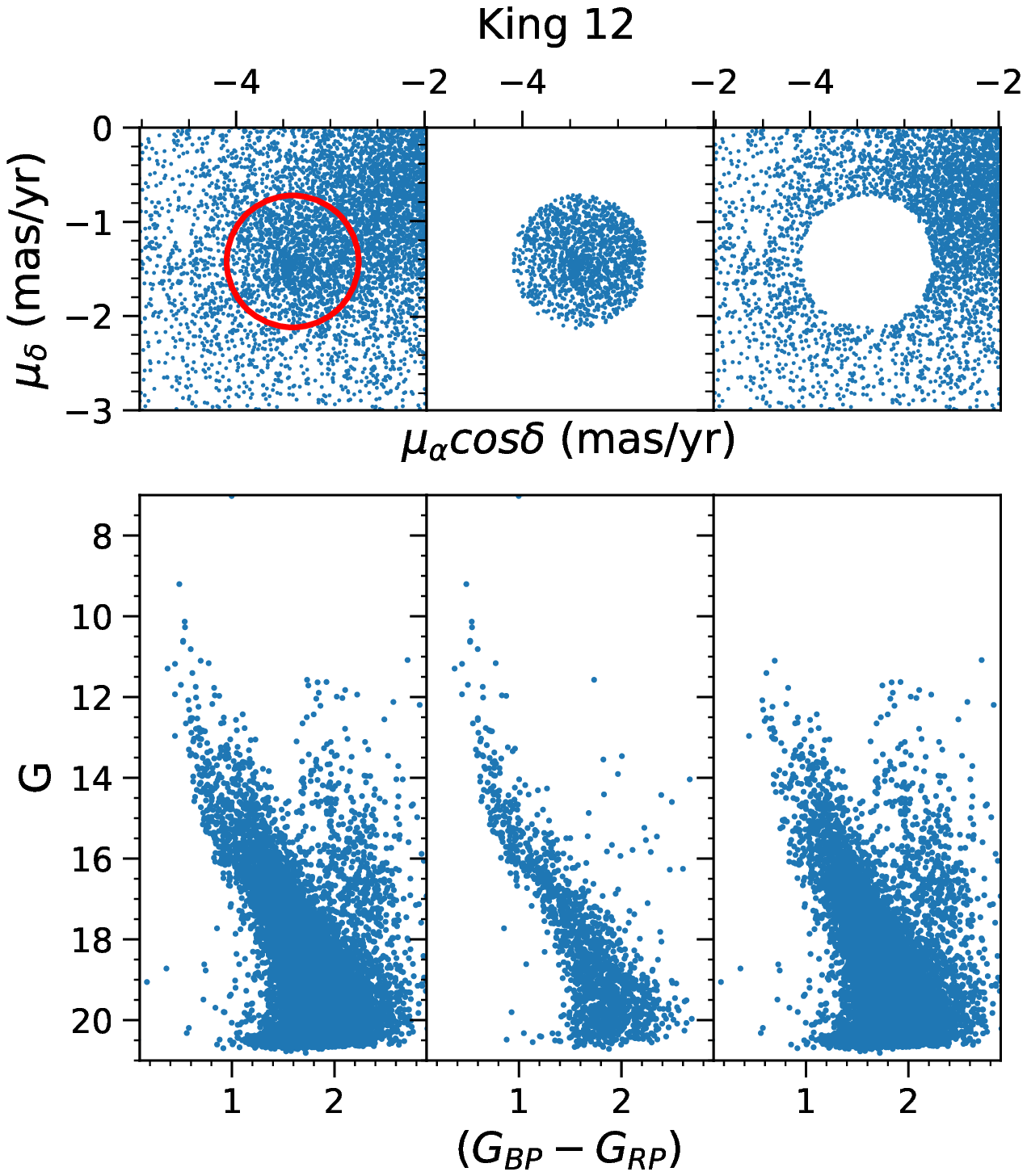}
	\caption{This figure shows the initial cluster member separation method using proper
    motion for the clusters NGC 1193, NGC 2355 and King 12. The top panels show the vector point
    diagrams and bottom panels show the respective colour-magnitude
    diagrams for the total stars, cluster members and field stars. 
 	}
	\label{vpd}
  \end{figure*}

Due to the proximity of the Galactic disc, the Galactic open clusters are highly degraded by the field
stars, which include both background and foreground stars. Hence the first measure of
our analysis is to decontaminate the sample of
cluster stars. For the present analysis, we selected the members of these clusters using the
proper motion from
the Gaia DR3 catalogue and then calculated their membership probability using
the method given by \citet{1998A&AS..133..387B} and described by \citet{2015A&A...584A..59S} and \citet{2017AJ....153..134S}. For the
initial separation of cluster stars from the field stars, the plot between proper motion 
in RA ($\mu_{\alpha}cos\delta$) and proper motion in DEC ($\mu_{\delta}$) known as
Vector Point Diagrams (VPDs) are used and shown in the top panels of
Fig. \ref{vpd}. In the first top panel
from the left, a compact clump of stars is differentiable from the
scattered field stars for all three clusters. The stars of these clumps
have similar proper motions and hence are selected as the most probable cluster
members by defining an eye-estimated circle (shown in red) around it.
The circle radius selection is performed to minimize the  contamination from the field
as well as save the faint members of the cluster.
The chosen circle radii are 1.2, 1.2, and 0.7 mas $yr^{-1}$ for NGC 1193,
NGC 2355 and King 12, respectively.
The lower first panel from the left shows the colour-magnitude diagram (CMD) of
total stars in the field of diameter 30$^\prime$.
The second panel shows VPD and CMD for the stars inside the red circle.
A precise sequence for all the clusters is visible in all three clusters.
The third panel shows VPD and CMD for the stars 
outside the circle in VPD.

\begin{figure*}
	\centering
    \includegraphics[width=13.5cm, height=6cm]{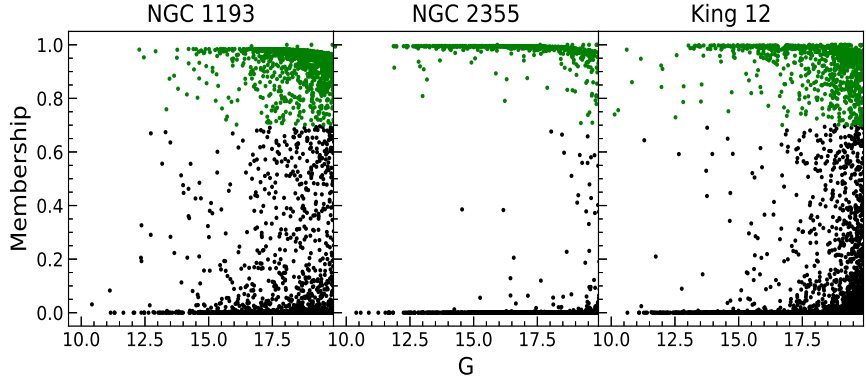}
    \caption{The membership of the stars in each cluster as a function of the $G$
     mag. The green points represent the stars selected for the analysis with a membership probability greater than 70 \%, and the black points are the rejected stars. 
   	}
	\label{membership}
  \end{figure*}

To calculate the membership probabilities of the stars in the clusters, we require the centres of the cluster as well as the field distributions. These are calculated by fitting the Gaussian
over the histograms of closely packed stars and other group of sparsely packed stars in VPD. The calculated proper motion centre
for the cluster distribution is (-0.16, -0.45), (-3.84, -1.05) and
(-3.32, -1.37) mas $yr^{-1}$, and the centre for the field distribution is (0.34, -1.63),
(-0.44, -1.20) and (-1.82, -0.69) mas $yr^{-1}$ for the clusters NGC 1193, NGC 2355, and King 12
respectively. The proper motion dispersion for the cluster distribution is calculated using the method given by \citet{1970AJ.....75..563J} and reported by \citet{10.1093/mnras/236.4.865} as 0.41, 0.18, and 0.09 mas $yr^{-1}$, and for the field
distribution, it is calculated as (3.81, 3.06), (3.23, 3.80), and (3.81, 3.06) mas $yr^{-1}$ for the
clusters NGC 1193, NGC 2355, and King 12, respectively. The membership probability
for the stars in the three clusters is plotted against their G
magnitude which is shown in the Fig. \ref{membership}. For this analysis, we only used the
stars having membership probability $\geq 0.7$ as shown with green colour and 
the black points signify the rejected stars.

We again selected the cluster members according to the parallax of the cluster members. 
We calculated the mean parallaxes
for the chosen stars above and then picked stars having
parallax less than $3 \sigma $ times the mean parallax.
So the final catalogue of cluster members consists of the stars
with membership probability $\geq 0.7$ and a parallax angle within
$3\sigma$ from the mean cluster parallax.

\begin{figure}
	\centering
	\includegraphics[width=9.5cm, height=5cm]{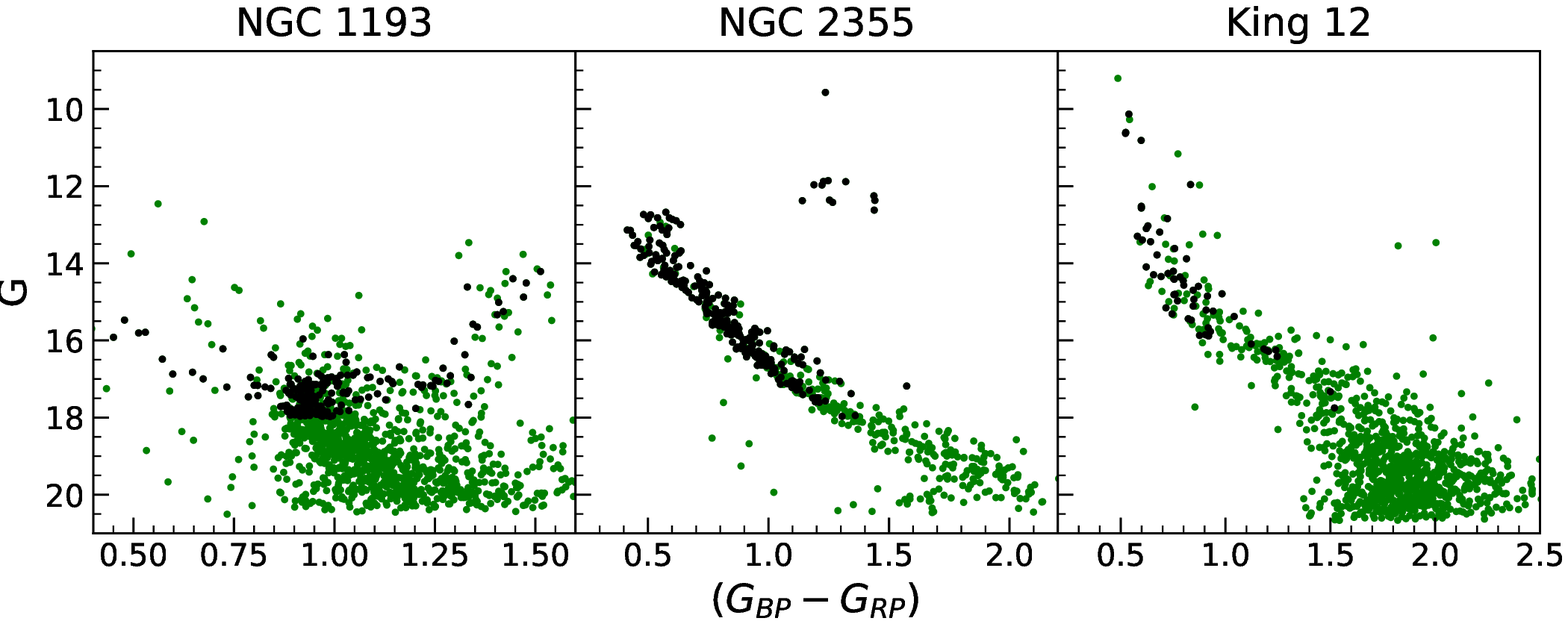}
	\caption{ Comparision of the members of the three clusters with the 
 cluster members given in \citet{2020A&A...640A...1C} having membership probability $\geq$ 0.7. The green points are from the present analysis, while the black points are from \citet{2020A&A...640A...1C}.
   	}
	\label{cg}
  \end{figure}

We also inspected the members of these clusters in \citet{2020A&A...640A...1C}. 
We found 215, 260, and 40 members having membership probability $\geq$ 0.7 for the
clusters NGC 1193, NGC 2355, and King 12, respectively. While in our analysis, we
identified 1432, 577, and 1107 stars having membership probability $\geq$ 0.7 for the 
clusters NGC 1193, NGC 2355, and King 12, respectively. 
The cluster members from both analyses are shown in Fig. \ref{cg}.

\begin{figure}
	\centering
	\includegraphics[width=4.2cm, height=4.5cm]{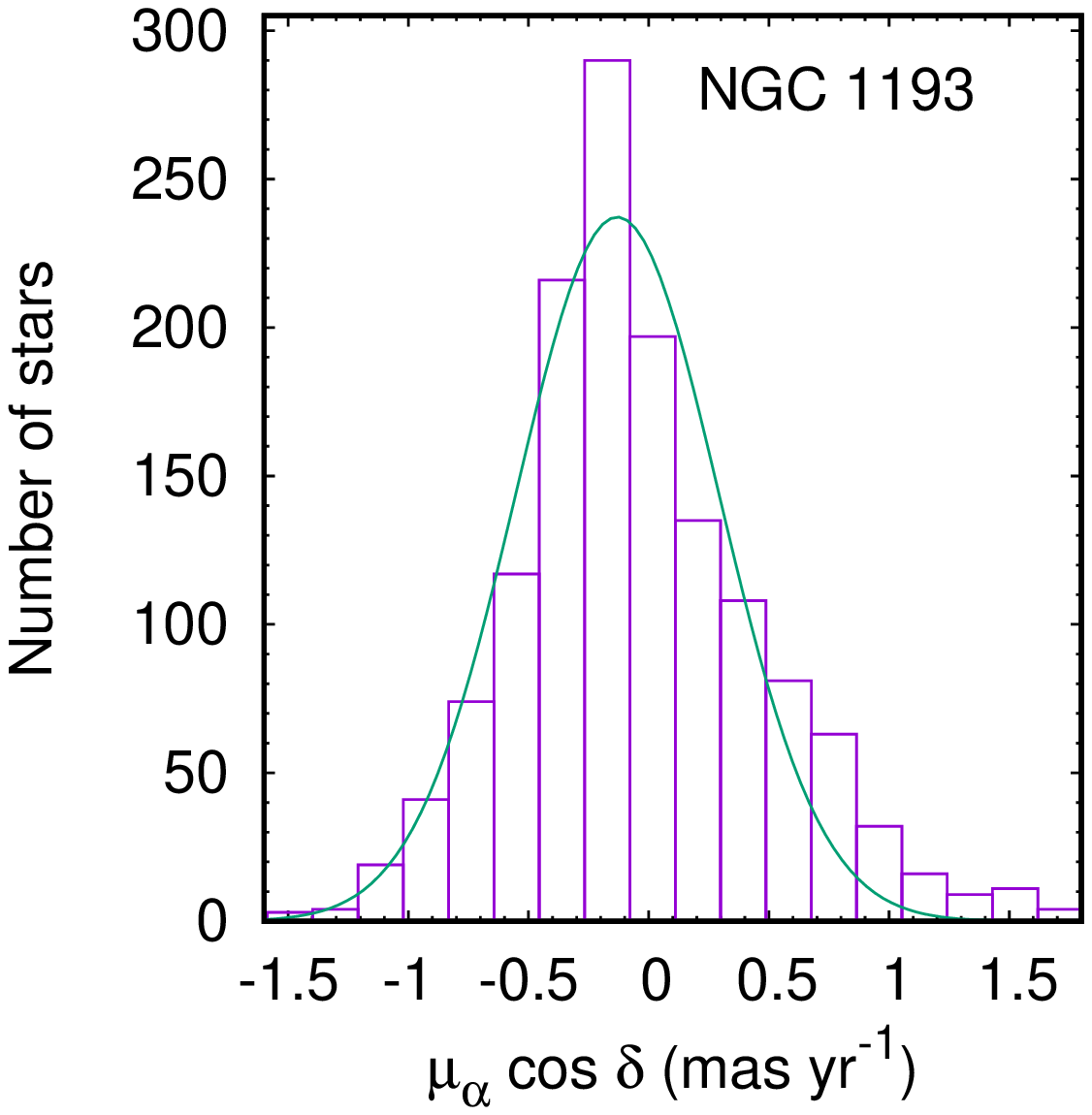}
	\includegraphics[width=4.3cm, height=4.5cm]{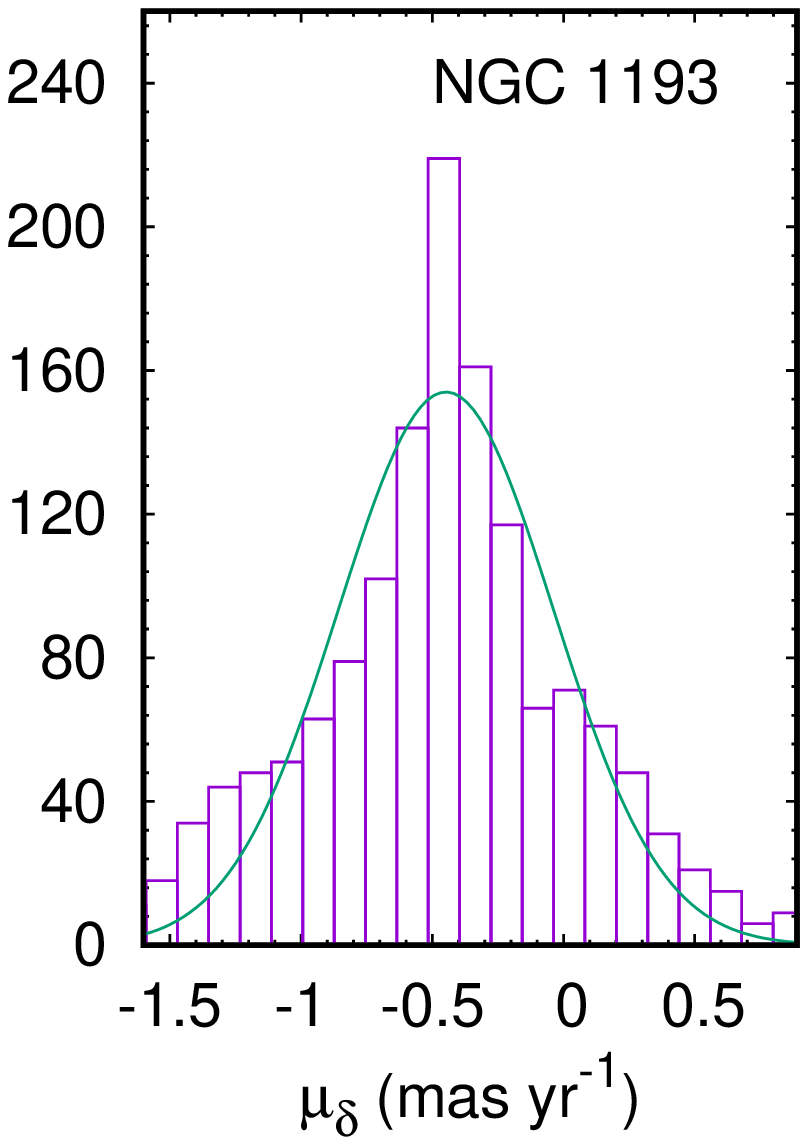}
	\includegraphics[width=4.2cm, height=4.5cm]{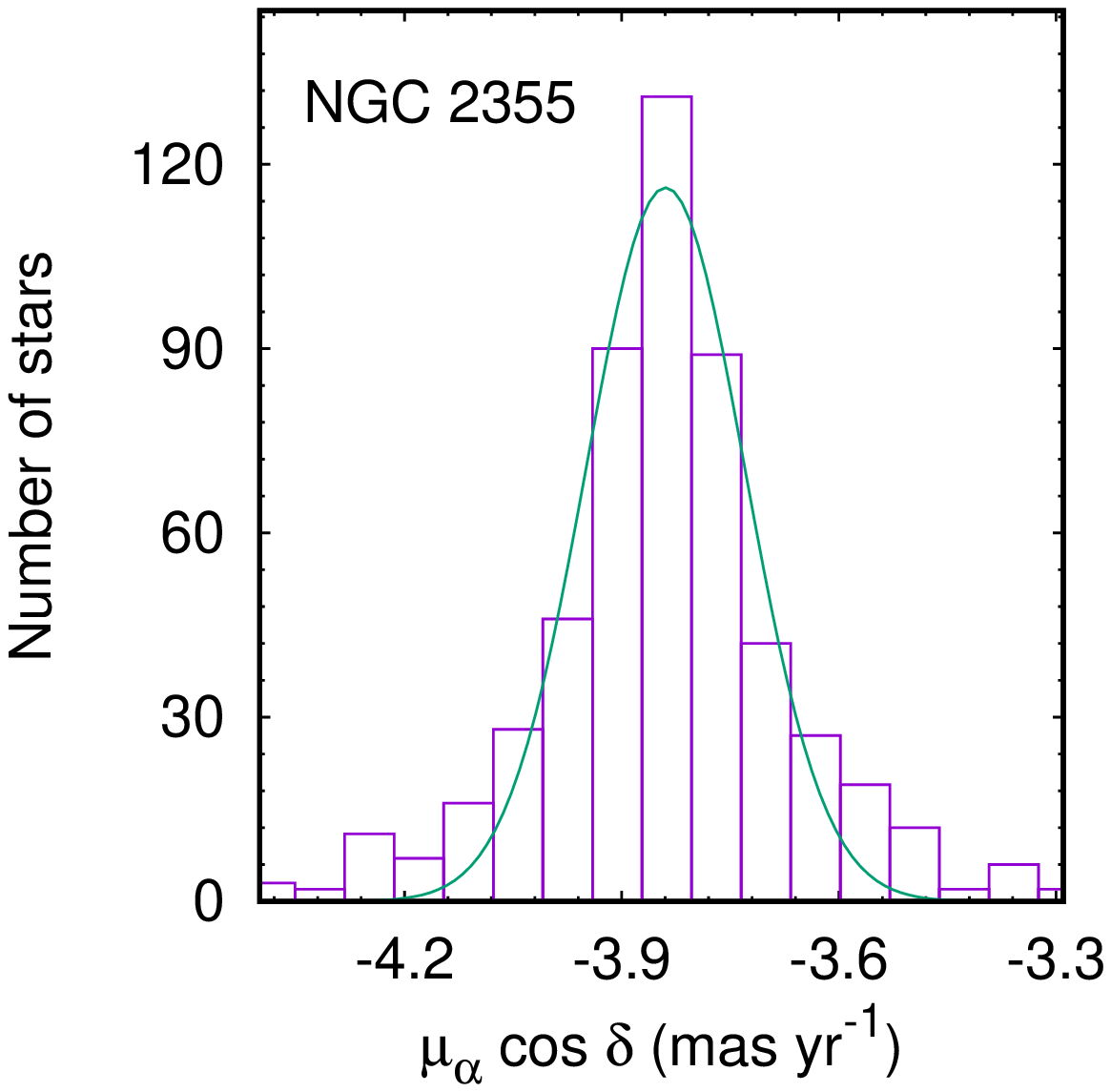}
	\includegraphics[width=4.3cm, height=4.5cm]{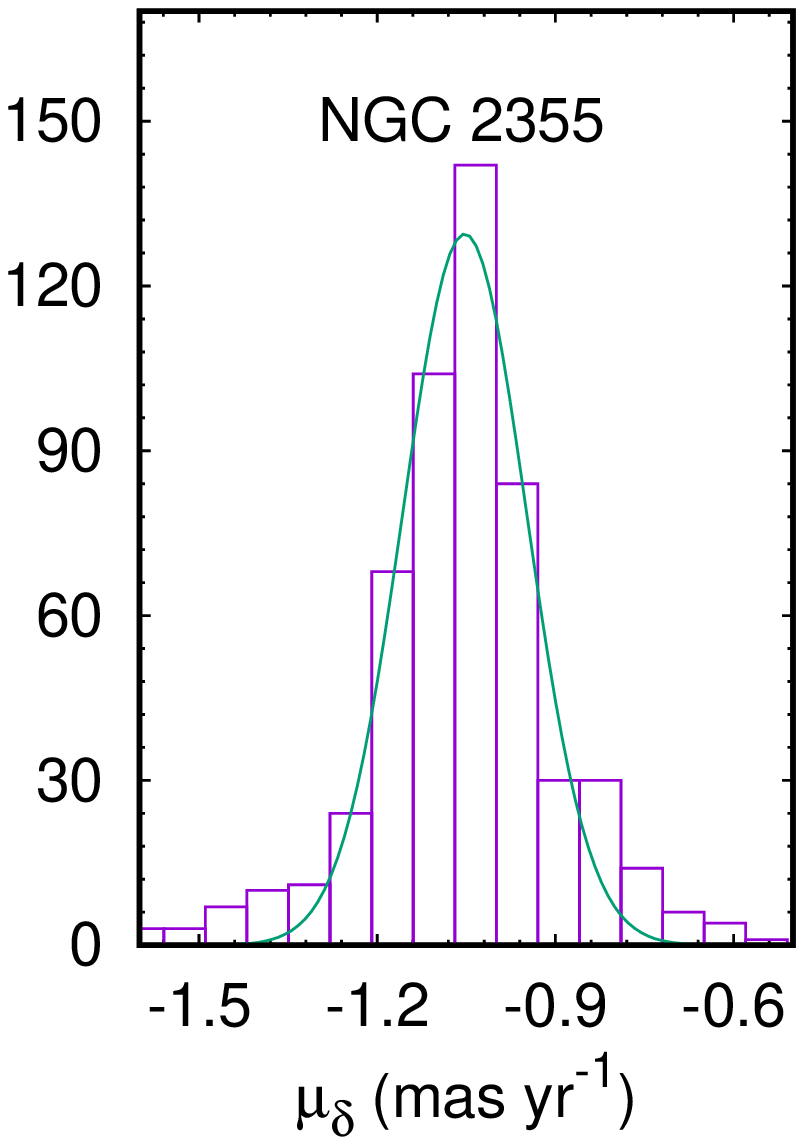}
	\includegraphics[width=4.2cm, height=4.5cm]{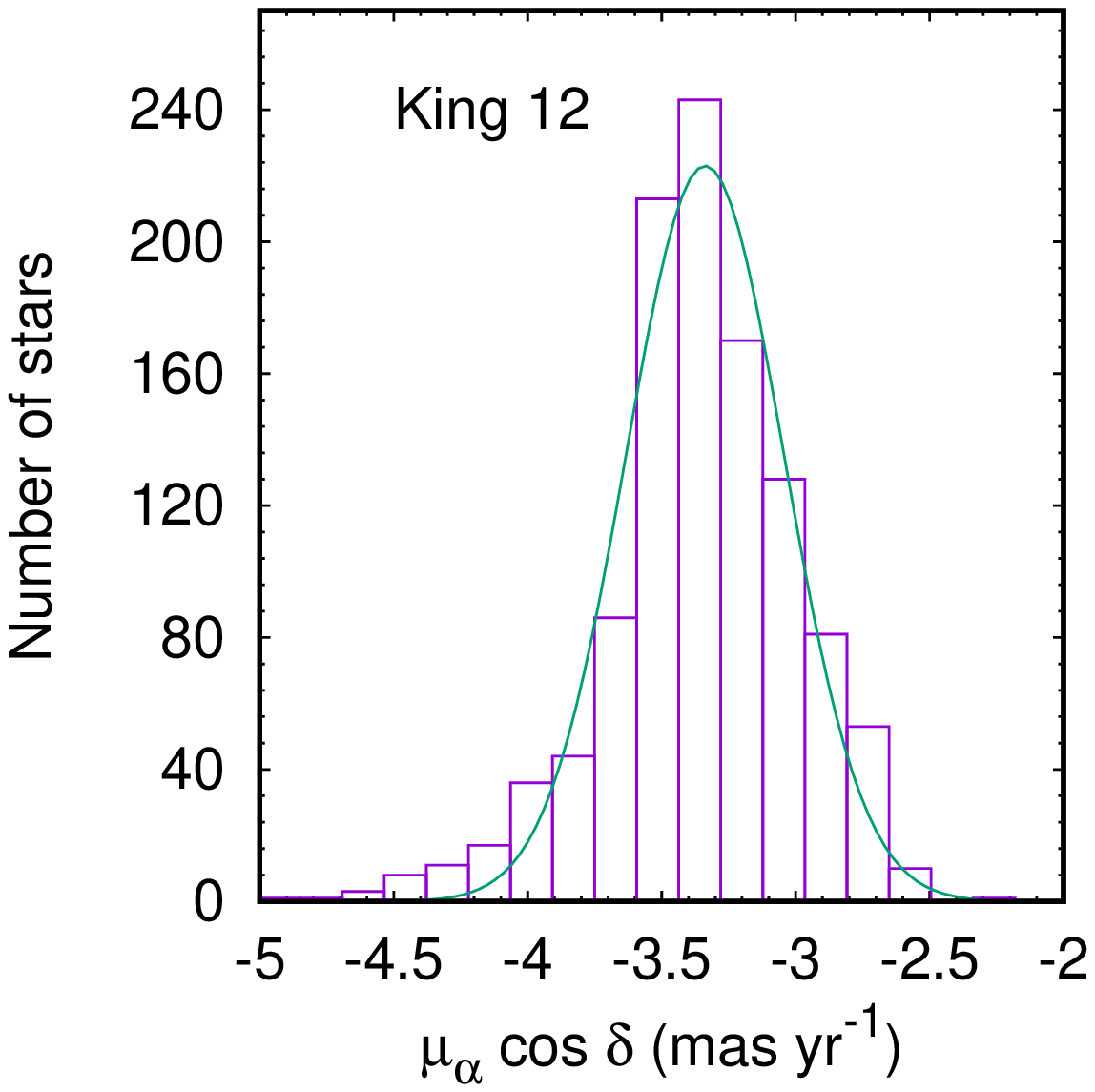}
	\includegraphics[width=4.3cm, height=4.5cm]{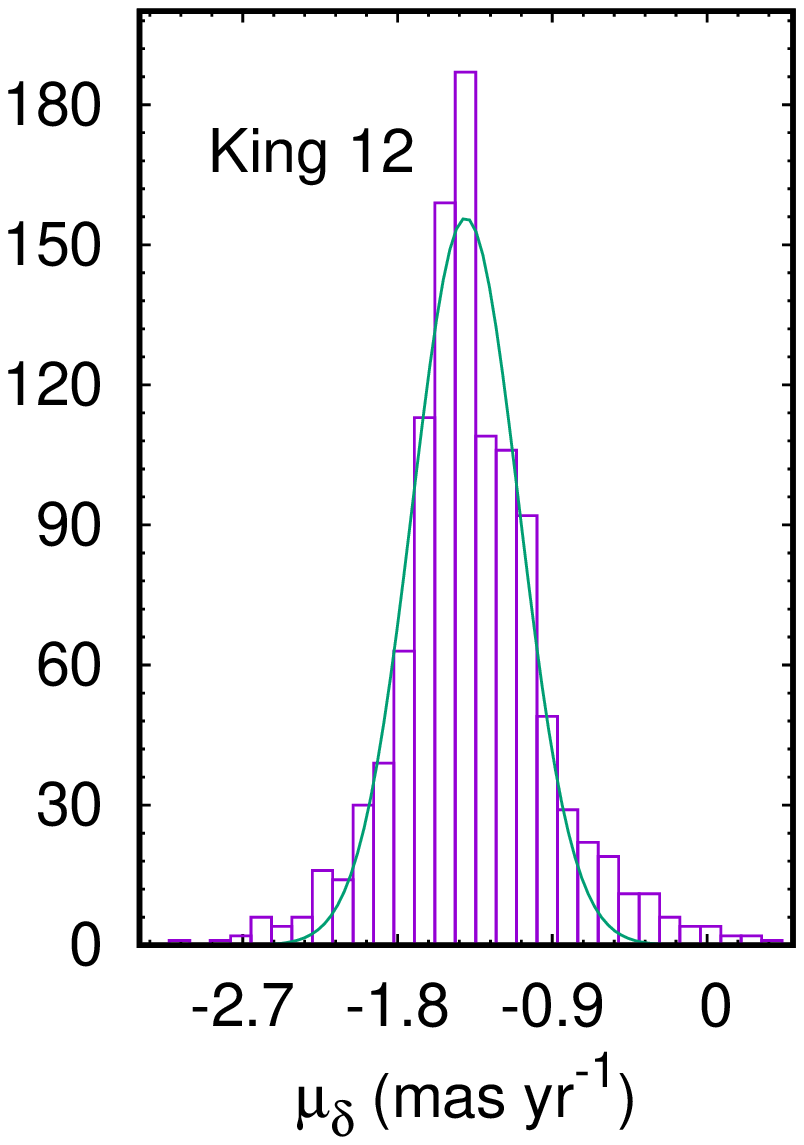}
	\caption{ Histograms in $\mu_{\alpha}cos\delta$ and $\mu_{\delta}$
	for the clusters NGC 1193, NGC 2355, and King 12.
 	The green curves represent Gaussian fitting.
   	}
	\label{pra}
  \end{figure}

The mean proper motions for the clusters are estimated by fitting a Gaussian over
histograms of $\mu_{\alpha}cos\delta$ and $\mu_{\delta}$ as shown in
Fig. \ref{pra}.
Hence, the mean proper motions of the clusters are estimated as
$ -0.13 \pm 0.03$ and $-0.48 \pm 0.04$ mas $yr^{-1}$ for NGC 1193;
$-3.84 \pm 0.01 $ and $-1.05 \pm 0.01$ mas $yr^{-1}$ for NGC 2355 and
$-3.34 \pm 0.02$,
and $-1.40 \pm 0.02$ mas $yr^{-1}$ for King 12.

\subsection{Distance estimation using parallax} \label{sec:parallax}

We utilised the approach given by \citet{2018A&A...616A...9L}, which suggests the estimation of the distance of a cluster using its average parallax value. Due to the associated errors in the parallax data from Gaia DR3, we calculated their
weighted mean using the probable cluster members selected above.
The mean parallax angles are $0.19 \pm 0.21$, $0.51 \pm 0.08 $ and $0.30 \pm 0.09$ mas
for the clusters NGC 1193, NGC 2355, and King 12, respectively. These values are corrected 
for the Gaia DR3 parallax zero-point (-0.021 mas)
offset suggested by \citet{2021A&A...654A..20G},

According to \citet{2015PASP..127..994B}, distance cannot be estimated by just
inverting the Gaia parallax. They suggested a probabilistic analysis to
determine distance and their uncertainties from the Gaia parallax data \citep{2015PASP..127..994B}.
By embracing the method illustrated by \citet{2018AJ....156...58B}, we determined the distances
of the clusters NGC 1193, NGC 2355, and King 12 as
$ 4.45 \pm 0.72 $, $1.97 \pm 0.15 $, and $3.34 \pm 0.29 $ kpc, respectively. For the clusters NGC 1193 and NGC 2355, these values are in good agreement 
with those in Table \ref{bpara}; however, for King 12, the distance obtained in the present analysis is higher than that given in WEBDA.


\section{Basic Parameters of the Clusters} \label{sec:ana}

\subsection{Cluster center and radius} \label{sec:cen}

\begin{figure}
	\centering
	\includegraphics[width=9.1cm]{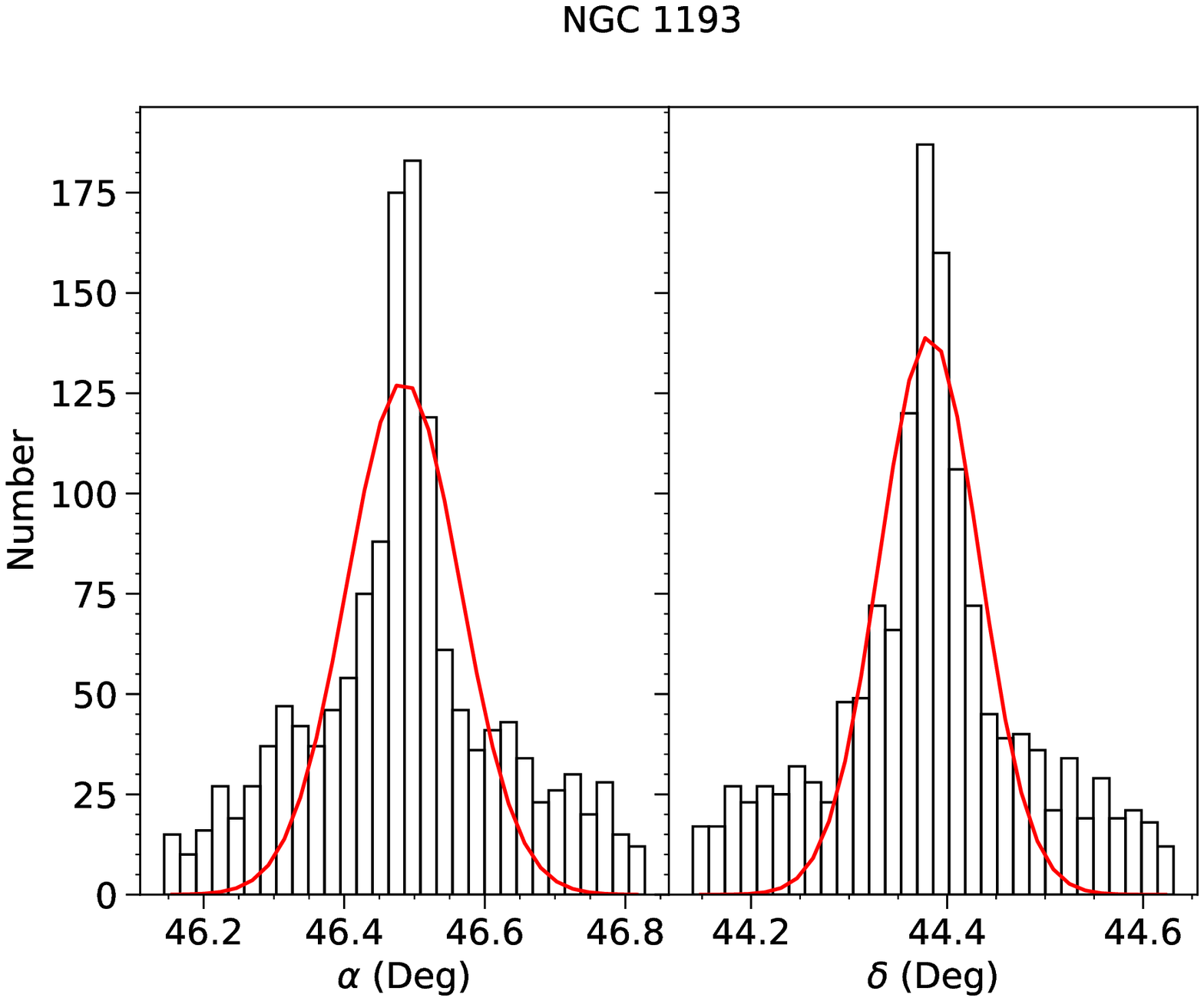}
	\includegraphics[width=9.1cm]{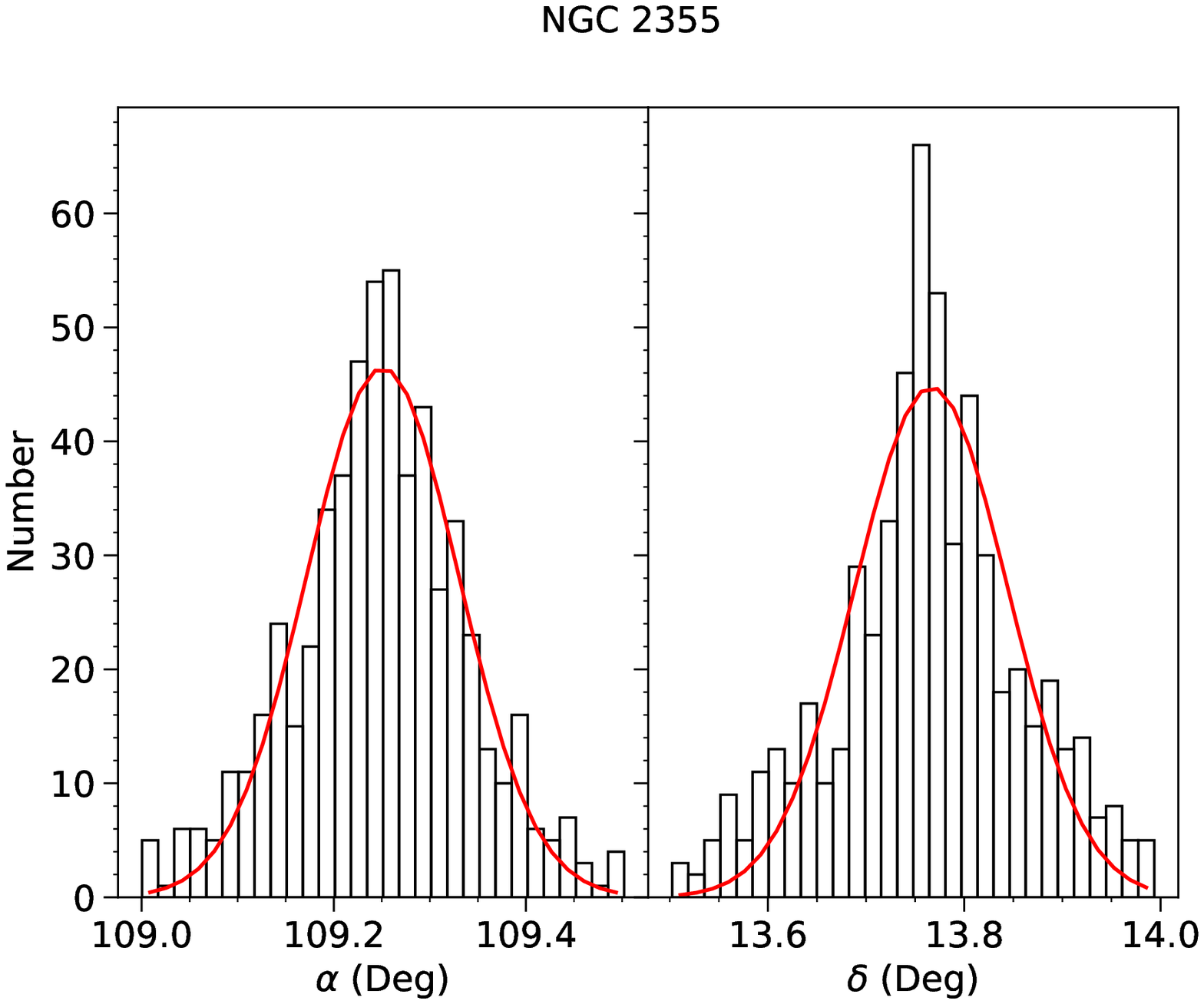}
	\includegraphics[width=9.1cm]{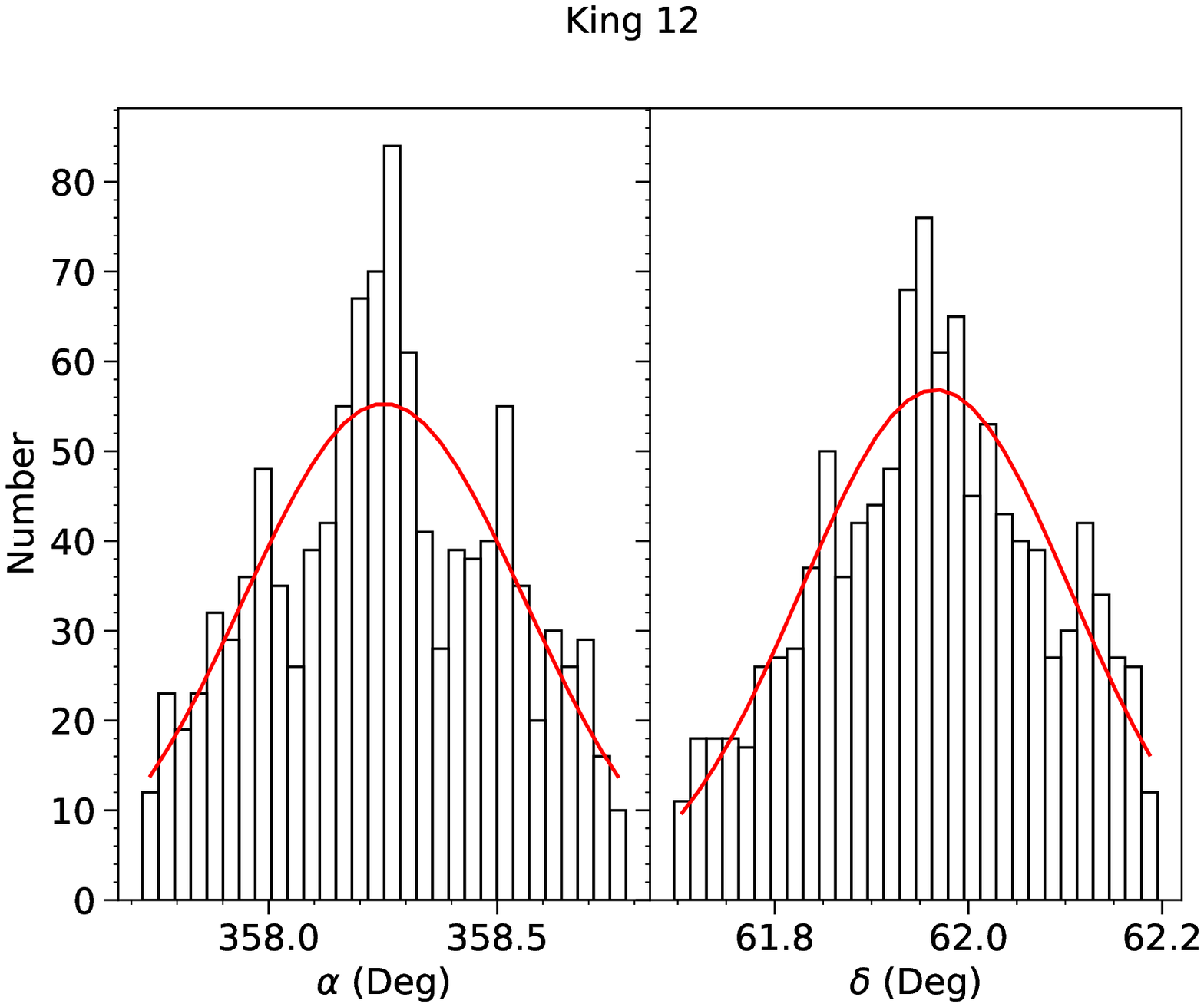}
	\caption{ Histograms in RA and DEC for the three clusters NGC 1193, NGC 2355, and King 12.
 	The red curves represent Gaussian fitting.
   	}
	\label{radec}
  \end{figure}

Generally, OCs are systems of thinly dispersed and loosely bound stars, but they maintain  
the highest stellar density at their centre. For meticulous determination of
cluster centre, we fitted Gaussian over histograms of right ascension (RA) and
declination (DEC), which is shown in Fig. \ref{radec}.
The estimated central coordinates for the clusters NGC 1193, NGC 2355, and King 12
are (46.484, +44.382), (109.251, +13.768), and (358.246, +61.967) degrees, respectively.

\begin{table*}
   \centering
    \caption{The values of central coordinates ($\alpha_{2000}$ and $\delta_{2000}$), parallax, distance ($d_{\odot}$),
    radial velocity ($V_{r}$) and proper motions ($ \mu_{\alpha} cos{\delta}$ and $ \mu_{\delta}$) determined in the current analysis
    except for radial velocity for King 12, for which there were not enough stars
    having radial velocity data in the Gaia DR3 catalogue. The radial velocity value
    mentioned in this table is taken from \citet{2007AN....328..889K} for King 12.
   }
   \begin{tabular}{rccccccc}
   \hline\hline
   Cluster   & $\alpha_{2000}$  & $\delta_{2000} $ &  Parallax & $ d_{\odot}$ & $V_{r}$ & $ \mu_{\alpha} cos{\delta}$ & $ \mu_{\delta}$   \\
   &  (deg) & (deg) & (mas) & (kpc)  & (km/sec) & (mas/yr) & (mas/yr)   \\
  \hline
   NGC 1193 & 46.484 & +44.382  & $ 0.19 \pm 0.21 $ & $4.45 \pm 0.72$ & $-82.64 \pm 3.35 $  &  $-0.13 \pm 0.03$ & $ -0.48 \pm 0.04 $	\\
   NGC 2355 & 109.251 & +13.768  & $ 0.51 \pm 0.08 $ & $1.97 \pm 0.15$ & $ +35.55 \pm 0.33 $ & $ -3.84 \pm 0.01 $ & $ -1.05 \pm 0.01 $ 	\\
   King 12  & 358.246 & +61.967   & $ 0.30 \pm 0.09 $ & $3.34 \pm 0.29$ &  $-38.00 \pm 6.80 $  & $ -3.34 \pm 0.02 $ & $ -1.40 \pm 0.02 $    	\\
  \hline
  \end{tabular}
  \label{vinp}
  \end{table*}

\begin{figure*}
	\centering
	\includegraphics[width=5.9cm,height=4cm]{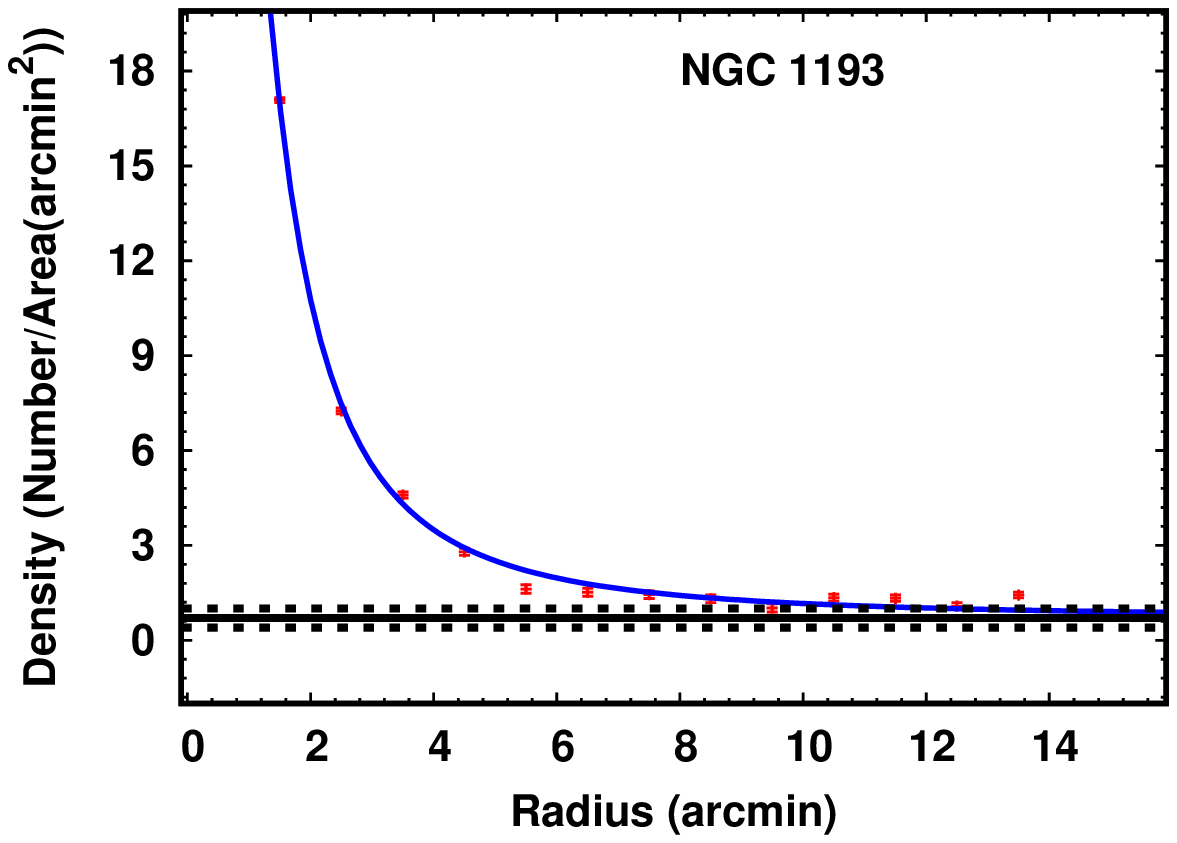}
	\includegraphics[width=5.9cm,height=4cm]{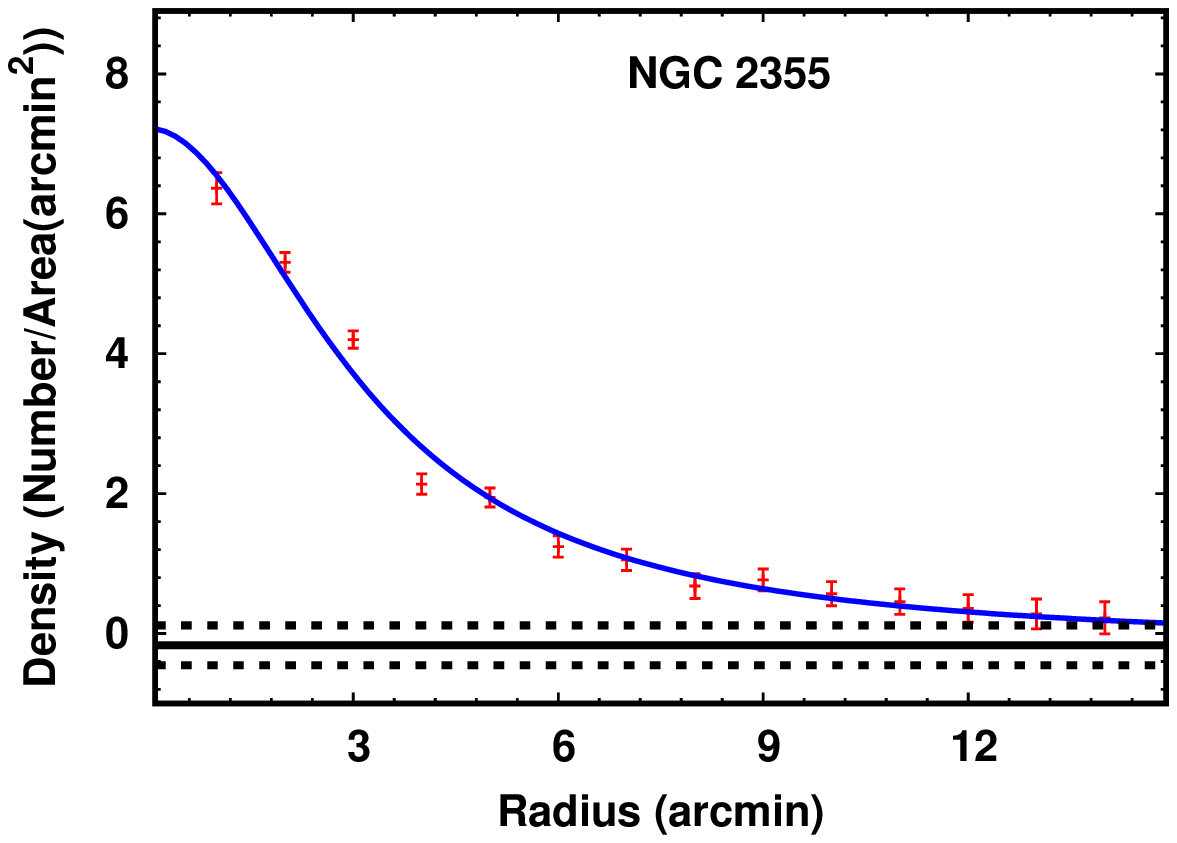}
	\includegraphics[width=5.9cm,height=4cm]{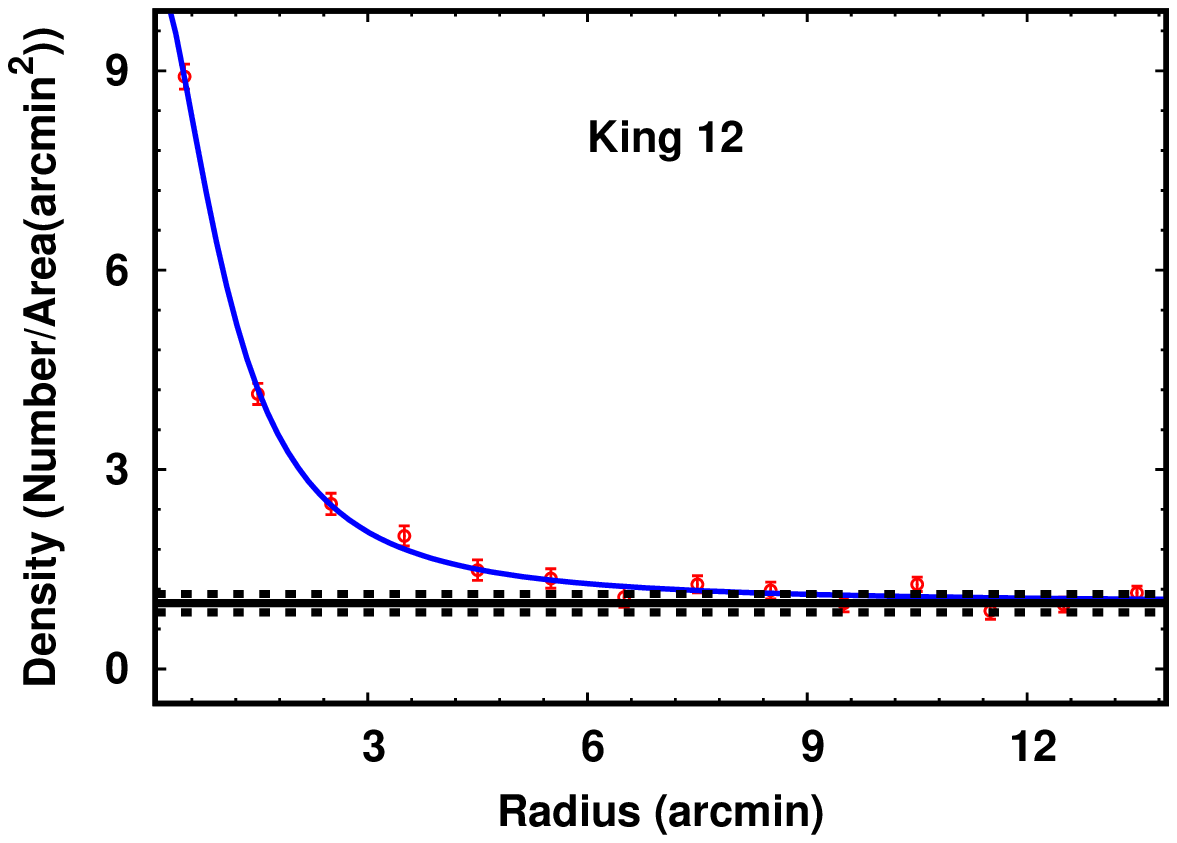}
	\caption{The number density of stars is plotted against the radius of stars
    and fitted the King's profile shown as a blue curve to calculate the
    structural parameters of the clusters. The horizontal continuous and
    dashed lines depict the background density and its error, respectively.
   	}
	\label{rdp_f}
  \end{figure*}
  
We have analyzed data of a larger area for each cluster as compared to their
visual size cited in the literature so that we can determine the precise sizes of the clusters and also
find the existence of corona.
To accomplish this goal, we plotted the radial surface density profile (RDP) for
each of the clusters under study.
We divided the cluster area into concentric
rings having the same thickness of $1^{\prime}$ and then computed the stellar density in each ring. The stellar densities are plotted in Fig. \ref{rdp_f}
against distance from the cluster centre.
We fitted King profile \citep{1962AJ.....67..471K} on the RDP of each cluster by adopting the least-square method, as shown 
by the blue curve in Fig. \ref{rdp_f}. 
The equation of the King profile is expressed as:   \\

\begin{equation}
f(r) = {f_{b} + \frac{f_{0}}{1+(r/r_{c})^{2}}}
\end{equation}
  
where $f_{b}$
is the background density, $f_{0}$ is the central density and $r_{c}$ is core radius. The core radius of a cluster can be defined as the
distance from the cluster centre where the stellar density becomes
half of the density at the cluster's centre.
The resultant parameters from this fitting for all the clusters are listed
in Table \ref{sp}. By visual examination of this fitting, the radius of clusters
is determined as 5 $\pm$ 0.4, 12 $\pm$ 0.3, and 6.5 $\pm$ 0.5 arcmin
for the clusters NGC 1193, NGC 2355, and King 12, respectively and there is no signature of corona in any cluster. We found that the central 
stellar density is more pronounced for cluster NGC 2355 than the other clusters. Hence the 
chances for survival of NGC 2355 are more as compared to 
NGC 1193 and King 12 based on their higher central concentration. The forthcoming sections will examine the factors impacting the clusters' survival.

\begin{table}
   \centering
	\small
   \caption{Structural parameters of the clusters, deduced from the fitting of King's profile on the
   radial density profile for the clusters under study. $r_{c}$, $f_{0}$ and $f_{b}$ are core radius, central star density and background star density, respectively}
   \begin{tabular}{lcccc}
   \hline\hline
  Cluster & Radius  & $f_{0}$ & $r_{c}$ & $f_{b}$   \\
   	& (arcmin) & ($\frac{number}{arcmin^{2}}$) & (arcmin) & ($\frac{number}{arcmin^{2}}$) \\
  \hline
  NGC 1193 & $05 \pm 0.4 $  &  $76.94 \pm 1.00 $ & $0.78 \pm 0.01 $  &  $0.70 \pm 0.10 $  \\
  NGC 2355 & $12 \pm 0.3 $  &  $07.39 \pm 0.29 $ & $3.16 \pm 0.24 $  &  $ -0.17 \pm 0.14 $  \\
  King 12  & $06.5 \pm 0.5 $  &  $06.03 \pm 0.31 $ & $1.84 \pm 0.17 $  &  $ 0.62 \pm 0.11 $  \\
  \hline
  \end{tabular}
  \label{sp}
  \end{table}


\subsection{Age, reddening, metallicity and distance of the clusters} \label{sec:iso}

The basic parameters of the clusters are determined by fitting
theoretical evolutionary isochrones from \citet{2017ApJ...835...77M}
on the colour-magnitude diagrams of the clusters. These isochrones are emanated from the evolutionary tracks computed with
PARSEC \citep{2012MNRAS.427..127B} and COLIBRI
\citep{2013MNRAS.434..488M} codes.
Details of these isochrones can be found in  \citep{2011SoPh..268..255C, 2019MNRAS.490.1383R}. In this section, we used the magnitude of stars in different filters
for a reliable determination of cluster properties through isochrone fitting. 
Along with data in the optical range from
Gaia DR3, and the observations from 104-cm ST, we also included near-IR data
from 2MASS; except for NGC 1193, we could not find enough member stars to fit isochrones. The uncertainty in the parameters
determined in this section is computed from the range in a good fit
of isochrones. The isochrones fitting and resulting parameters for each
cluster are discussed as follows:

\begin{figure*}
	\centering
	\includegraphics[width=4cm]{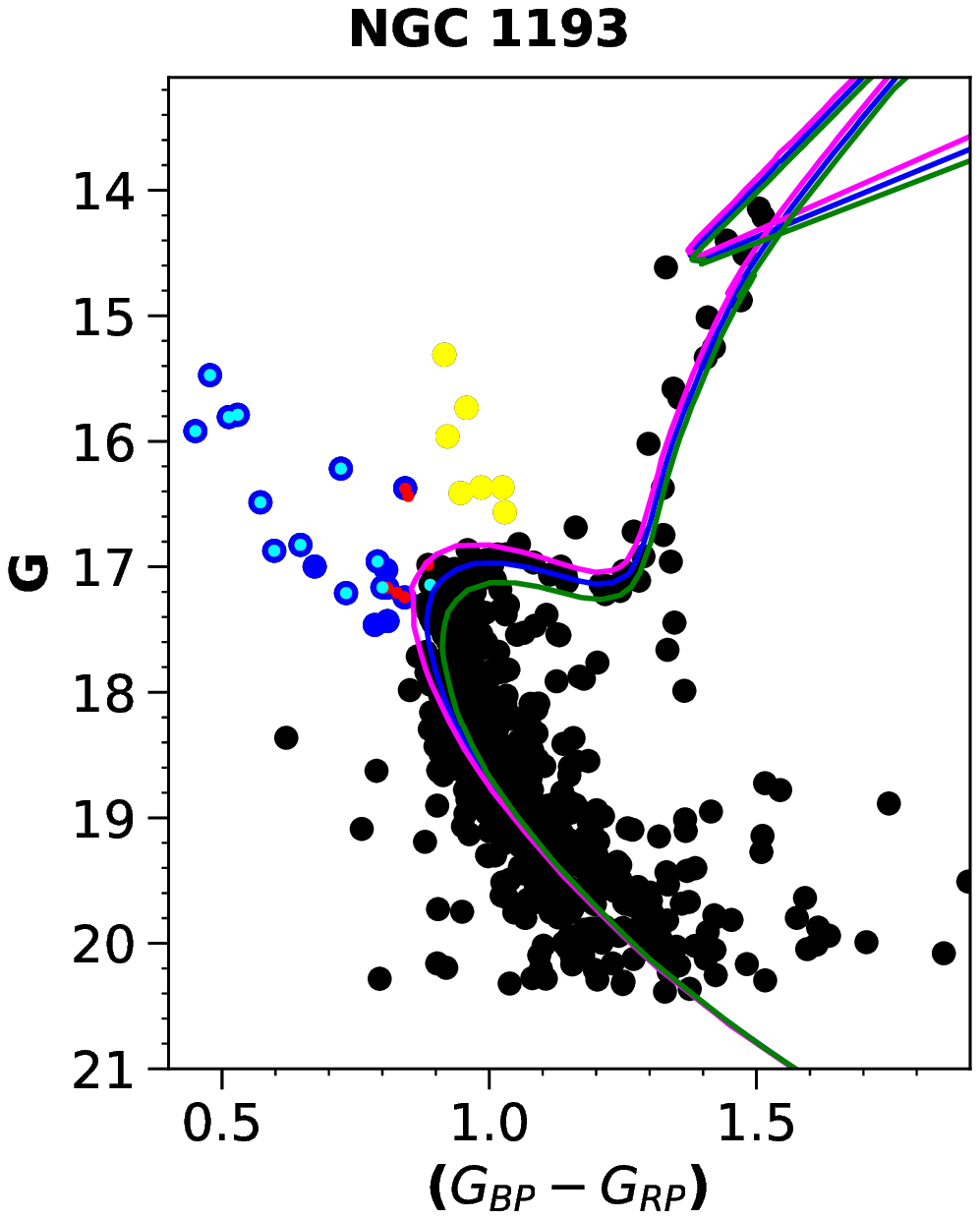}
	\includegraphics[width=8cm]{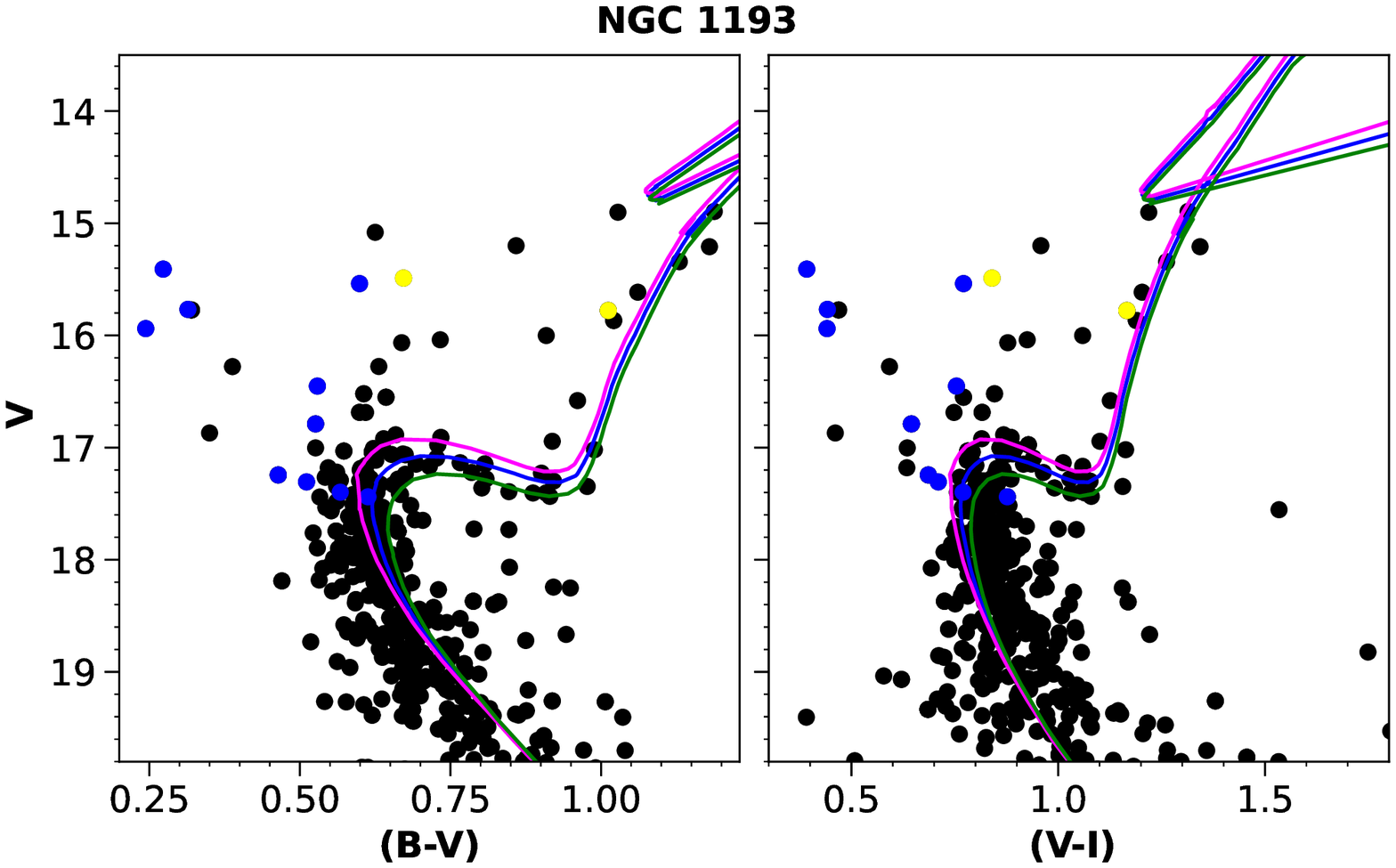}
    \caption{The color-magnitude diagrams in $(G, G_{BP}-G_{RP})$, $(V,B-V)$
    and $(V, V-I)$ frames for cluster 1193 using the cluster
    members. \citet{2017ApJ...835...77M} isochrones
    of log(age) 9.61, 9.68, and 9.75 are fitted on cluster sequence shown with green, blue and magenta colours,
    respectively. The blue straggler stars and yellow straggler stars identified in the present analysis 
    are shown with blue and yellow colours, respectively. The stars, which are taken from \citet{2021yCat..36500067R} and \citet{2021yCat..75071699J}, are shown by red and cyan colours, respectively.}
	\label{iso1193}
  \end{figure*}
  
{\it NGC 1193}: For this cluster, we plotted three colour-magnitude diagrams
i.e. $(G,G_{BP}-G_{RP})$, $(V,B-V)$, and $(V,V-I)$ and
shown in Fig. \ref{iso1193}. After performing several iterations with
isochrones of different ages and metallicity, we found the best fitted
isochrones of $Z= 0.008$ and age ( $4.79 \pm 0.58$ ) Gyr.
The fitting gives the distance modulus as $14.1 \pm 0.2$ mag. From this, 
the heliocentric distance of the cluster is found to be $5.26 \pm 0.20$ kpc.
This distance is proximate to the distance calculated by the mean parallax of the cluster.
We have also calculated the values of colour excess
$E(G_{BP}-G_{RP})$, $E(B-V)$ and $E(V-I)$ as 0.26, 0.16 and 0.20 mag respectively.

The stars which are bluer and brighter than the turn-off stars in the CMD of a star cluster 
are known as Blue straggler Stars (BSS), and evolved counterparts of these stars are 
known as Yellow Straggler stars (YSS) \citep{2021MNRAS.507.1699J}.
Based on their positions in the CMD in Gaia filters, we found seventeen blue straggler
stars with membership probability $\geq$ 95 \% and seven yellow straggler stars with membership probability $\geq$ 88 \% in this cluster, as shown in blue and yellow, 
respectively. We cross-matched these blue stragglers with the catalogue of blue stragglers by \citet{2021yCat..36500067R} and 
\citet{2021yCat..75071699J}. \citet{2021yCat..36500067R} and \citet{2021yCat..75071699J} catalogued
twelve and eighteen BSS, respectively, for this cluster. Among the twelve common stars between the
two catalogues, as shown by the cyan points in the cluster CMD, we confirm eleven stars as BSS, and the 
twelfth star lies on the cluster main-sequence hence not a BSS. The remaining six BSS claimed by \citet{2021yCat..75071699J} are also 
plotted in the same CMD with red colour, three of which we confirm as cluster BSS one lie on the 
main-sequence hence not BSS and the remaining two located beyond the cluster region hence are
field BSS. In addition, we found four more BSS
not compiled in the catalogues mentioned above.
The yellow stragglers found in our analysis are shown by yellow colour in the cluster CMD, and none of these YSS is
catalogued in these catalogues. 

\begin{figure*}
	\centering
	\includegraphics[width=4cm]{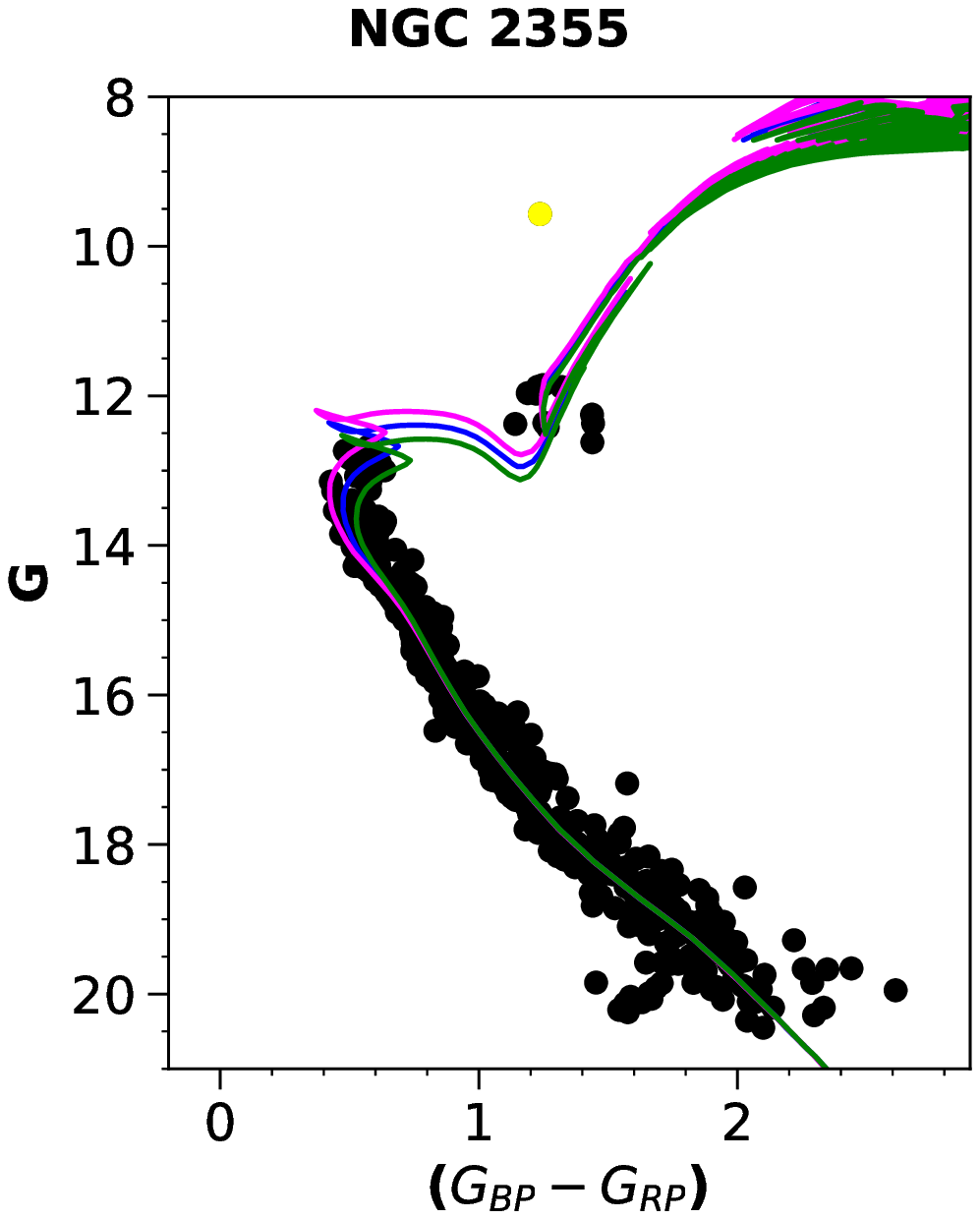}
	\includegraphics[width=8cm]{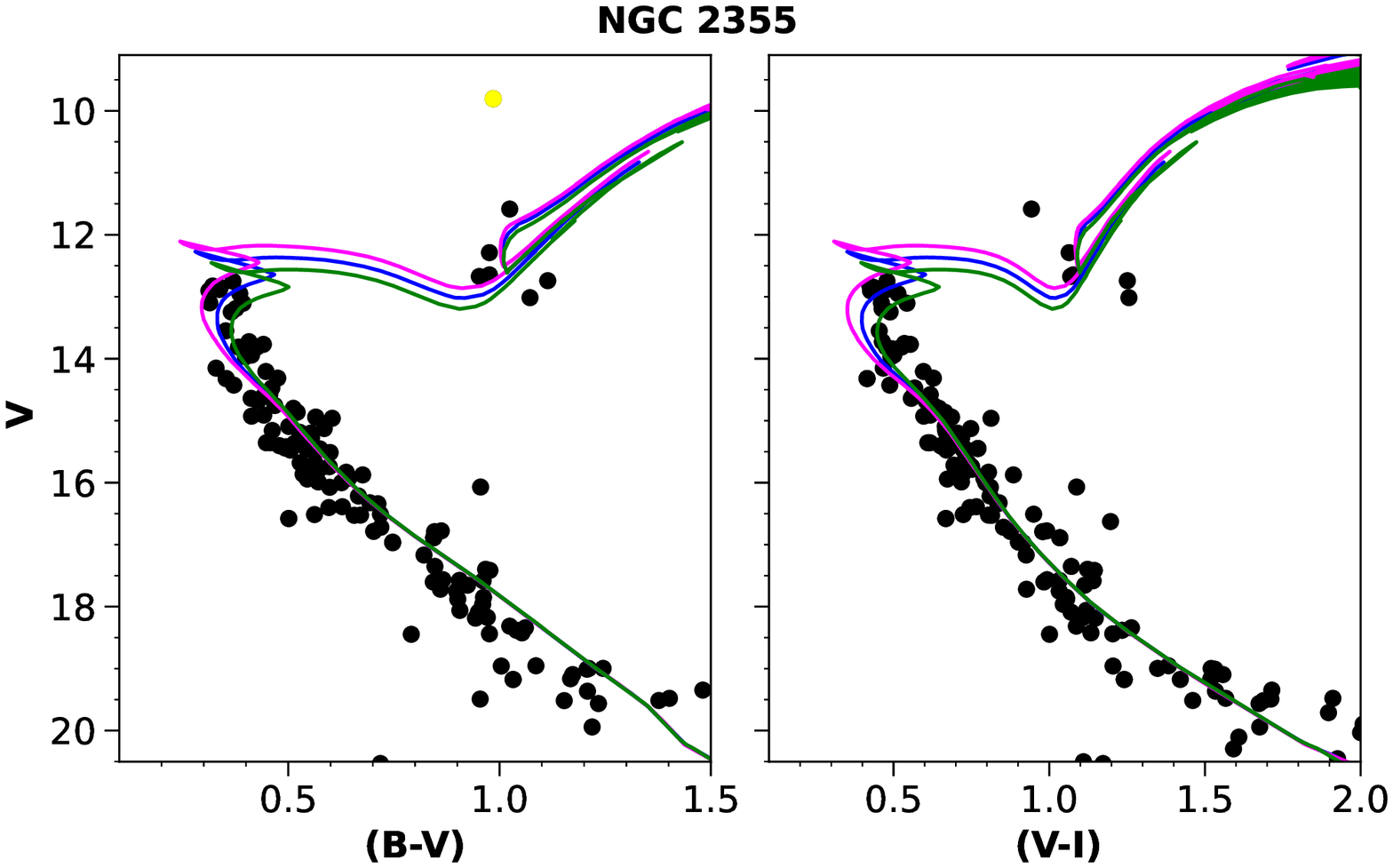}
	\includegraphics[width=8cm]{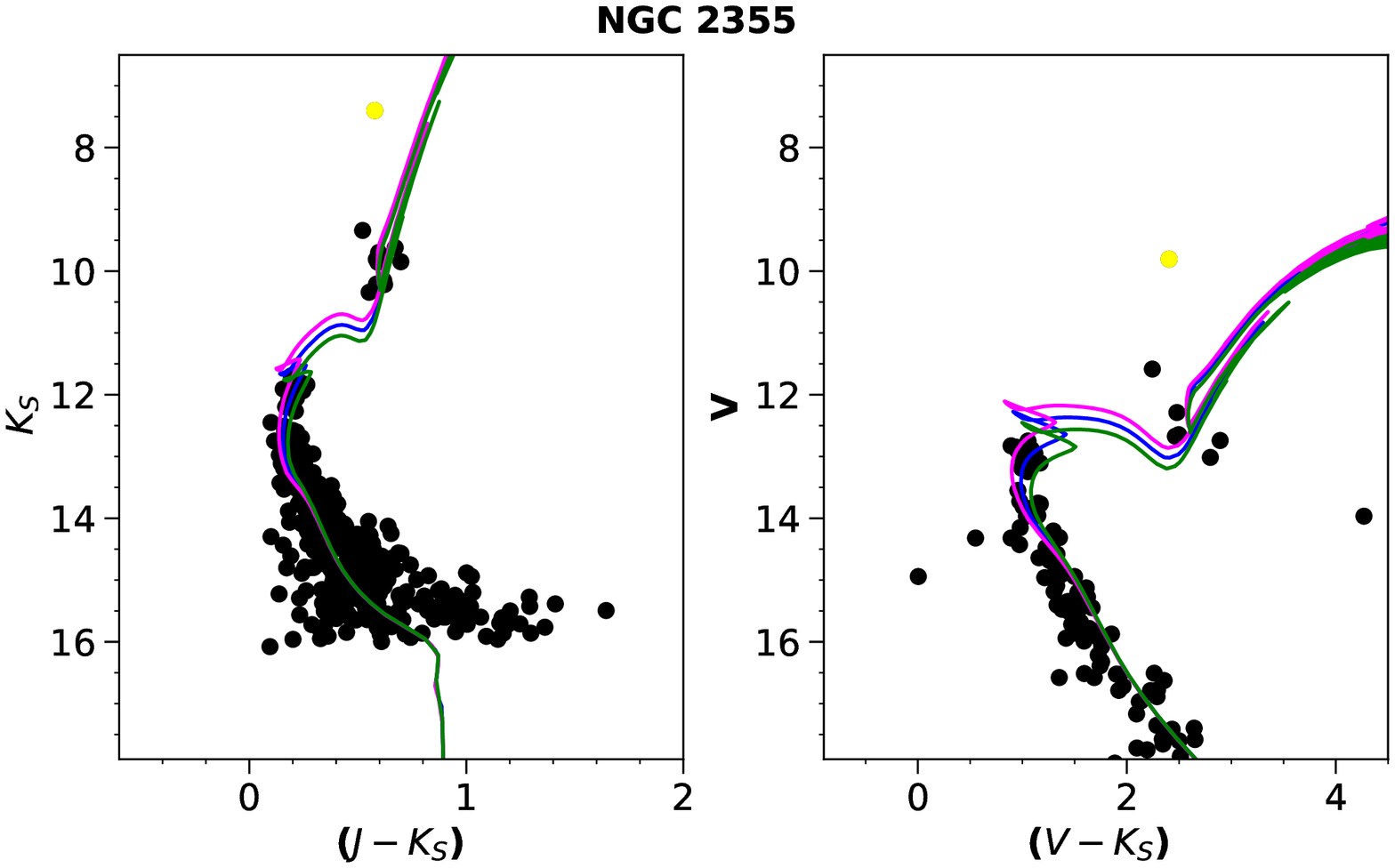}
	\includegraphics[width=8cm]{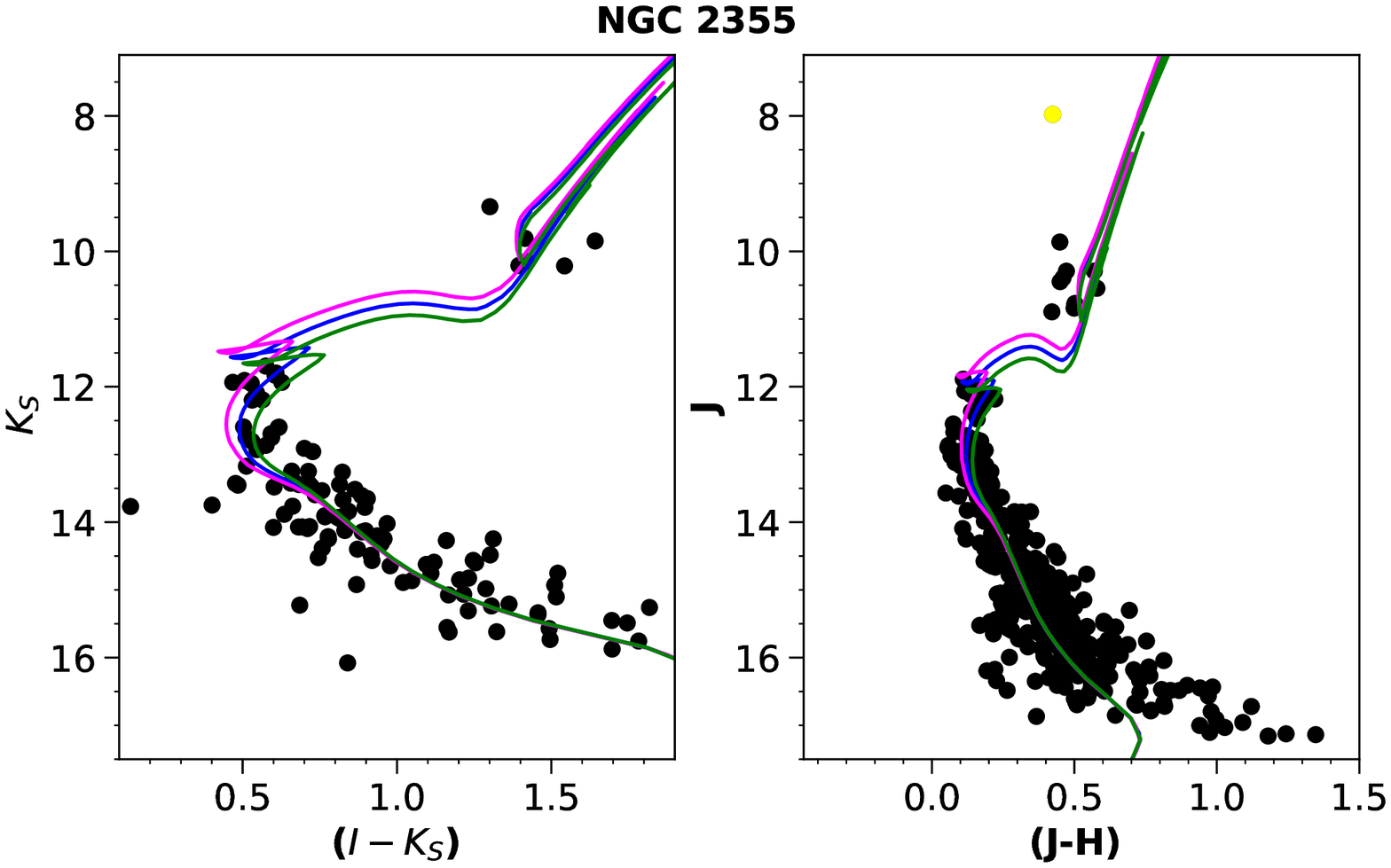}
	\caption{The color-magnitude diagrams in $(G, G_{BP}-G_{RP})$, $(V,B-V)$, $(V,V-I)$), $(K_{s},J-K_{s})$,
  	$(V,V-K_{s})$, $(K_{s},I-K_{s})$ and $(J,J-H)$ planes for the cluster NGC 2355
	with the best fitted \citet{2017ApJ...835...77M} isochrones
	of log(age) 9.05, 9.1, and 9.15 on
	the cluster sequence with green, blue and magenta colours,
    respectively. The star shown with yellow colour is a Yellow straggler.  
	}
	\label{iso2355}
  \end{figure*}
  
We also inspected the location of these BSS and YSS and found that all the BSS 
are located near to the cluster centre with a maximum radius of 1.7 arcmin except for one BSS, which is located at a distance of 4.5 arcmin from the centre of the cluster, while only three YSS are located close to the 
cluster centre (having a maximum distance of 1.5 arcmin), and four are located on the outskirts of the cluster with a maximum distance of 5 arcmin.
Since BSS are more massive than the MS stars hence the location of BSS
towards the cluster's centre might be a result of the mass segregation. According to \citep{2021MNRAS.507.1699J}, the low mass BSS found in the clusters older than 2 Gyr are 
most likely formed by mass transfer between the companion binary stars. Hence, we expect a mass transfer mode of
formation for BSS found in this cluster.

\begin{figure*}
	\centering
	\includegraphics[width=4cm]{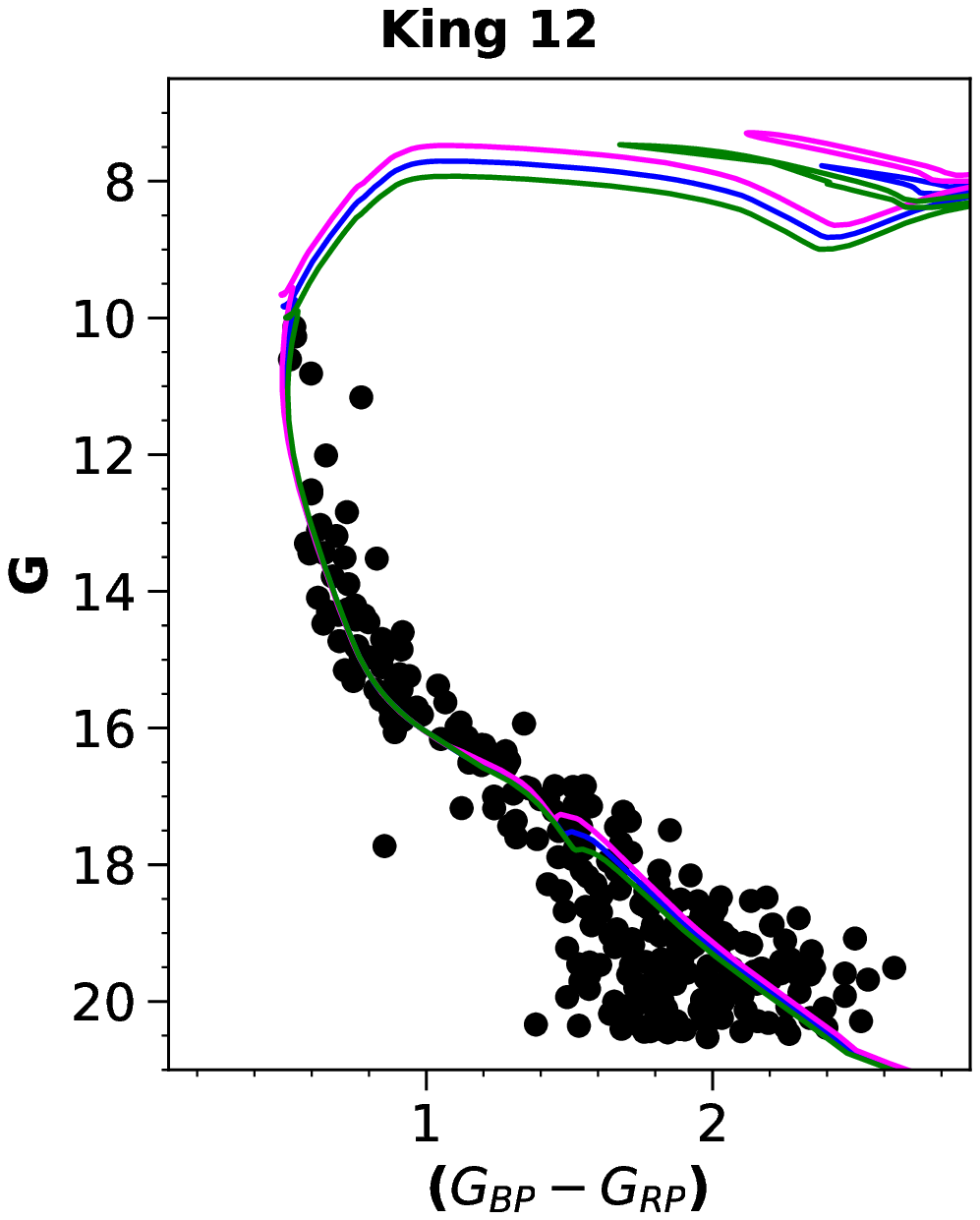}
	\includegraphics[width=8cm]{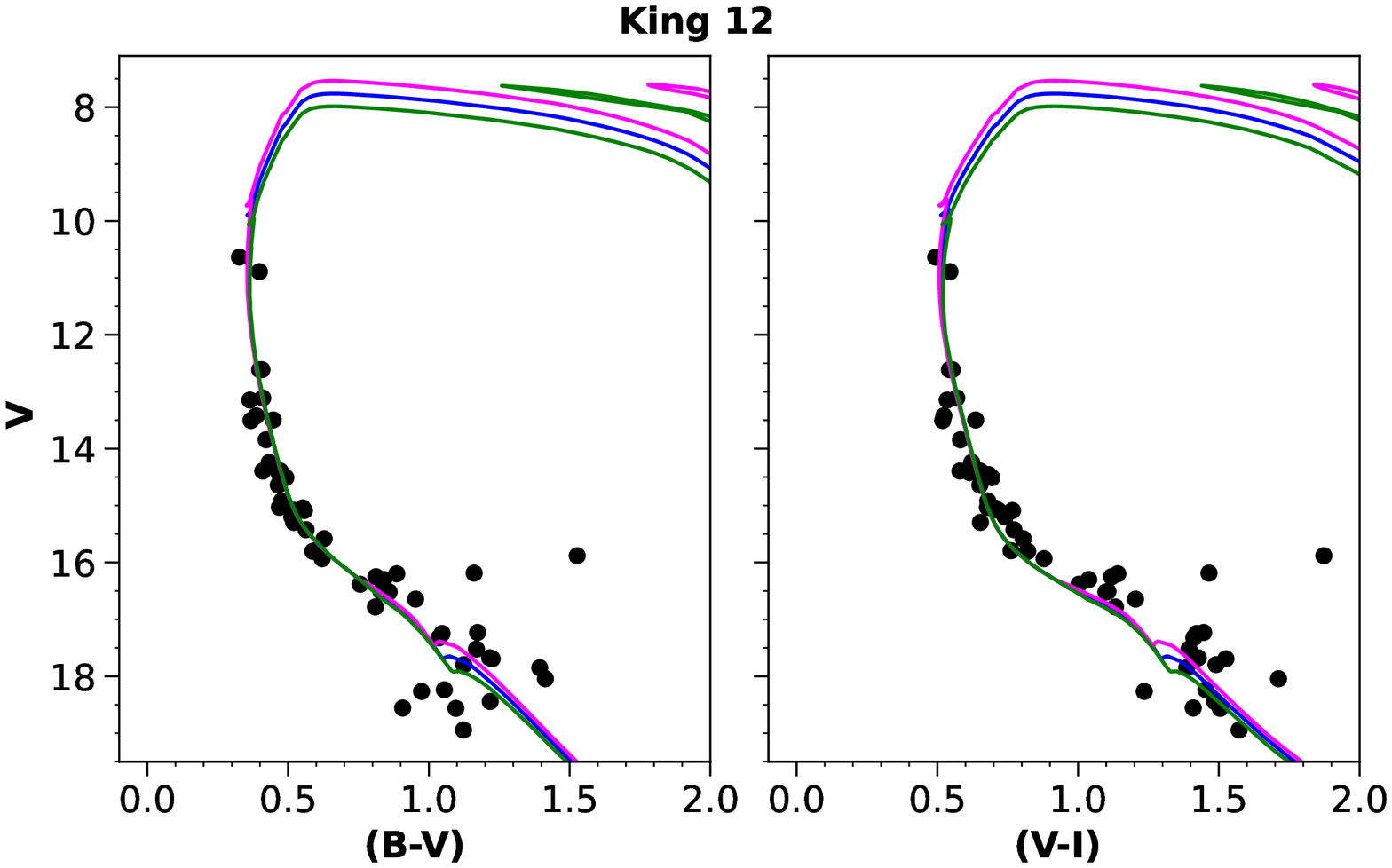}
	\includegraphics[width=8cm]{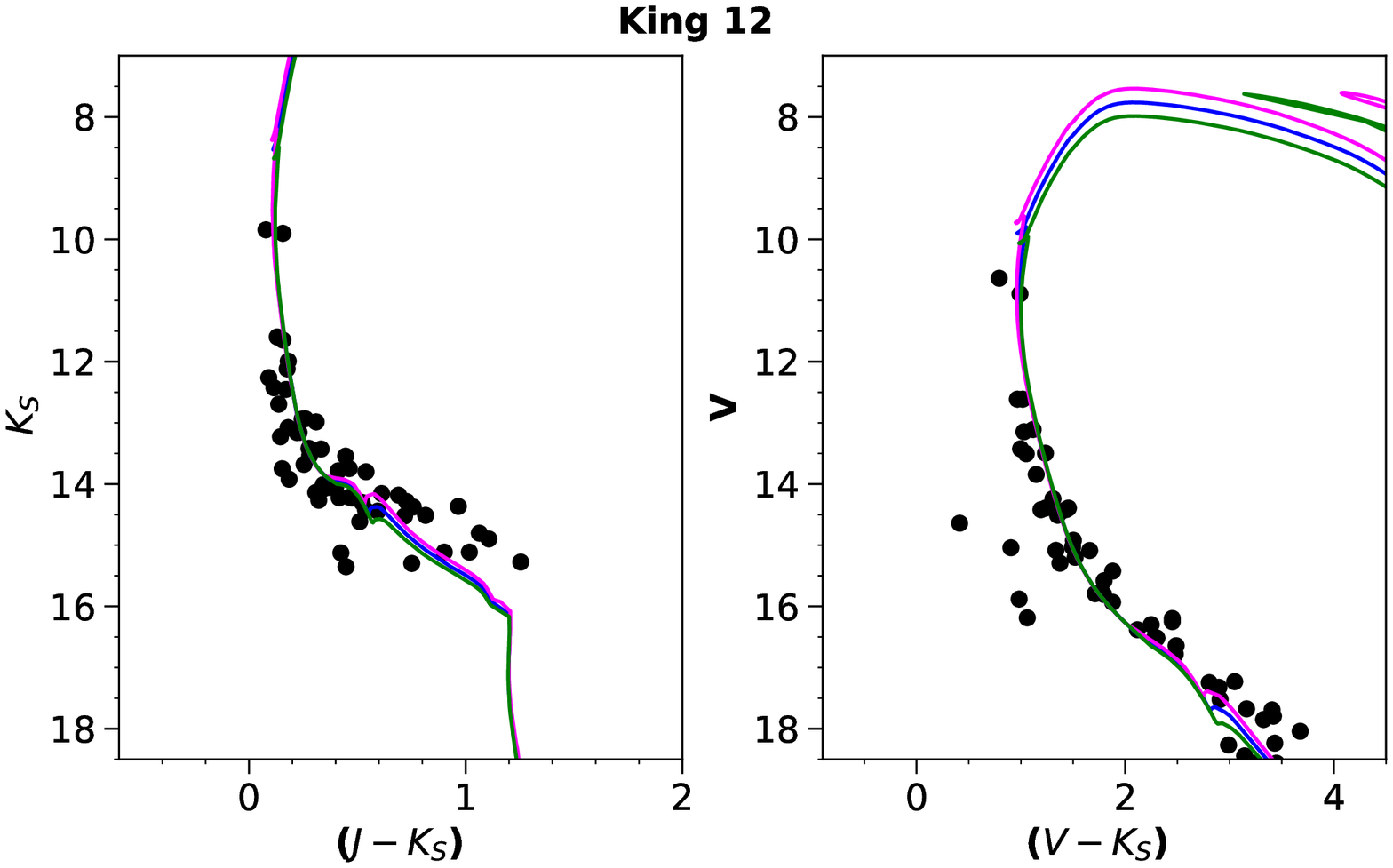}
	\includegraphics[width=8cm]{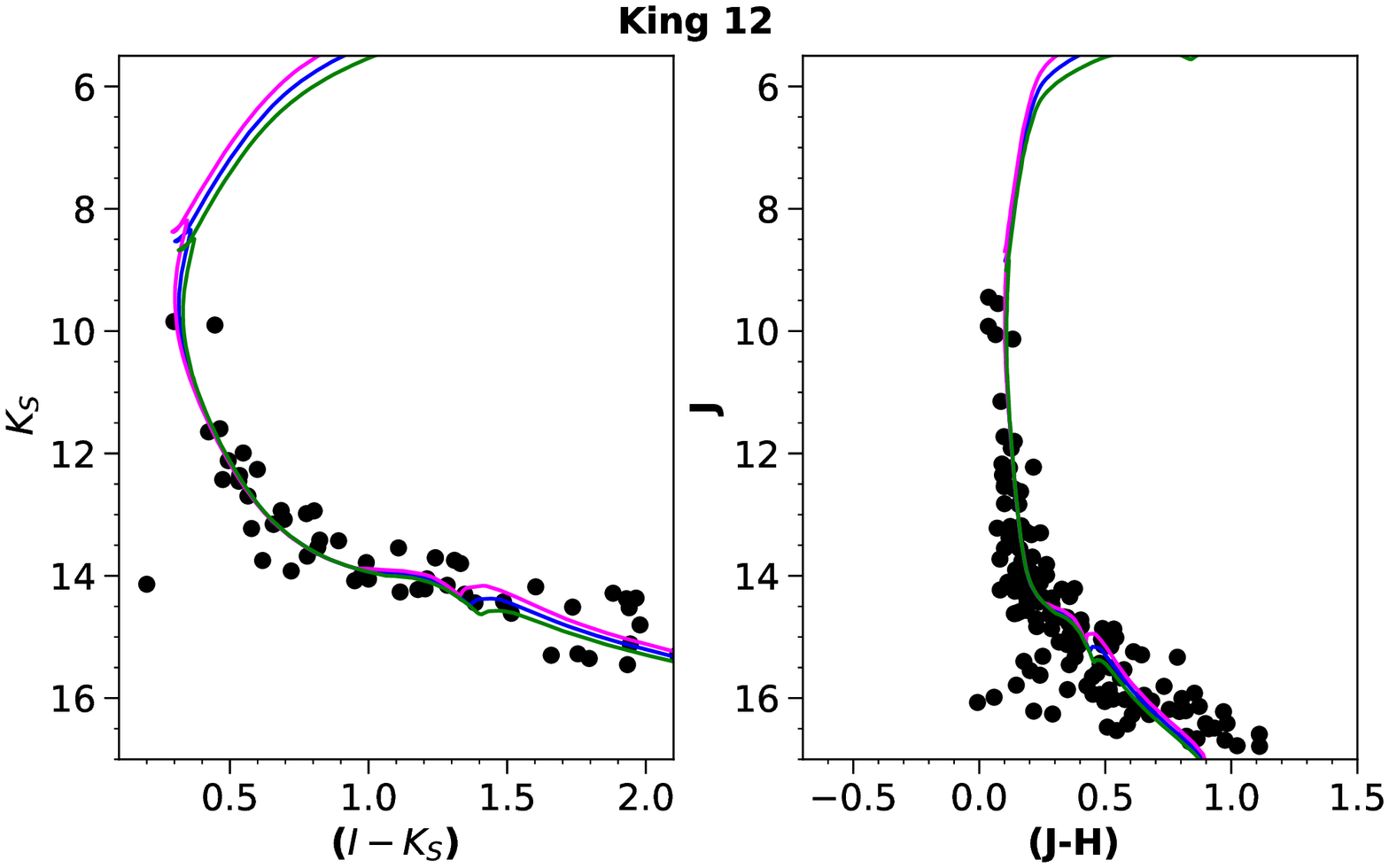}
	\caption{The color-magnitude diagrams in $(G, G_{BP}-G_{RP})$, $(V,B-V)$,
    $(V,V-I)$, $(K_{s},J-K_{s})$, $(V,V-K_{s})$,
	$(K_{s},I-K_{s}$ and $(J,J-H)$ planes for the cluster King 12. The best-fitted isochrones of log(age)
 	7.24, 7.29, and 7.34 are indicated by magenta, blue, and green colours, respectively.
	}
	\label{isok12}
  \end{figure*}
  
{\it NGC 2355}: We plotted seven CMDs
($(G,G_{BP}-G_{RP})$, $(V,B-V)$, $(V,V-I)$, $(K_{s},J-K_{s})$,
$(V,V-K_{s})$, $(K_{s},I-K_{s})$ and $(J,J-H)$) for this cluster and shown
in Fig. \ref{iso2355}.
From the best-fitted isochrone in all these CMDs, we found
$Z = 0.008$, age as $1.26 \pm 0.1$ Gyr and distance modulus as $11.5 \pm 0.1$ mag. The heliocentric
distance is derived as $1.83 \pm 0.11$ kpc. The distance derived here agrees with the distance derived using parallax.
Values of colour-excess $E(G_{BP}-G_{RP})$, $E(B-V)$, $E(V-I)$, $E(V-K_{s})$, 
$E(J-H)$, $E(J-K_{s})$ and $E(I-K_{s})$ are estimated as 0.23, 0.15, 0.20, 0.60, 0.03, 0.075, and 0.33 mag respectively.

We identified an unusual star above the red clump of the cluster, which was also reported 
by \citet{Soubiran2000} as a straggler star. This might be a probable Yellow Straggler star and is shown in the CMDs with yellow colour. The membership probability of this star is more than 90 \%. This star is saturated in the $I$ filter.

{\it King 12}: Fig. \ref{isok12} shows CMDs in $(G,G_{BP}-G_{RP})$, $(V,B-V)$,
$(V,V-I)$, $(K_{s},J-K_{s})$, $(V,V-K_{s})$, $(K_{s},I-K_{s}$
and $(J, J-H)$ planes for the cluster King 12 with best-fitted isochrones.
From the best-fitted isochrones, the cluster's age, metallicity, and distance modulus values are determined
as $19 \pm 2$ Myr, 0.0152 and $13.90 \pm 0.12$ mag, respectively. The distance
corresponding to the distance modulus is 2.63 $\pm$ 0.06 kpc.
The isochrone fitting gives the value of $E(G_{BP}-G_{RP})$, $E(B-V)$,
$E(V-I)$, $E(V-K_{s})$, $E(J-H)$, $E(J-K_{s})$ and
$E(I-K_{s})$ as 0.87, 0.58, 0.75, 1.70, 0.20, 0.29, and 0.80 mag, respectively.

\begin{table}
   \centering
	\small
   \caption{The colour-excess ratios in different filters for each cluster. The normal values are taken from \citet{2002A&A...381..219D, 2012A&A...539A.125D, 2019ApJ...877..116W}.}
   \begin{tabular}{lcccc}
   \hline\hline
  Cluster & $\frac{E(B-V)}{E(G_{BP} - G_{RP})}$ & $\frac{E(V-I)}{E(B-V)}$ & $\frac{E(J-H)}{E(B-V)}$ & $\frac{E(J-K_{s})}{E(B-V)}$    \\
   	& (mag) & (mag) & (mag) & (mag)   \\
  \hline
  Normal   & 0.76 & 1.25 & 0.33 & 0.49 \\
  NGC 1193 & 0.62 $\pm$ 0.10 & 1.25 $\pm$ 0.05  &  -   &   -   \\
  NGC 2355 & 0.65 $\pm$ 0.12 & 1.33 $\pm$ 0.11 & 0.33 & 0.50 $\pm$ 0.10  \\
  King 12  & 0.66 $\pm$ 0.03 & 1.29 $\pm$ 0.03 & 0.34 $\pm$ 0.02 & 0.50 $\pm$ 0.02 \\
  \hline
  \end{tabular}
  \label{cexcess}
  \end{table}
  
We calculated the color-excess ratios as $E(B-V)/E(G_{BP} - G_{RP})$, $E(V-I)/E(B-V)$, $E(J-H)/E(B-V)$ and 
 $E(J-K_{s})/E(B-V)$ and tabulated in Table \ref{cexcess} in different filters for each cluster. The normal values are taken from \citet{2002A&A...381..219D, 2012A&A...539A.125D, 2019ApJ...877..116W} following the general interstellar law for $R_{V}$ = 3.1. 
 All the values are comparable with the normal values suggesting the presence of general interstellar dust in the direction of these clusters. 
A comparison between parameters calculated in the present analysis and literature
values is given in Table \ref{ctab}. The table displays a sound consensus
of parameters determined in current studies with the literature.

\begin{table*}
   \centering
   \small
   \caption{A comparison of the parameters of the clusters with
   literature where DM is the distance modulus of the clusters.}
   \begin{tabular}{lcrll}
   \hline\hline
  Author & Age & DM  &  $E(B-V)$   \\
    	&   (Gyr)  & (mag) &  (mag)  \\
  \hline
  &{\bf NGC 1193}&&   \\
  \citet{1988AcA....38..339K} & 8  &  13.8 &	0.33   \\
  \citet{2005AN....326...19T}  & 8 &  - &  0.10 $\pm$ 0.06  \\
  \citet{2008JKAS...41..147K}   & 5  & $13.30 \pm 0.15$ &  $ 0.19 \pm 0.04$   \\
  Present study & $ 4.79\pm 0.58 $  &  $14.1 \pm 0.2 $ &  0.16   \\
 \hline
  &{\bf NGC 2355}&&  	\\
  \citet{1991AcA....41..279K} & -  &  -  &  0.12 \\
  \citet{Soubiran2000}  & 1   & 12.5  & 0.16 \\
  Present study   & $ 1.26 \pm 0.10 $  & $11.90 \pm 0.10 $ &  0.15  \\
  \hline
  &{\bf King 12}&&  	\\
  \citet{2013BASI...41..209T}  & 0.01 & 12.00 $\pm$ 0.15  &   $0.63 \pm 0.05$\\
   \citet{2012ApJ...761..155D} & 0.02 & -  & -  \\
   \citet{2013MNRAS.429.1102G} & 0.07 & 12.05 &  $ 0.51 \pm 0.05 $ \\
   \citet{2014NewA...26...77L} & 0.01 to 500 & 14.1 &  0.58 \\
  Present study & $0.019 \pm 0.002 $  & $13.75 \pm 0.12 $   & 0.58   \\
  \hline
  \end{tabular}
  \label{ctab}
  \end{table*}


\section{Luminosity and mass function of the clusters} \label{sec:luminosity}

\subsection{Photometric completeness and contamination from field} \label{comp}
\begin{figure}
	\begin{center}
	\centering
	\includegraphics[width=9.8cm,height=8.5cm]{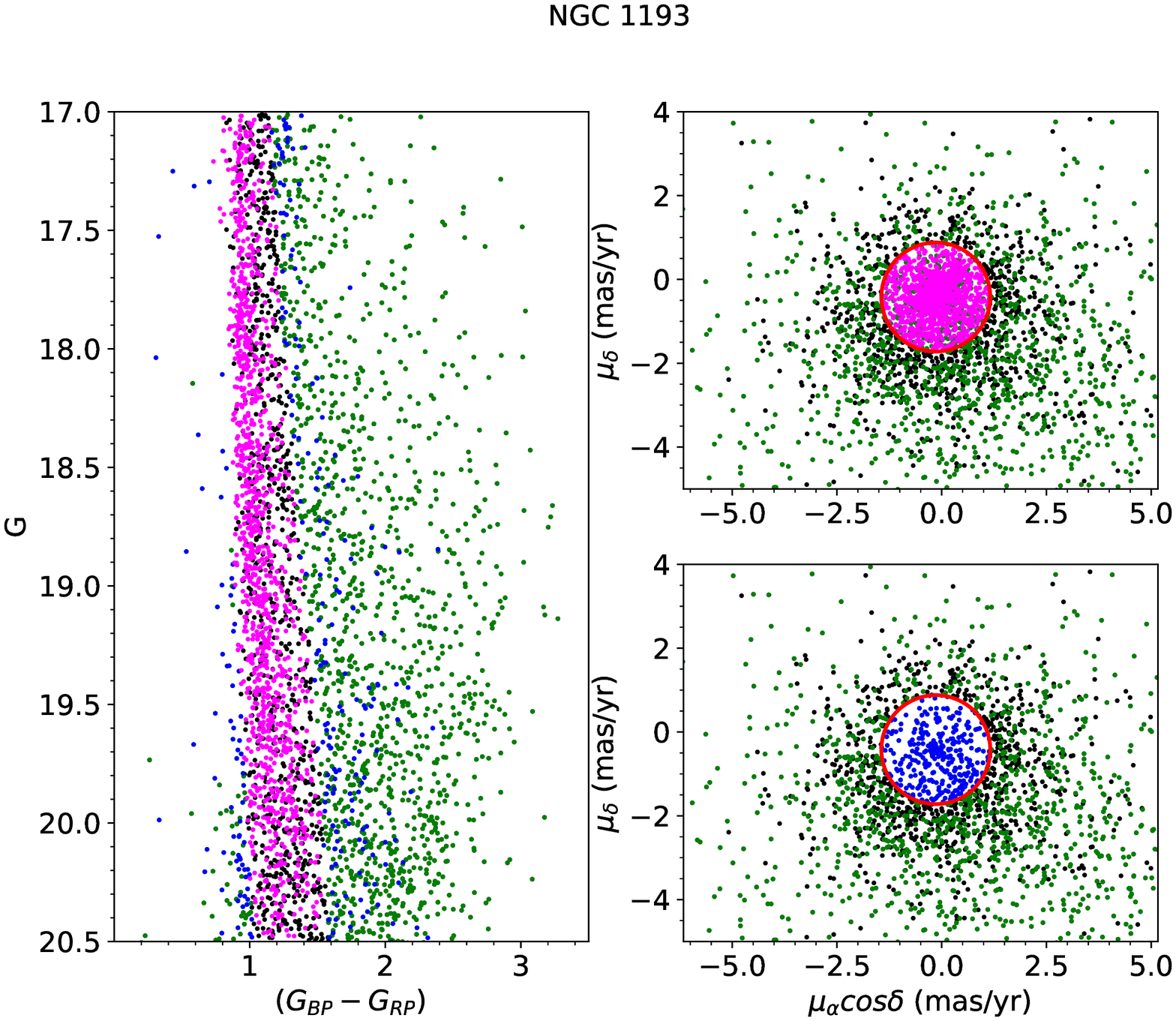}
	\caption{The CMD and VPD for the cluster NGC 1193 to calculate the
    	number of field stars in every magnitude bin. The magenta colour represents
    	stars which are on the main sequence of the CMD as well as inside the
    	chosen circle in the VPD, the blue colour represents the stars which are
    	not on the cluster main-sequence but inside the circle in VPD. Green
    	colour represents the stars that are outside of the main sequence, as well
    	as the circle in VPD}
	\label{contamination1}
	\end{center}
  \end{figure}

The true number of cluster members is required to generate the luminosity function for a cluster. The estimation of the number of cluster members can be
affected by two factors: the completeness of the data set and the inclusion of field stars. This analysis uses stars found in the Gaia DR3 catalogue.
The Gaia DR3 catalogue has the same sources with kinematical data and photometry in three broad bands as in Gaia EDR3.
Gaia EDR3 is complete for the sources having $12 < G mag < 17$. For the
sources fainter than 17 mag, completeness depends on the density. Since the open clusters under study
are not such a dense and crowded field, data incompleteness may not be affected. For Gaia DR2 stars,
\citet{2020MNRAS.497.4246B} computed a completeness of more than 90 percent upto
20 G mag stars in an uncrowded field. Since the completeness of EDR3 is higher than DR2 in fainter 
regions \citep{2021gdr3.reptE....V}, we expect
more than 90 per cent completeness for the stars up to 20 mag. To construct the luminosity
function, we only included stars brighter than 20 mag for the current analysis.

  \begin{figure*}
	\begin{center}
	\centering
	\includegraphics[width=8.8cm,height=8.8cm]{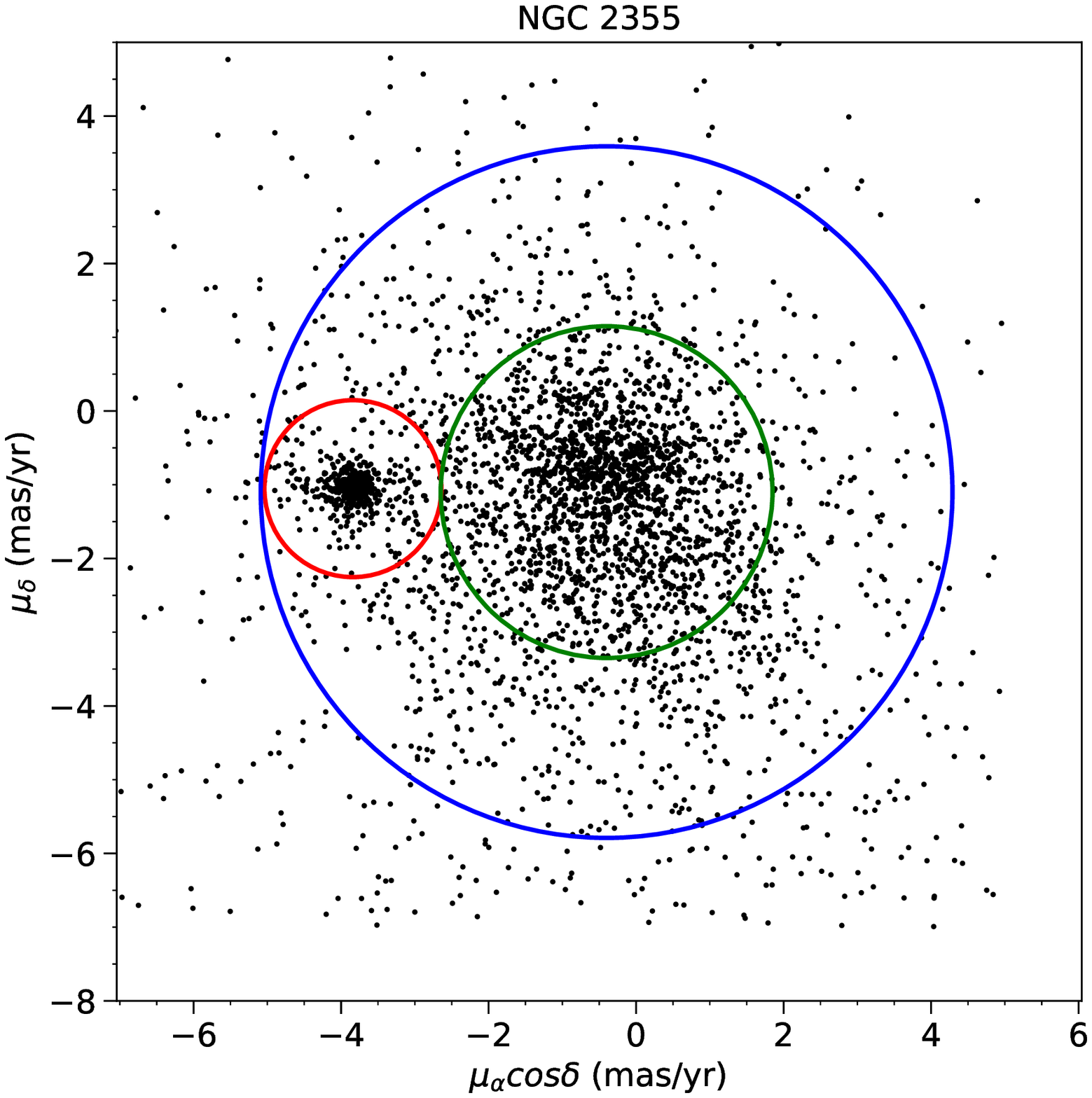}
	\includegraphics[width=8.8cm,height=8.8cm]{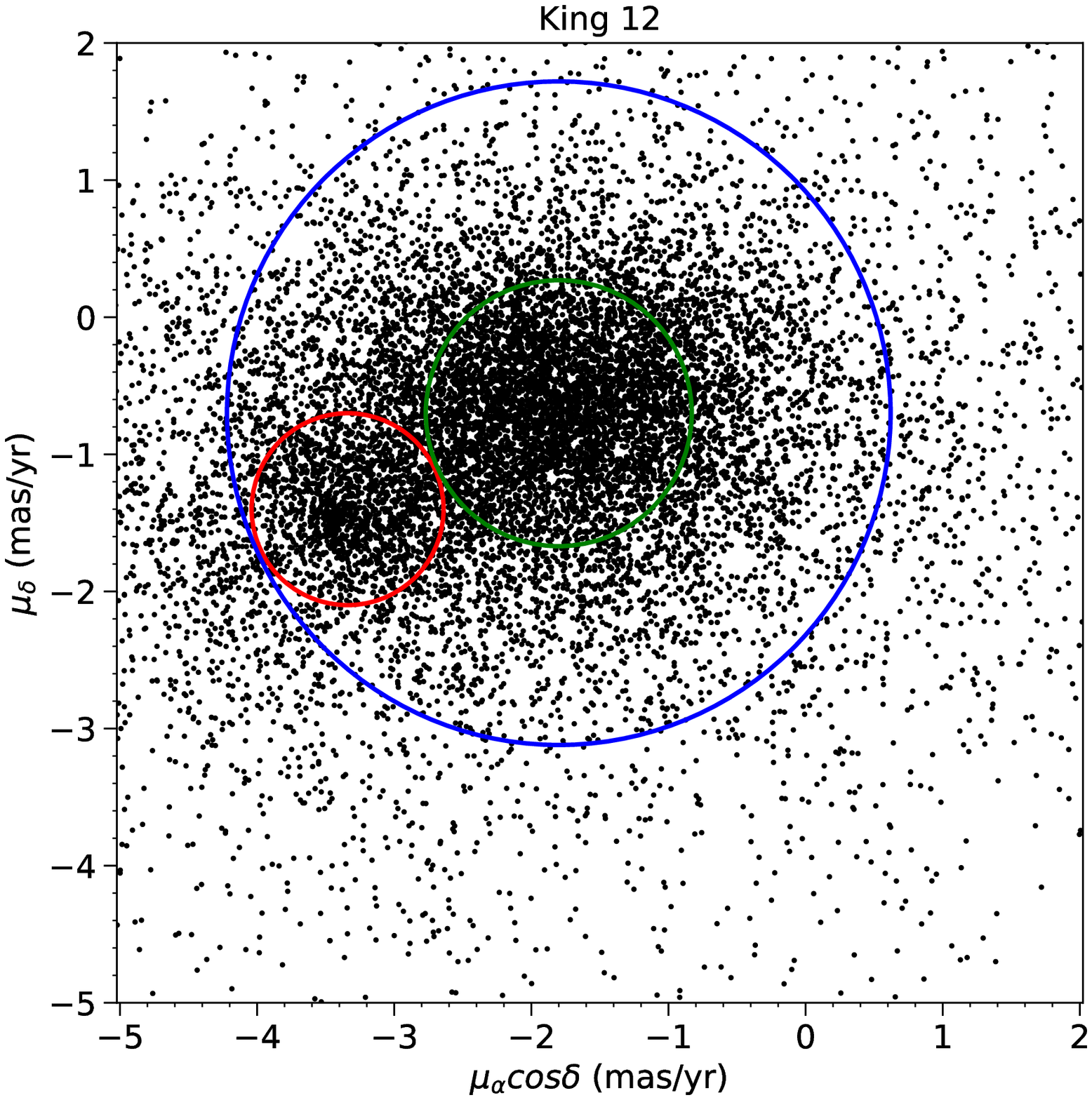}
	\caption{The VPD of the clusters NGC 2355 and King 12 for which the cluster
    	distribution and field distributions are well separated from each other.
    	The distribution inside the red circle is cluster distribution while
    	distribution inside the green circle is the field region, and 
    	a concentric circle around the field region shown in blue.
	}
	\label{contamination2}
	\end{center}
  \end{figure*}

We have selected the cluster members using Gaia DR3 proper motion and parallax to minimize the field star contamination. Field stars are not entirely
separated from the cluster region even after involving these criteria. Also, some field stars having similar motion as the cluster stars might 
be included as cluster members.
We adopted the method used by \citet{2012A&A...540A..16M} to overcome these possibilities.
The VPDs of the three clusters shown in Fig. \ref{vpd} have two types of distributions. One is that the cluster distribution is
not separated from the field stars, which is valid for the cluster NGC 1193
and the other is that the cluster distribution is entirely different from the
field star distribution, which is true for the clusters NGC 2355 and King 12.
So \citet{2012A&A...540A..16M} suggested two different approaches for these
different distributions.

The first approach is shown in Fig. \ref{contamination1}. For this, only main-sequence
stars have been used. The figure shows CMD and VPD of the stars in cluster NGC 1193.
This method uses CMD and VPD to calculate the number of field stars.
There are three populations of stars in this diagram: the first population is
inside the VPD circle as well as on the main-sequence (magenta), the second population is located inside the VPD circle but outside the main-sequence (blue), and the third population is outside the VPD circle as well as the main-sequence (green). So the stars in blue are field stars but got selected as cluster stars; similarly, the main-sequence may contain field stars.
The number of field stars can be calculated as follows:

\begin{equation}
N_{field} = n_{field} \times \frac{n_{in}}{n_{out}} \\
\end{equation}

where, $N_{field}$ is total number of field stars, $n_{field}$ is the total number
of stars which are outside the circle in VPD, $n_{in}$ is the total number of stars
inside the circle in VPD and outside of the CMD, and $n_{out}$ is the total stars
outside of the CMD. Using this method number of the field stars is calculated in every magnitude bin
and then calculated the number of cluster members in each magnitude bin.

The second approach for the distribution of VPD of clusters NGC 2355 and King 12
is shown in Fig. \ref{contamination2}. Two different populations for cluster and
field are visible in the VPDs, but the possibility of field stars with similar
proper motion stays. To calculate the number of field stars, three circles
are drawn on the cluster VPD. One is drawn around the cluster population having the same radius as used in section \ref{contamination2} shown in red colour, the second is around the field population shown
in green colour, and the third one is a concentric circle
around the field population in blue colour. The two concentric circles should touch the red circle. So now the VPDs are divided into three regions,
cluster region ($A$), field region ($B$), and outer annulus ($C$). So the number
of field stars inside the cluster region can be calculated as follows:

\begin{equation}
N_{field} = N_{C} \times \frac{Ar_{A}}{Ar_{C}} \\
\end{equation}

where, $N_{field}$ is the number of field stars in region $A$, $N_{C}$ is the
number of stars in region $C$ and, $Ar_{A}$ and $Ar_{C}$ are the area of the
regions $A$ and $C$ respectively. We calculated the field stars and actual cluster members in each magnitude bin using this method.

The number of field stars inside the cluster region
in each magnitude bin is listed in Table \ref{cl_members}. The cluster member selection using proper motion eliminates all the bright
field stars but includes several faint field stars. The number of cluster members calculated in
this section produces the luminosity function, mass function, and mass segregation in the open clusters under study.

\begin{table}
   \centering
   \caption{The number of field stars calculated in each magnitude bin for
   the clusters under study using the method described in the text.
   }
   \begin{tabular}{ccccccccc}
   \hline\hline
 $ G $ (Mag) &  NGC 1193 & NGC 2355 & King 12   \\
  \hline
   10 - 11 &  -  &  -  &  0  \\
   11 - 12 &  -  &  0  &  0   \\
   12 - 13 &  -  &  0  &  0	\\
   13 - 14 &  -  &  1  &  1	\\
   14 - 15 &  -  &  2  &  2	\\
   15 - 16 &  -  &  3  &  3	\\
   16 - 17 &  -  &  5  &  5	\\
   17 - 18 & 09  &  6  & 10	\\
   18 - 19 & 21  & 10  & 21 	\\
   19 - 20 & 29  & 12  & 40 	\\
  \hline
  \end{tabular}
  \label{cl_members}
  \end{table}

\subsection{Luminosity function} \label{subsec:l}

\begin{figure*}
	\begin{center}
	\centering
	\includegraphics[width=13.5cm,height=6cm]{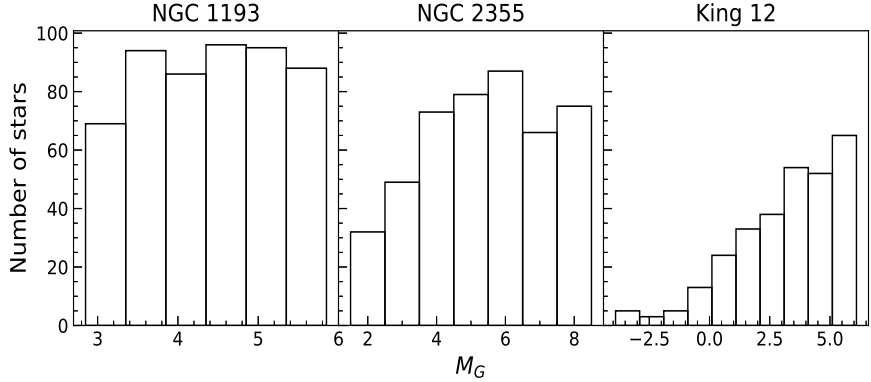}
    	\caption{Histograms for the main sequence stars in each magnitude bin. $M_{G}$
    	is the absolute magnitude in the Gaia filter.}
	\label{lf}
	\end{center}
  \end{figure*}

The luminosity function reflects the distribution of stars in a cluster according
to the luminosity. As stated above, the luminosity function for the main-sequence stars in each cluster is
generated from true cluster stars calculated above.
The apparent $G$ magnitude of these stars is converted into the absolute magnitude
using the distance modulus calculated in section \ref{sec:iso}. The number of
stars in each magnitude bin was calculated and then corrected for the field star contamination and plotted in Fig.\ref{lf} for the
clusters NGC 1193, NGC 2355, and King 12.

From Fig. \ref{lf}, it is concluded that NGC 1193 has a flat luminosity function which implies that with age 
brighter stars of the cluster are evolved, but it still holds its fainter stars. 
The luminosity function of NGC 2355 increases up to $M_{G} \sim $ 6 mag but decreases later. For King 12, an increasing luminosity function is observed. This
implies that the cluster retains most of its low mass
stars due to its very young age, hence not much dynamically evolved. One dip between -2.8 to -1.9 is also visible for this cluster which was also observed by \citet{2013MNRAS.429.1102G}.

\subsection{Mass function and Mass segregation} \label{subsec:m}

\begin{figure}
	\centering
	\includegraphics[width=9.0cm,height=7.2cm]{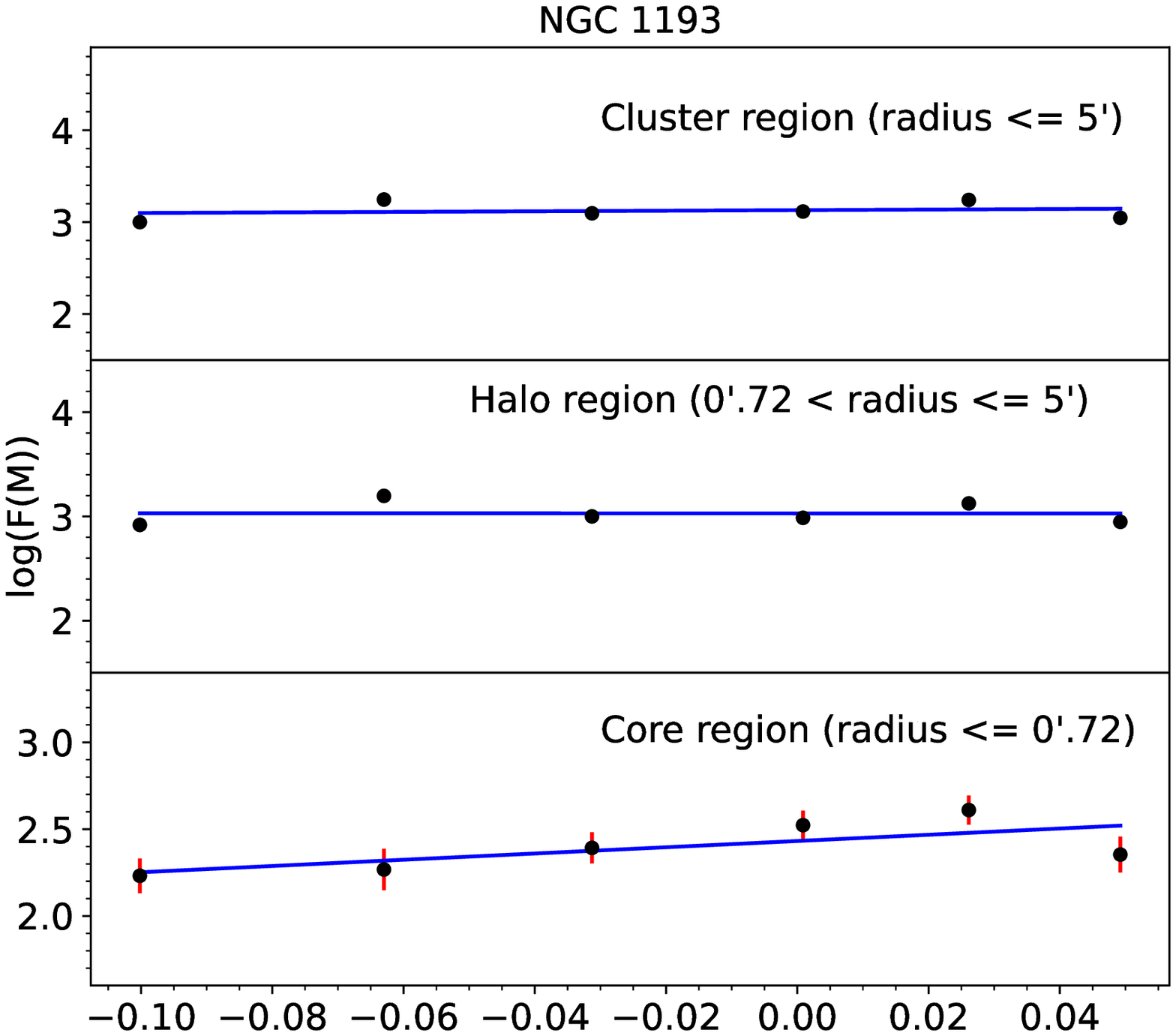}
    \includegraphics[width=9.0cm,height=7.2cm]{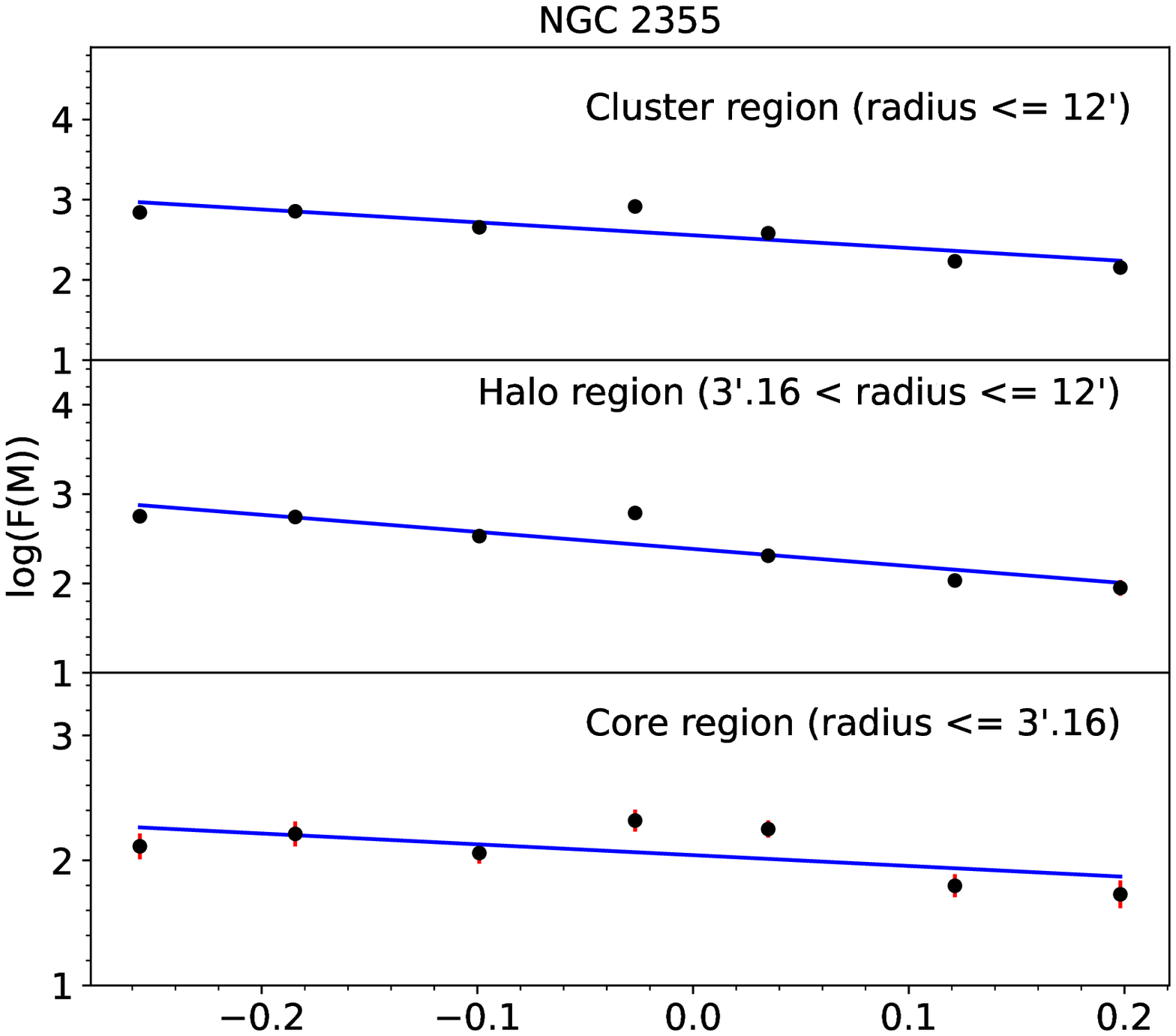}
    \includegraphics[width=9.0cm,height=7.2cm]{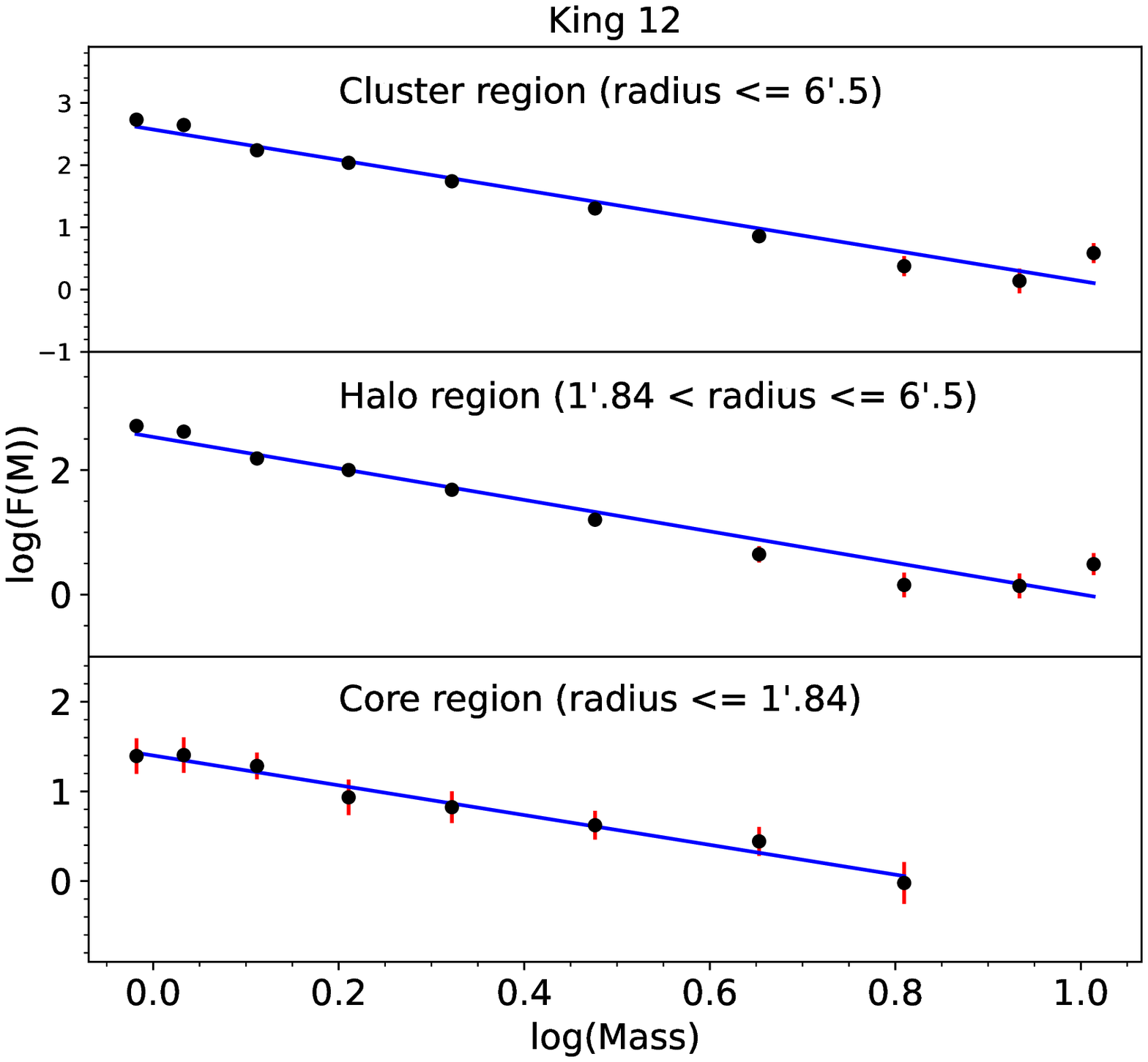}
	\caption{The slope of the mass-function for NGC 1193, NGC 2355, and King 5 in the three regions, namely core,
 	halo and the entire cluster region. The slope of the mass function is calculated
    by least square fitting, as shown by the blue line.
 	}
	\label{mf1193}
\end{figure}
  
In this section, the luminosity function of a cluster is converted into the mass
function with the help of best-fitted theoretical evolutionary tracks, and shown in
Fig. \ref{mf1193} for the clusters NGC 1193, NGC
2355 and King 12, respectively.
For this conversion, theoretical models given by \citet{2017ApJ...835...77M} have
been used as discussed in section \ref{sec:iso}.
By least-square fitting, we calculated the slope of the distribution by the fitting of the following equation:   \\

\begin{equation}
log \frac{dN}{dM} = -(1+x) \times log(M) + constant  \\
\end{equation}

where $dN$ is the number of stars in a particular mass bin $dM$, and $M$ is the
central mass of that bin. The mass function
slope $x$ can be determined from the above equation.
The mass function slope is calculated in three regions for each cluster, i.e., core, halo, and the entire cluster region. 
The radii values of these cluster regions are taken from section \ref{sec:cen}.
The values of mass function slopes for all the regions in each cluster are
listed in Table \ref{mftab}.

\begin{table}
   \centering
   \caption{ The mass-function slopes for each region for the clusters under study.
   }
   \begin{tabular}{lcrcr}
   \hline\hline
  Cluster & Mass  &  	\multicolumn{3}{c} {Mass function slope ($x$)}   \\
    	& range $(M_{\odot})$	& Core  & Halo  & Entire region  \\
  \hline
  NGC 1193 & 0.75 - 1.15 & $-0.99 \pm 0.10$ & $ -2.80 \pm 0.10$ & $ -1.31 \pm 0.09 $  \\
  NGC 2355 & 0.50 - 1.69 & $ -0.14 \pm 0.16$ & $ 0.91 \pm 0.15$ &  $ 0.61 \pm 0.15$  \\
  King 12  & 0.90 - 10.97 & $ 2.18 \pm 0.07$ & $ 0.80 \pm 0.10$ & $ 1.71 \pm 0.07 $ \\
  \hline
  \end{tabular}
  \label{mftab}
  \end{table}
 
For NGC 1193 and NGC 2355, the mass-function slopes for all three regions are
less than the Salpeter value $x = 1.35$ \citep{Salpeter1955ApJ...121..161S}. The mass function
slope for the core and entire cluster region of King 12 is higher than the Salpeter
value, while it is lower for the halo region.
Mass function slopes are becoming flatter towards the outer region as compared to
the core region in all the clusters.

\begin{figure*}
	\centering
	\includegraphics[width=13.5cm,height=6cm]{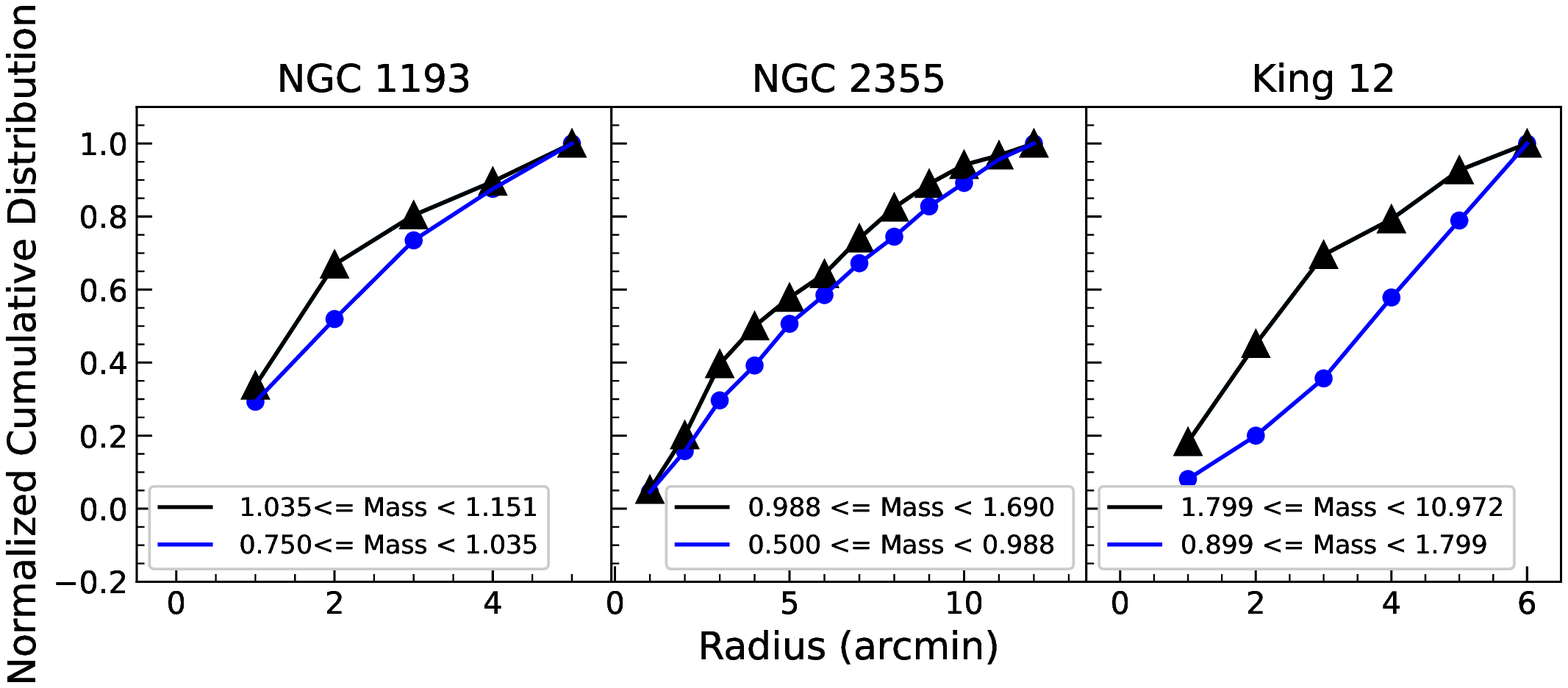}
	\caption{Normalized cumulative distribution of stars in the clusters NGC 1193,
    NGC 2355 and King 12. The innermost curve represents the
	massive stars, while the outer curve represents a lighter one.
	}
	\label{ms}
  \end{figure*}

The cumulative radial distribution in two mass ranges are shown in Fig. \ref{ms}
for the clusters NGC 1193, NGC 2355, and King 12. Using this figure, we have
studied the signature of mass segregation in the sample clusters.
By the visual examination of the Fig. \ref{ms}, it is evident that the two
populations are mixed for the clusters NGC 1193 and NGC 2355 but for King 12
it looks like the high-mass stars are more centrally located than the
fainter population. We then examined these populations statistically using
the Kolmogorov-Smirnov (K-S) test. We find a very low percentage of confidence level for these populations to be distinct. So we conclude that mass-segregation is not present in these clusters.

\section{Kinematics and dynamics of the clusters}\label{dynamics}

\subsection{Internal kinematics}\label{kinematics}
\begin{figure*}
	\centering
	\includegraphics[width=8.8cm,height=8.0cm]{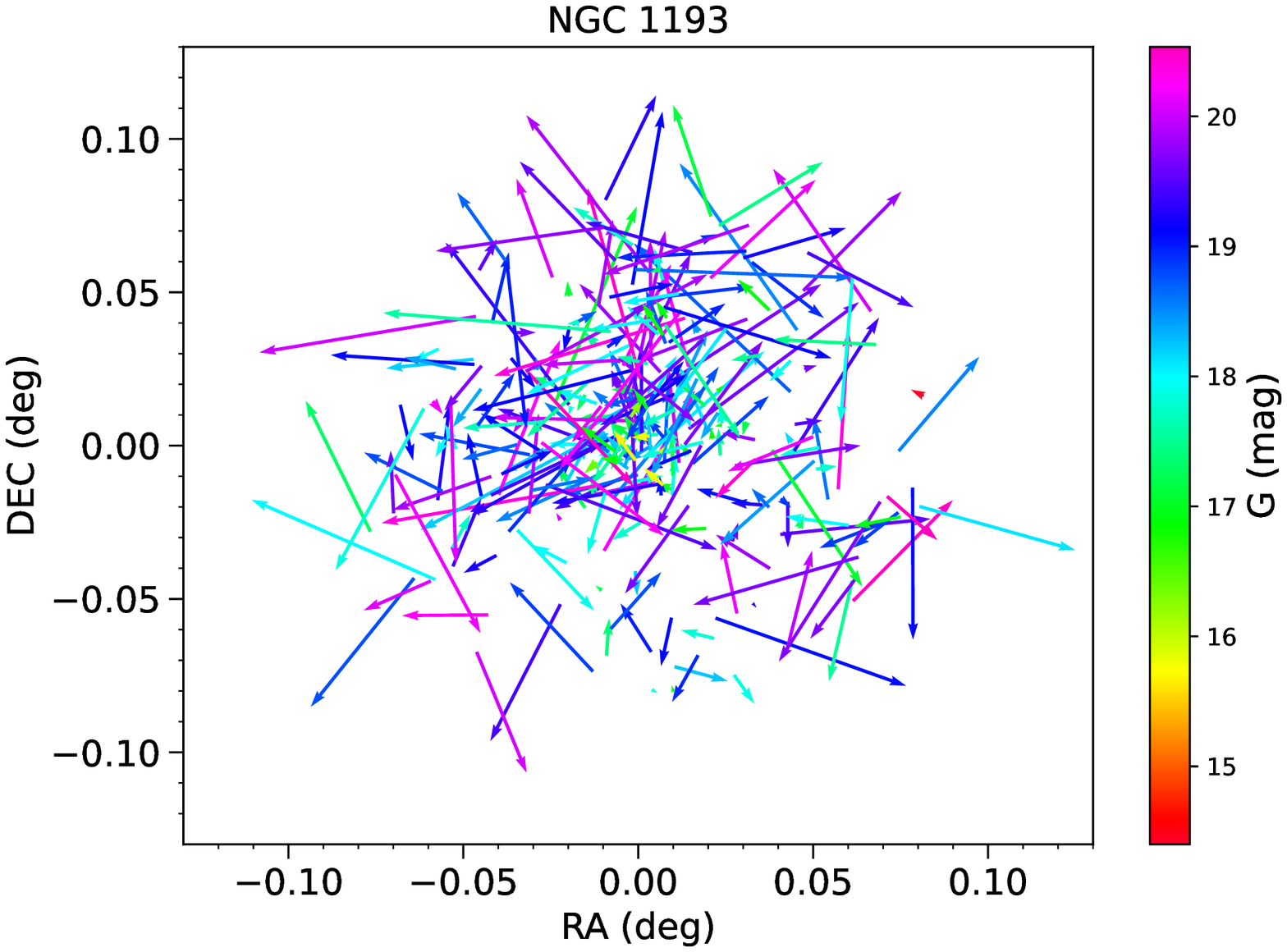}
	\includegraphics[width=8.8cm,height=8.0cm]{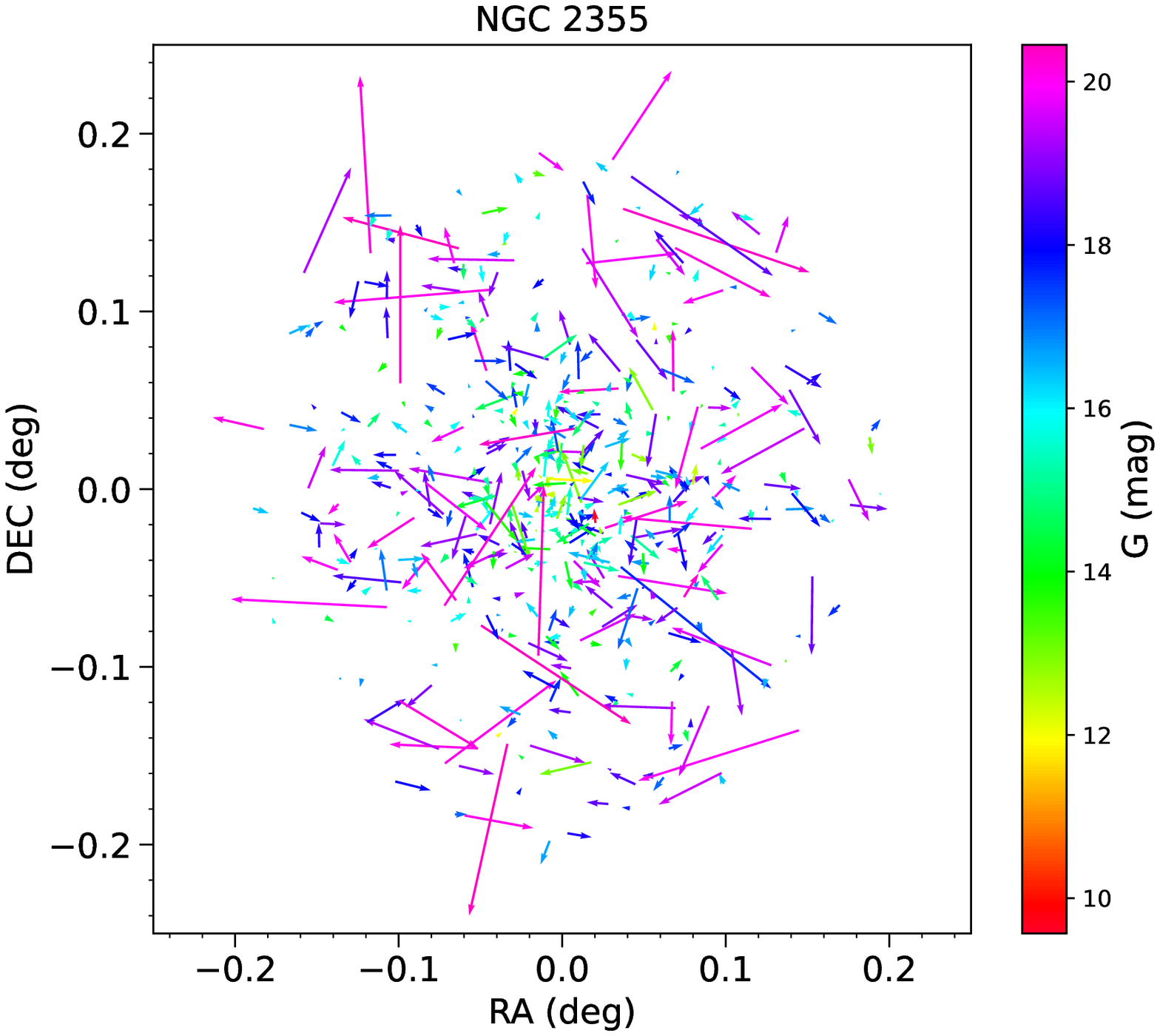}    
	\includegraphics[width=8.8cm,height=8.0cm]{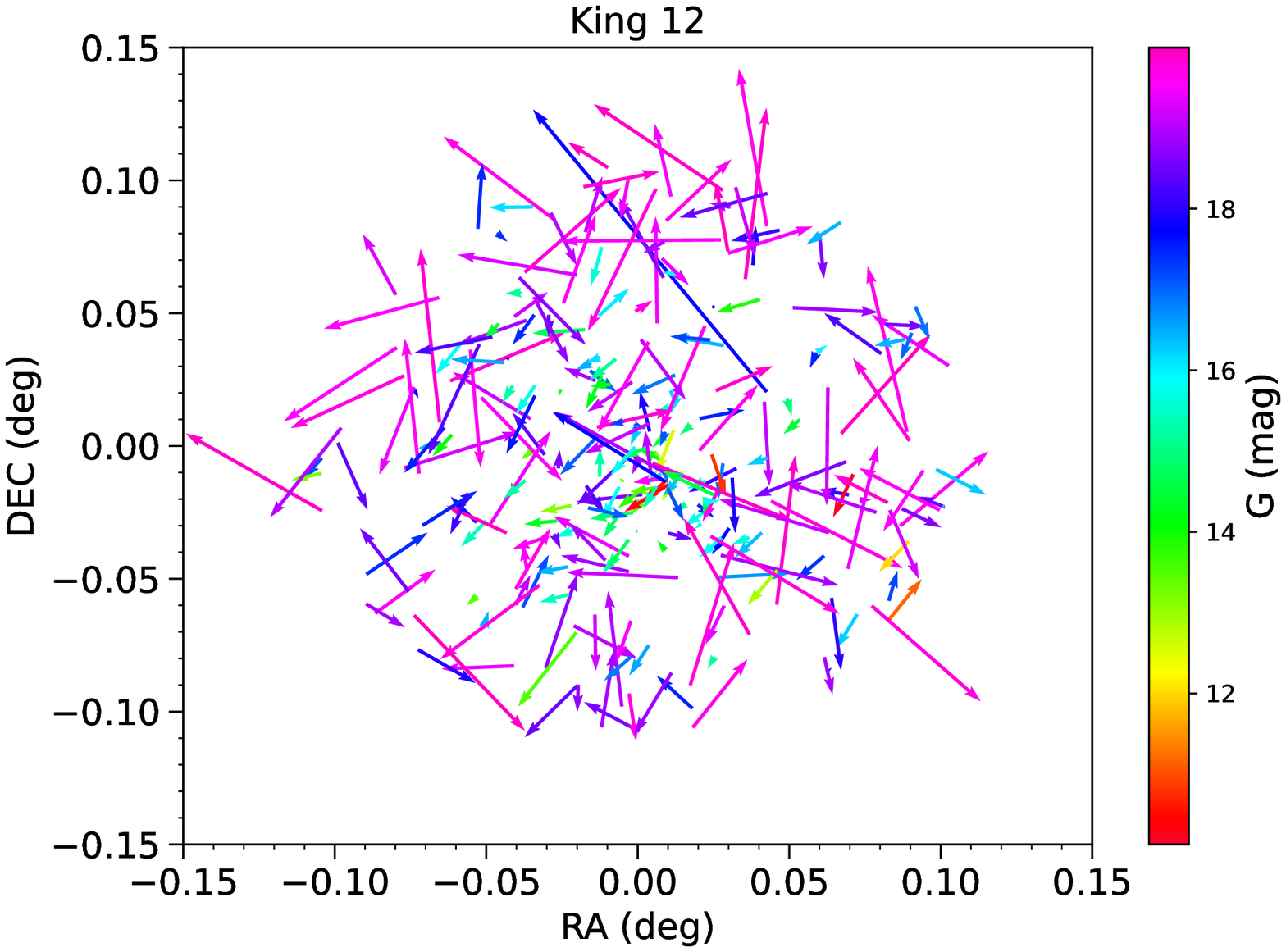}
	\caption{This figure shows the intracluster motion of the stars in the cluster
    region. The length of the vectors is the magnitude, and its direction is
    the direction of stellar motion. The vectors are also colour-coded
    according to the colour map shown on the right.
	}
	\label{motion}
  \end{figure*}

\begin{figure*}
	\centering
	\includegraphics[width=5cm,height=4cm]{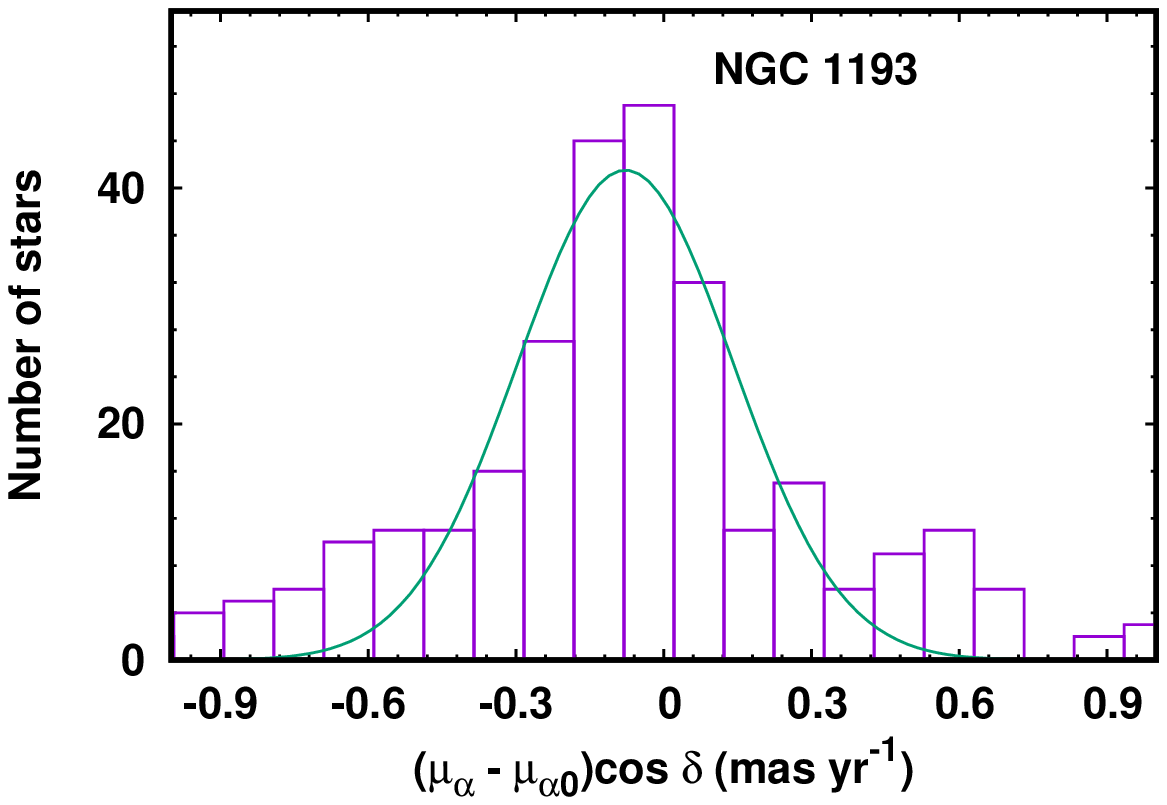}
	\includegraphics[width=5cm,height=4cm]{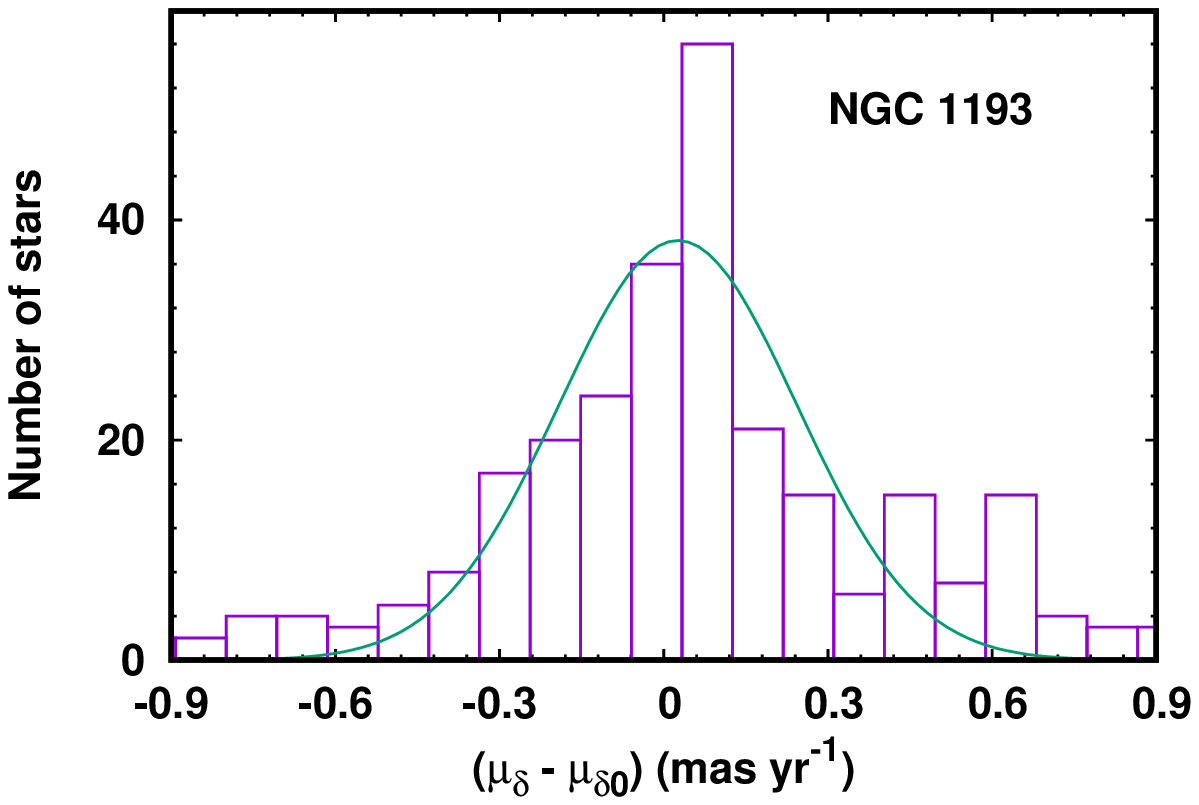}
	\includegraphics[width=5cm,height=4cm]{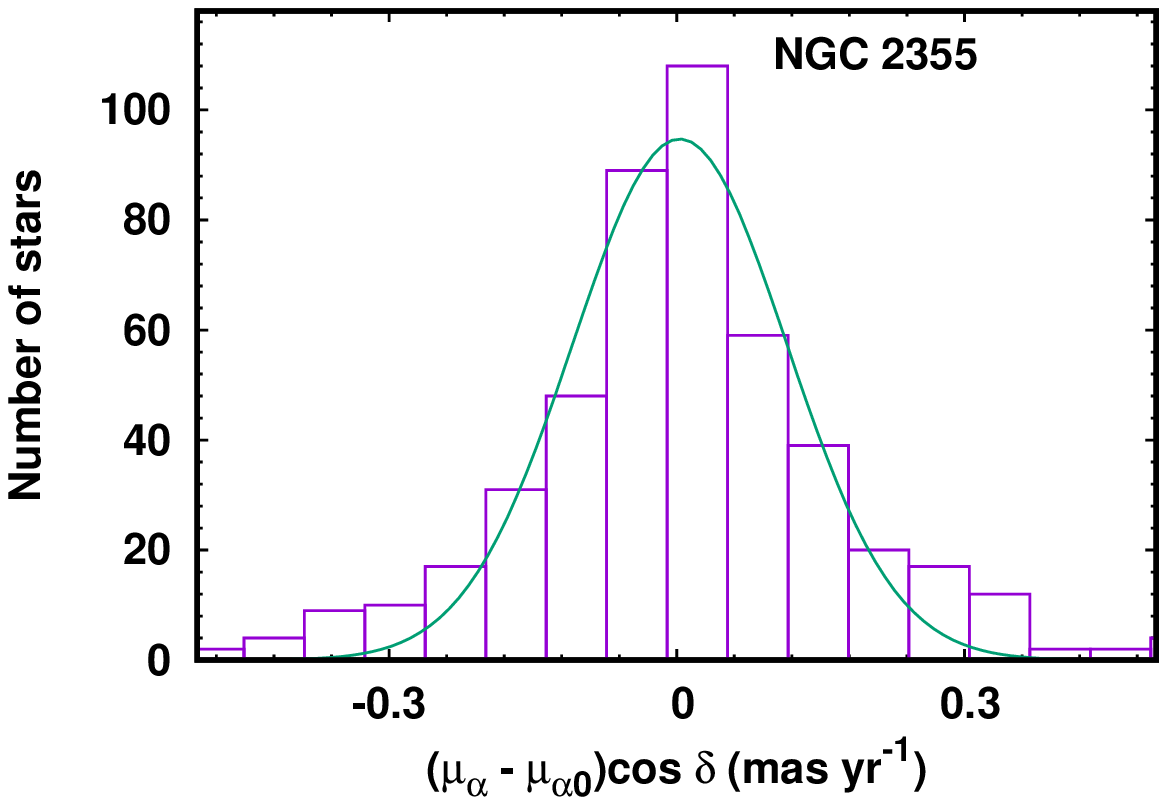}
	\includegraphics[width=5cm,height=4cm]{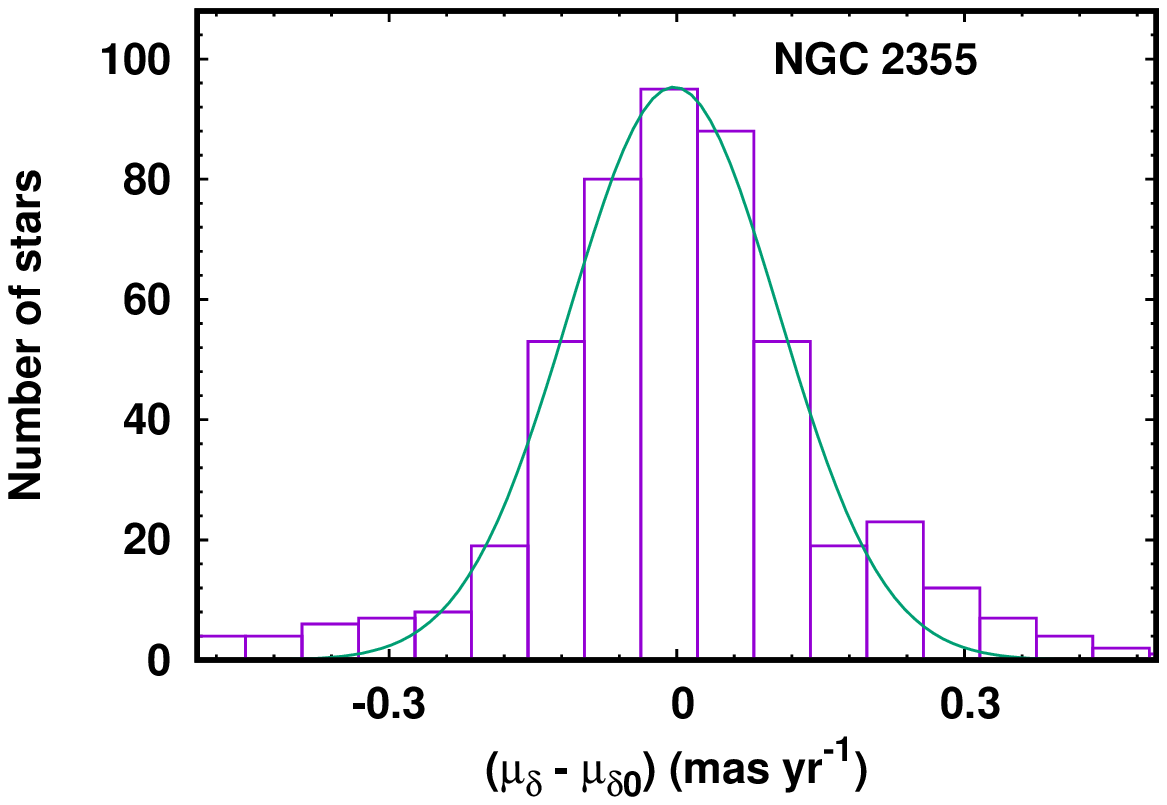}
	\includegraphics[width=5cm,height=4cm]{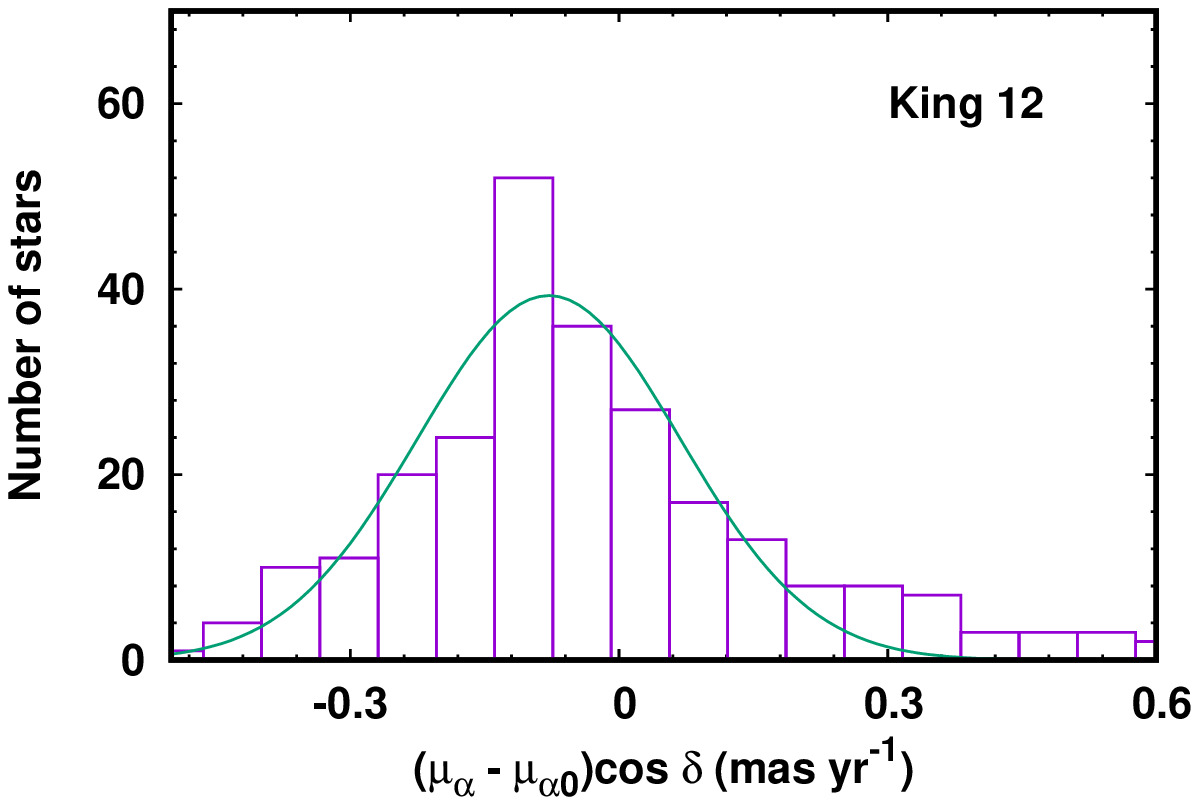}
	\includegraphics[width=5cm,height=4cm]{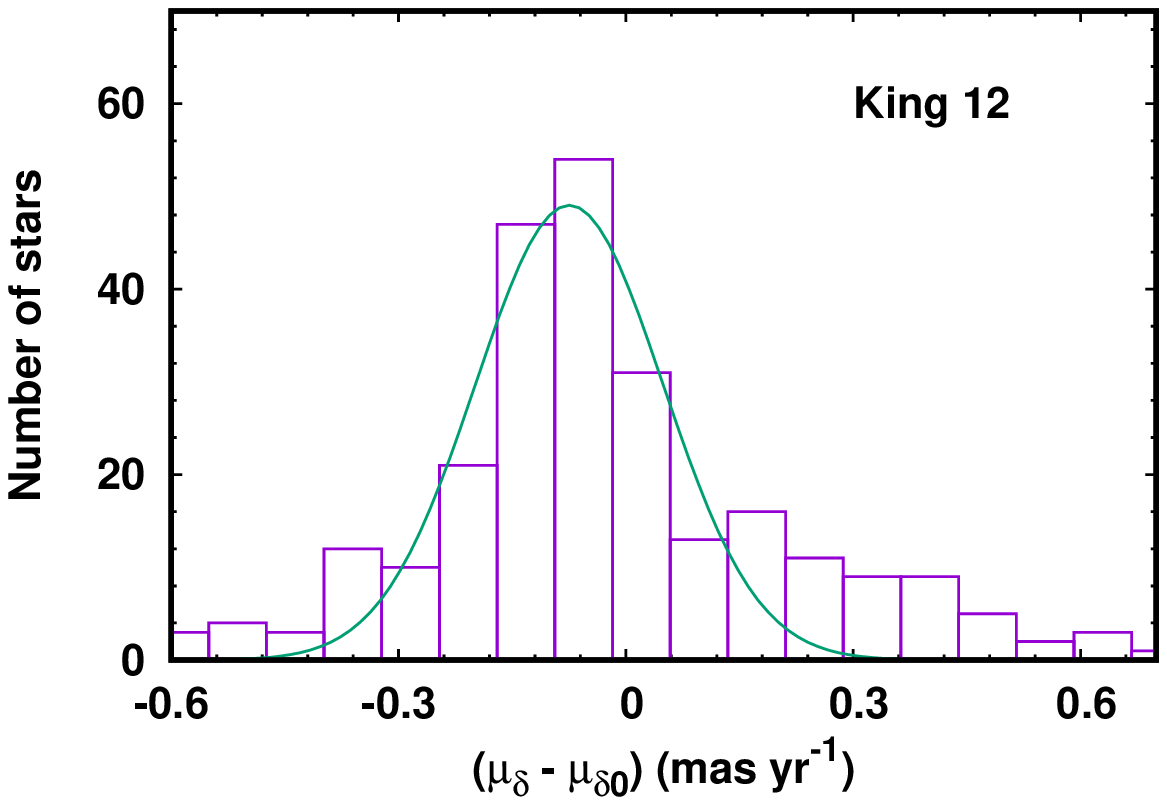}
	\caption{The figure shows the distribution of velocities in two directions.
    The gaussian curve is fitted over the distribution to calculate the mean
    motion as shown by the red colour.
    }
	\label{motion2}
  \end{figure*}

Fig. \ref{motion} shows the internal kinematics of the three studied clusters.
In this figure, the motion of the individual star relative to the mean cluster
motion is plotted in cluster space within the cluster radius. The length
of the arrows represents the magnitude, and the direction of the arrows shows the
direction of the relative motion. The colour of the arrows represents the magnitude in
$G$ filter. 
These figures show that all the stars are moving randomly, stars in NGC 1193 are moving with larger velocities than the other two clusters. 
The mean relative motion of stars in the clusters is calculated using the histogram shown in Fig. \ref{motion2}.
In this figure, we plotted the histograms of the relative motion of stars in the
clusters and fitted a Gaussian over it. From this histogram, mean relative motion
is calculated as ($-0.08 \pm 0.02, 0.03 \pm 0.03$), ($0.00 \pm 0.01, 0.00 \pm 0.01$)
and ($-0.08 \pm 0.01, -0.07 \pm 0.01$) for the clusters NGC 1193, NGC 2355 and
King 12, respectively. Based on these values, it can be deduced that the overall relative motion for NGC 2355 is nearly zero in both directions, whereas stars in NGC 1193 move outward in the RA direction only. The stars in King 12 have negative relative motion in both directions,
indicating that the stars of this cluster are moving outward and, thus cluster is expanding.

\subsection{Dynamics in the Galaxy} \label{sec:or}

\begin{table*}
   \centering
   \caption{Position and velocity coordinates of the three clusters in Galactocentric
	coordinate system. Where $R$ is the radial distance of the cluster from
    the Galactic centre, $\phi$ is the position angle of the clusters
    relative to the sun's direction, $Z$ is the vertical distance of the
    clusters from the Galactic disc and, $U$, $V$ and $W$ are radial,
   tangential and vertical components of velocity for the clusters, respectively.
   }
   \begin{tabular}{lcccccccc}
   \hline\hline
   	Cluster   & $R$ & $\phi$ &  $Z$ & $U$  & $V$  & $W$	\\
   	& (kpc) & (kpc) & (radians) & (km/sec) &  (km/sec) & (km/sec)   \\
  \hline
   	NGC 1193 & 12.20 & 0.20 & -0.92 & $36.74 \pm 1.29 $  & $-219.95 \pm 11.00 $ &  $-14.87 \pm 6.12$	\\
   	NGC 2355 &  10.13 & 0.08 & 0.42 & $ -12.70 \pm 3.20 $ & $ -246.64 \pm 2.10 $ & $ 21.07 \pm 2.48 $ 	\\
   	King 12  & 10.25 & 0.30 & 0.01 & $ 02.76 \pm 6.00 $ & $ -259.12 \pm 6.73 $ & $ 02.37 \pm 6.20 $  	\\
\hline
  \end{tabular}
  \label{inp}
  \end{table*}

We studied the dynamics of the three clusters in the Galaxy by deriving their orbits with the help of the Galactic model given by \citet{1991RMxAA..22..255A}.
For this analysis, we followed the procedure discussed in \citet{2019MNRAS.490.1383R}. For orbit calculation,
we have included the updated values of constants in Galactic potentials
given by \citet{2017AstL...43..241B}. The position and velocity components for all the
clusters are taken from Table \ref{vinp}.
The radial velocity data for NGC 1193 and  NGC 2355 come from Gaia DR3 while, for King 12, it is adopted from \citet{2007AN....328..889K}.
To transform the clusters' heliocentric position and velocity coordinates
into Galactocentric coordinates, transformation matrices given by
\citet{1987AJ.....93..864J} were used.
For this conversion, the coordinates of the Galactic centre and Galactic North pole
are taken from \citet{2004ApJ...616..872R} as (17:45:32.224, -28:56:10)
and (12:51:26.282, 27:7:42.01)
respectively. The position and velocity of the Sun are taken as (8.3,0,0.02) kpc
and (-10.4, +237, +7.3) km/s.
The resulting position ($R, \phi, Z$) and velocity ($U, V, W$) components are
listed in Table \ref{inp}. The radial component is positive towards the Galactic
centre, the tangential component is positive along the Galactic rotation, and the
vertical component is positive towards the Galactic north pole.

  \begin{figure*}
	\centering
	\includegraphics[width=5.9cm,height=5.9cm]{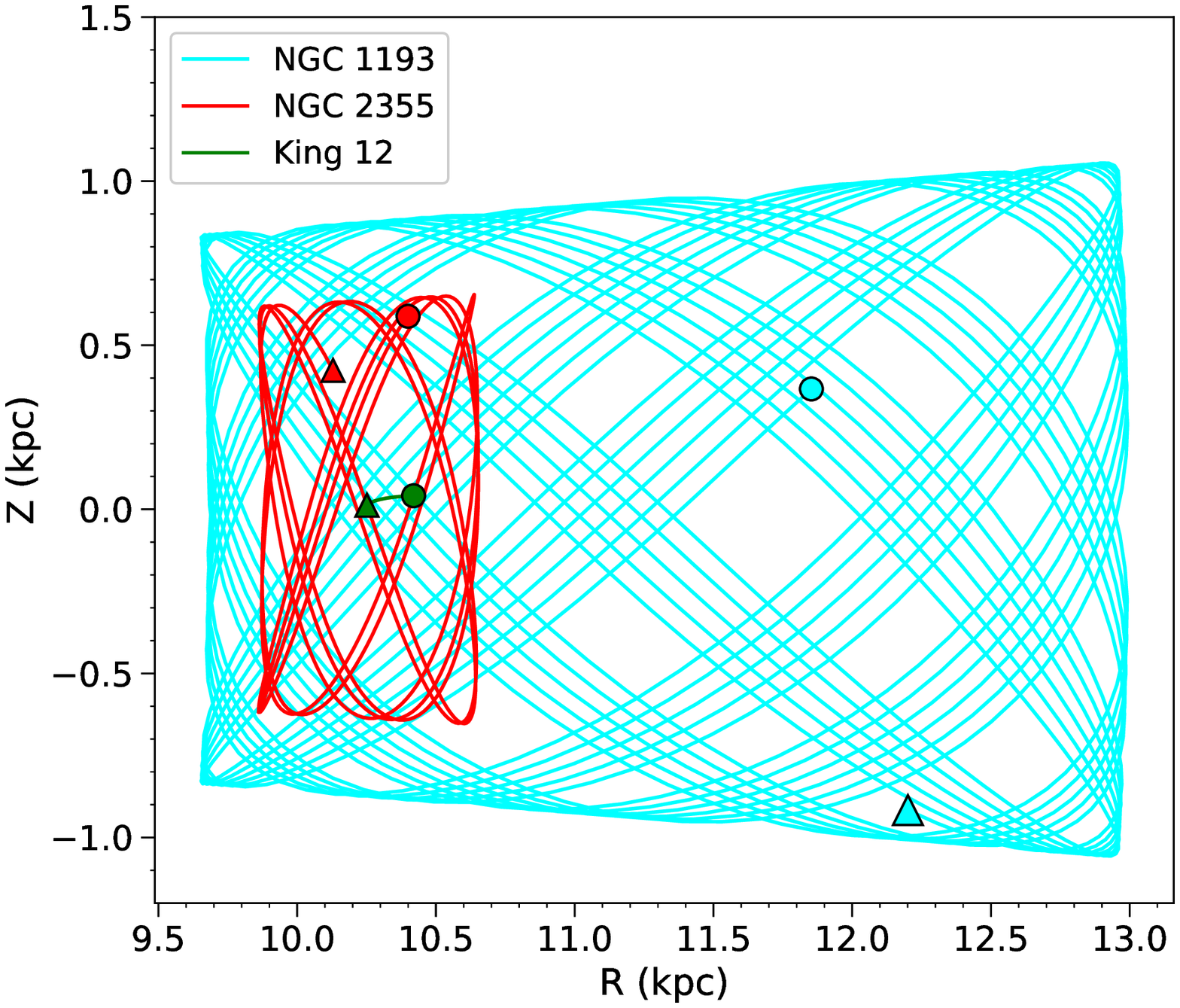}
	\includegraphics[width=5.9cm,height=5.9cm]{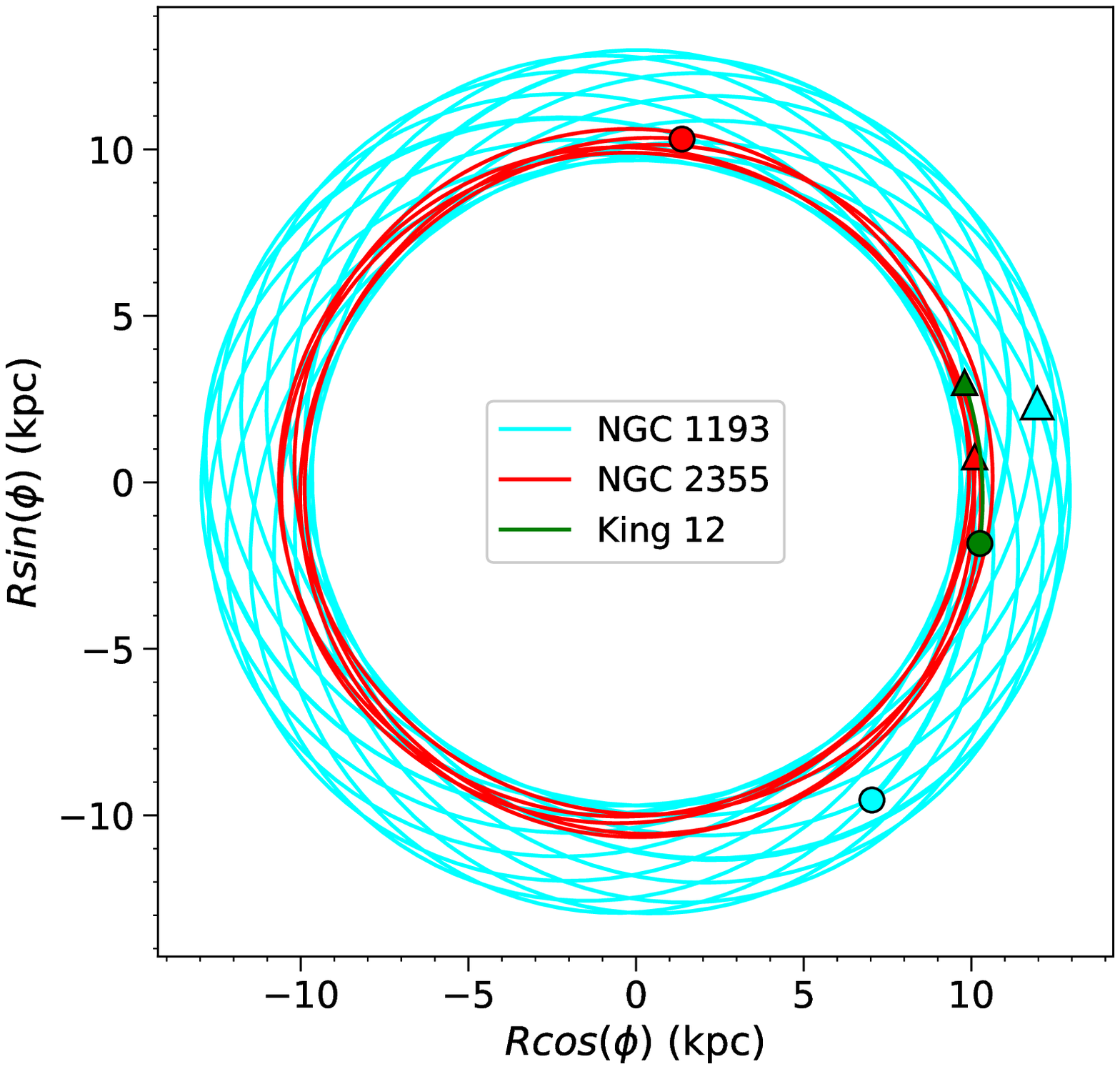}
	\includegraphics[width=5.9cm,height=5.9cm]{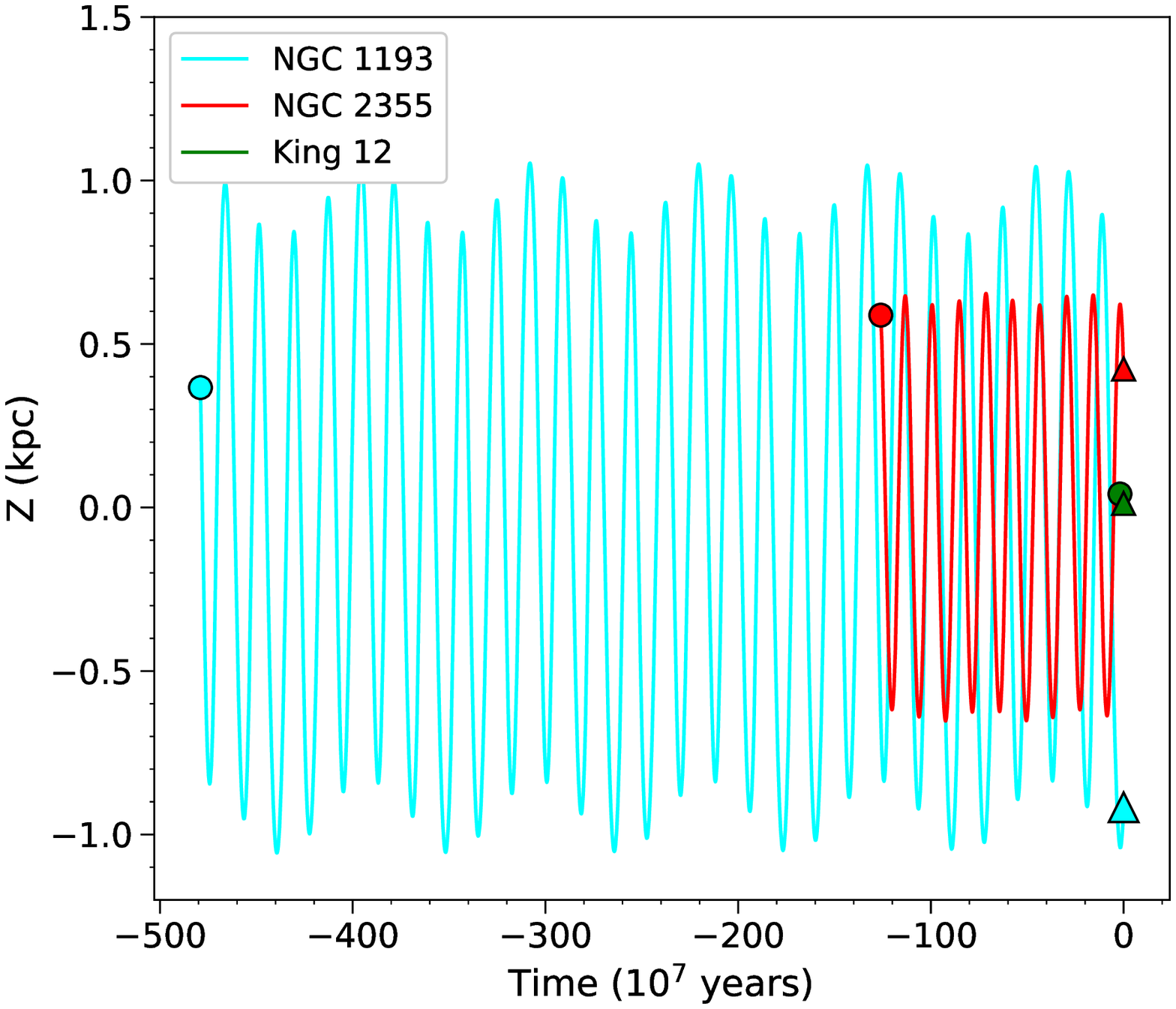}
	\caption{
 	Galactic orbits of the clusters NGC 1193, NGC 2355, and King 12
 	determined using Galactic potential models shown by the cyan, red, and green colour 
    respectively. The left panel shows the side picture; the middle panel shows the top 
    picture of the orbits, and the right panel shows the motion of the clusters in $Z$ 
    direction as a function of time. The triangles and circles denote the clusters' birth 
    and present-day position, respectively.
 	}
	\label{6nm}
  \end{figure*}

The path of the clusters in the Galaxy is integrated backwards in time equal to
the age of the clusters from their current positions and shown in Fig. \ref{6nm} with cyan, red and green colours for the clusters NGC 1193, NGC 2355 and King 12 respectively.
The left panel of this figure shows a plot of the distance from the Galactic centre
and the height from the Galactic disc. The middle panel shows a plot between the two
components of the radial distance from the Galactic centre. The right panel shows a plot between the time the cluster takes to reach its present-day position from its birth and the vertical
distance from the Galactic disc. Here, the present-day position is denoted by zero, and negative time denotes its past lifetime. The filled circles denote the present-day positions, and the filled triangles denote the birth positions of the
clusters in all the panels. The orbital parameters of all three clusters
are tabulated in Table \ref{orpara}. As shown in Fig. \ref{6nm}, King 12
have not completed even one revolution; hence to calculate the orbital parameters,
we again integrated its motion for a time period of 800 Myr. In this table, $e$ is the eccentricity of
the orbits, $R_{a}$ and $R_{p}$ are the apogalactic and perigalactic distances
respectively, $E$ is the energy, $J_{z}$ is the third component of the momentum,
$T_{R}$ is the time period of radial motion.

\begin{table*}
   \centering
   \caption{Orbital parameters of clusters calculated
	using the Galactic potential model for the three clusters under study.
   }
   \begin{tabular}{lcccccccc}
   \hline\hline
Cluster  & $e$  & $R_{a}$  & $R_{p}$ & $Z_{max}$ & $E$ & $J_{z}$ & $T_{R}$   \\
  	&	& (kpc) & (kpc) & (kpc) &  $(100 km/sec)^{2}$ & (100 kpc km/s) & (Myr)  \\
   \hline\hline
   NGC 1193 & 0.001 & 12.90 & 12.98 & -1.06 & -9.17 & -26.84 & 346 \\
   NGC 2355 & 0.001 & 10.64 & 10.61 & 0.66 & -9.72 & -24.98 & 256   \\
   King 12  & 0.000 & 11.59 & 11.58 & -0.05 & -9.39 & -26.56 & 247   \\
  \hline
  \end{tabular}
  \label{orpara}
  \end{table*}

The left panel of Fig. \ref{6nm} infers that the side view of the orbits looks like a box, and
the middle panel infers that all the clusters are moving in a circular orbit around
the Galactic centre. All three clusters' birth and present-day positions are located towards the Galactic anticenter. The eccentricity for the clusters NGC 1193 and NGC 2355 is nearly
zero and completely zero for King 12, indicating that the clusters' orbit is circular. Comparing the $Z_{max}$ values from Table
\ref{orpara}, it is evident that the cluster NGC 1193 is tracing the highest
distance from the Galactic disc. In the right panels of Fig. \ref{6nm}, the maximum distance in
the Z direction is constant with time for NGC 2355, while it changes periodically for NGC 1193. A similar change in Z is visible in the left panels that the
maximum value of Z is constant for NGC 2355, but it decreases on going towards
the Galactic centre. It may be due to the birth positions of the clusters, as
the birth position of NGC 2355 is closer to the Galactic disc, while the birth location of
NGC 1193 is more than double. So NGC 2355 is experiencing almost constant force from the Galactic
disc while this force is increasing towards the Galactic centre for the cluster NGC
1193. King 12 is a very young cluster; hence not completed even one circle. This
cluster was born even closer to the Galactic disc, moving very immediately
to the disc, leading to a shorter life for this cluster. The comparatively small
force from the Galactic disc on NGC 1193 is reflected as longer velocity vectors
in Fig. \ref{motion} compared to the other two clusters. We compared the 
orbital parameters calculated in this analysis with \citet{2009MNRAS.399.2146W} in which we found data only for NGC 2355 and King 12. We found that the orbits in \citet{2009MNRAS.399.2146W} are more
eccentric than in the present study. It is because \citet{2009MNRAS.399.2146W} have taken the same integration time for all the clusters, whereas we have taken it equal to the cluster's age. Also, they adopted a slightly different approach for calculating the eccentricity than the present study.

\begin{figure*}
	\centering
	\includegraphics[width=7cm,height=6cm]{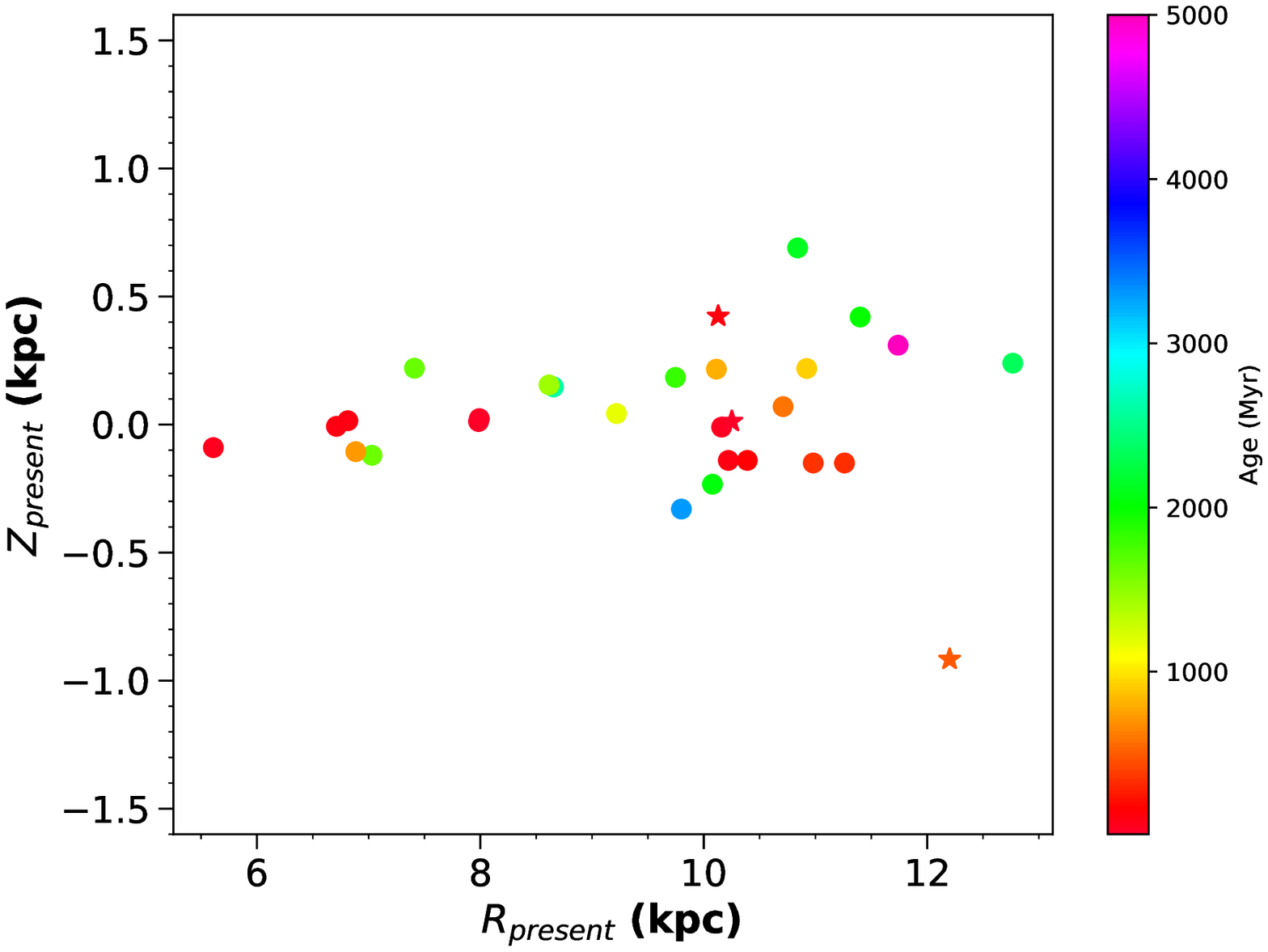}
	\includegraphics[width=7cm,height=6cm]{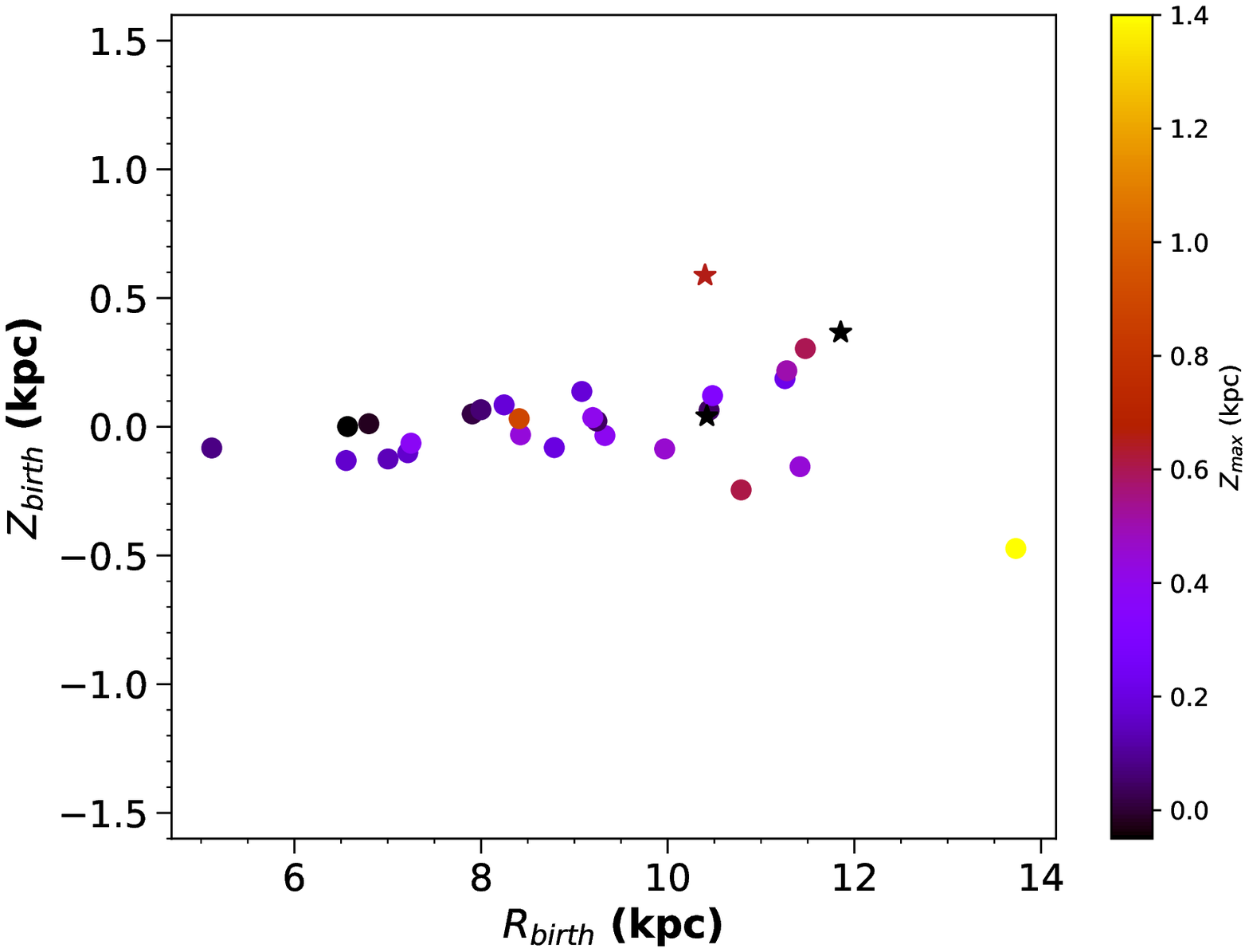}
	\caption{The left panel of this figure shows the present-day position
    of the clusters as a function of their age, while the right panel shows
    birth locations of the clusters as a function of the maximum distance travelled
    by the clusters away from the Galactic disc. The circles represent
    the clusters from our previous studies, while the asterisk are clusters
    from the present study.
 	}
	\label{allcls}
  \end{figure*}

\begin{figure*}
	\centering
	\includegraphics[width=7cm,height=7cm]{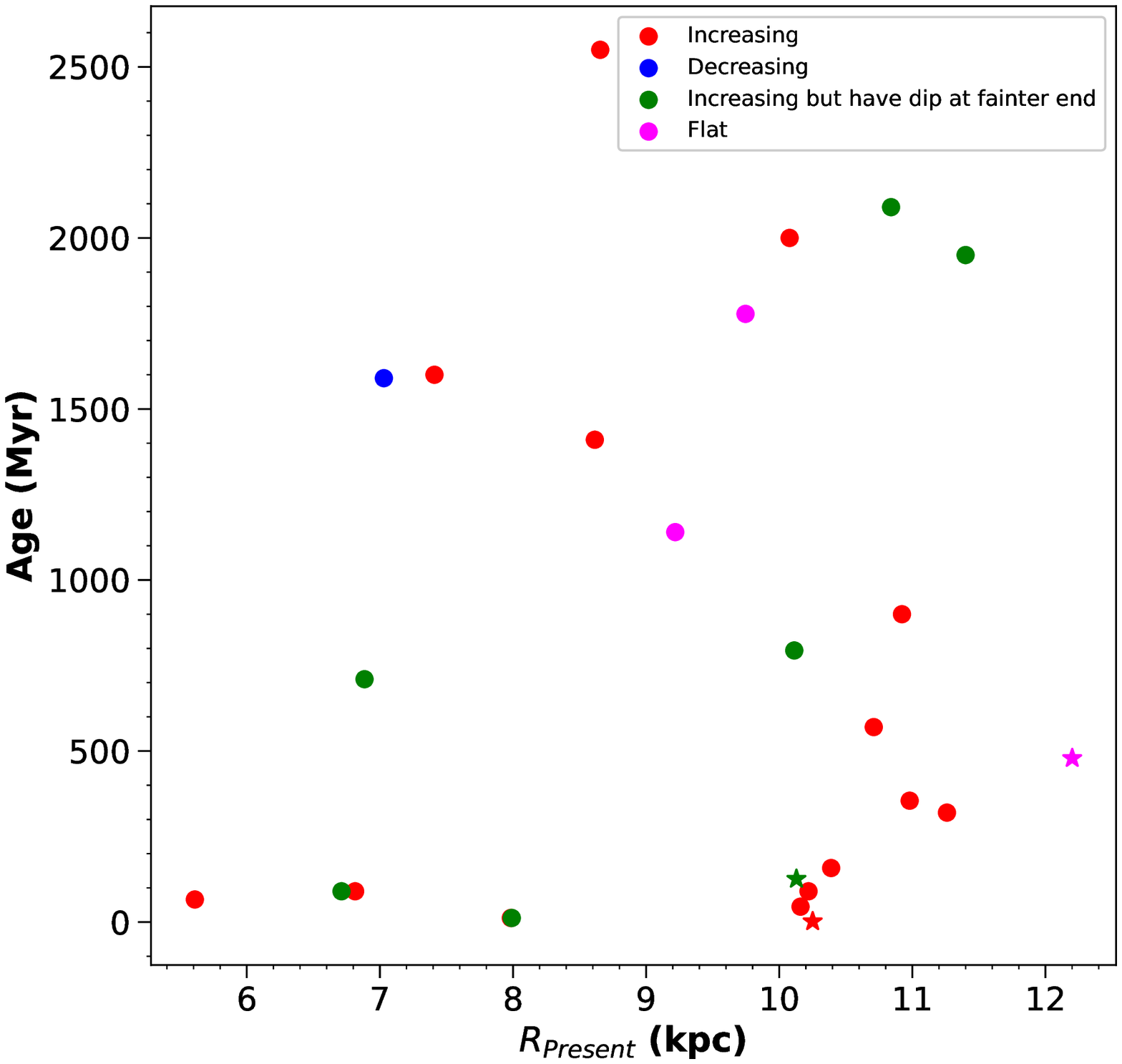}
	\includegraphics[width=7cm,height=7cm]{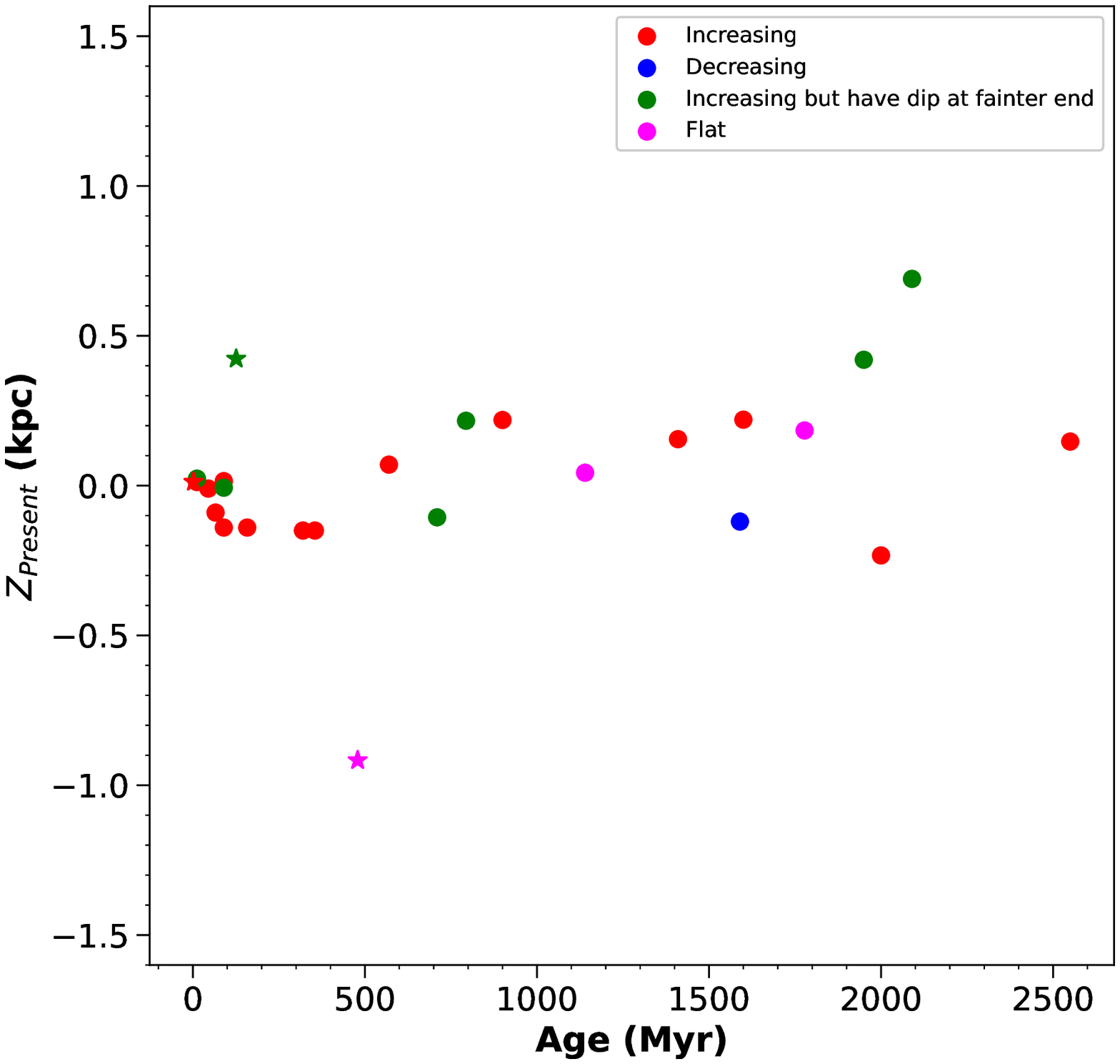}
	\caption{The left panel shows the distance of the clusters from the Galactic
    centre according to their age and luminosity function, while the right
    the panel shows the distance of the clusters from the Galactic disc
    according to their age and luminosity function. The circles represent
    	the clusters from our previous studies, while the asterisk are clusters
    	from the present study.
 	}
	\label{allcls2}
  \end{figure*}

To investigate how orbital parameters behave with the cluster properties we created two diagrams that are displayed in Fig \ref{allcls} and \ref{allcls2}. To produce these figures, we included a total of 29 open clusters, in which 26 open clusters are included from our past studies 
\citep{2019MNRAS.490.1383R, 2020MNRAS.494..607B, 2020AJ....160..119B, 2021AJ....161..102S, 
2021AJ....161..182B, 2021MNRAS.503.5929B, 2021PASJ...73..677B, 2022PASP..134d4201B, 2022AJ....164..171B, 2023NewA...9801938S}
and three clusters are from the present analysis. The three clusters from the current analysis are shown by asterisk symbols while the clusters from the earlier analysis are represented by filled circles.
The present-day positions of the clusters in the Galaxy as a function of their ages are plotted
in the left panel of Fig. \ref{allcls}, which shows that all the young clusters
are quite near to the Galactic disc. In contrast, older clusters are spread
over comparatively at a longer distance. In the right panel of this figure, the birth
positions of the clusters are plotted as a function of the maximum distance travelled
by the clusters in the Z direction. This figure shows that the clusters 
born close to the Galactic disc are orbiting closer to it while those born
at a larger distance from the Galactic core and Galactic disc are tracing higher scale height.

\citet{2022A&A...659A..59T} have studied 389 open clusters in the
Galaxy and they found that as a cluster gets older, its halo becomes less
populated and core size decreases, resulting from mass segregation and evaporation of stars from the cluster. They also
concluded that there are several physical processes along with mass segregation, which led to the disruption of the cluster. The other factor which affects the cluster disruption might be their closeness to the
galactic disc is visible in Fig. \ref{allcls2}. The left panel of Fig. \ref{allcls2} shows the
relationship between distance from the Galactic centre, age, and luminosity function
of the clusters. There is no strong relationship evident from this diagram
except a few young clusters located towards the Galactic centre have
lost their faint stars. For more clear inference, a large sample is required.
The right panel of this figure shows the relationship between age,
distance from the Galactic disc and the luminosity function of the clusters. The young
clusters that are very close to the Galactic disc have lost their fainter stars
while the clusters, those are located far from the disc, have their fainter stars
intact, even a few intermediate-age clusters have increasing luminosity functions.


\section{Discussion and Conclusions} \label{con}

This article presents a detailed study of the three open clusters NGC 1193, NGC 2355
and King 12 using photometric and kinematic data. We selected the cluster members from the field population by calculating their membership probability.
We analyzed data covering large areas to determine the true extent of the
clusters and to check if any signatures of the corona were present in these
clusters. We found that among the three clusters, NGC 1193 is a small and compact
cluster, while NGC 2355 is a sparsely populated cluster populating a large area in the sky having a highly dense central region. We found no sign of corona in these clusters.

The fundamental parameters for the clusters were determined by fitting the
isochrones on cluster CMDs. To get more reliable results, the CMDs were constructed
in several wavebands using data from Gaia DR3 in $G, G_{BP}, G_{RP}$ filters,
104-cm ST in $V, B, I$ filters and 2MASS in $J, H, K$ filters. From these diagrams,
we found that NGC 1193 is an old cluster of 4.79 Gyr, NGC 2355 is an
intermediate age cluster of age 1.26 Gyr, and King 12 is a young open cluster of
age 19 Myr. NGC 1193 consists of a thick main sequence than a usual main sequence
of an open cluster. This cluster has several giant stars, blue stragglers and yellow stragglers. The members of NGC 2355 have very different kinematics
from the field stars, showing a clean cluster population in CMDs.

The CMD of King 12 shows that the cluster has a handful of stars at its brighter
end. As mentioned in the introduction of this article, \citet{2012ApJ...761..155D} and \citet{2013MNRAS.429.1102G} found a gap in
the main sequence, which reflected a gap in the luminosity function of the cluster.
In the current analysis, we found a gap in the main-sequence of the cluster in
11 to 12 mag in $V$ filter, but the gap is not there in CMDs constructed using
Gaia and 2MASS filters. The number density of stars dropped in this region, which is reflected as a dip in the luminosity function of the cluster.

We have studied the internal and external dynamics of the clusters using the
photometric and kinematical data from Gaia DR3. The influence of dynamical evolution
on the clusters can be studied in terms of luminosity functions. We found a flat
luminosity function for NGC 1193, which reveals that the high-mass stars
evolved over time, and low-mass stars are still bound to the clusters. The luminosity
function of NGC 2355 increases up to $\sim$ 6 mag, and a dip is observed afterwards. It is
because the cluster has lost a few of its fainter members. We found an increasing
luminosity function for cluster King 12, which indicates that the fainter stars bind this cluster due to its very young age. The mass function slopes for
all the clusters are different from the solar neighbourhood for NGC 1193 and
NGC 2355, the slope is flatter than the Salpeter \citep{Salpeter1955ApJ...121..161S} value, and for King 12, it is comparatively higher.

The two-dimensional velocity diagram shows the internal kinematics of the clusters
plotted in Fig. \ref{motion}. These figures show that the stars are moving randomly inside the clusters.
We also calculated the clusters' relative mean motions, as shown in Fig. \ref{motion2}. The dynamics of the clusters
in our Galaxy are studied by determining their orbits using the Galactic potential models.
From these plots, we can conclude that NGC 1193 is born at a more considerable distance from
Galactic centre as well as Galactic disc and moving up to larger distances in
both radial and perpendicular directions. It is moving up to a scale height of
1.21 kpc from the Galactic disc. NGC 2355 is moving at a comparatively low scale
height, having a maximum value of 0.65 kpc from the Galactic disc and tracing approximately
equal distance in every orbit in a perpendicular direction. King 12 was born very
close to the Galactic disc hence going only upto 0.03 kpc distance from the
Galactic disc. These orbits indicate that King 12 is highly affected by the
Galactic tidal forces, therefore it is evaporating at a faster rate. The cluster's higher negative mean relative motion in both 
directions can confirm this. The mean relative motion of NGC 2355 is zero, possibly because the cluster
is comparatively larger in size, has a dense central region, and the stars have smaller velocities. NGC 1193 has
negative relative mean motion in the RA direction only, and this cluster is least
affected by the Galactic tidal forces; hence still bound by its fainter members
despite its older age.

We tried to find a relation between the positions of the open clusters in the Galaxy
and their age and orbits, for which we have included the results from our previous analyses and shown
in Fig \ref{allcls}. From these plots, we found that the younger clusters
are close to the Galactic disc, while the older clusters show a spread in
their positions. The clusters born close to the disc are 
not tracing a larger scale height from the Galactic disc.
We also investigated the relations of cluster positions with the luminosity
functions in Fig. \ref{allcls2}. These show that the open clusters, 
located at a more considerable distance from the Galactic centre, have increasing luminosity
functions. Even young clusters close to the galactic disc have lost their fainter stars, and the older clusters located at a certain height from the Galactic disc have increasing luminosity functions.

\section*{Acknowledgements}

We would like to thank the anonymous reviewer for taking the time and effort to review this manuscript and giving very valuable suggestions to improve the quality of this manuscript. 
We thank Prof. Annapurni Subramaniam and Dr Vikrant V. Jadhav for having such
constructive discussions.
This work has made use of data from the European Space Agency (ESA) mission
{\it Gaia} (\url{https://www.cosmos.esa.int/gaia}), processed by the {\it Gaia}
Data Processing and Analysis Consortium (DPAC,
\url{https://www.cosmos.esa.int/web/gaia/dpac/consortium}). Funding for the DPAC
has been provided by national institutions, in particular, the institutions
participating in the {\it Gaia} Multilateral Agreement.

\section*{DATA AVAILABILITY}

We have used two publicly available data sets: Gaia DR3 and 2MASS for the clusters NGC 1193, NGC 2355 and King 12. Data from Gaia DR3 can be accessed from:
https://gea.esac.esa.int/archive/

Data from 2MASS can be accessed from: 
https://vizier.cds.unistra.fr/viz-bin/VizieR?-source=II/246

The photometric data of the three clusters NGC 1193, NGC 2355 and King 12 can be requested to Geeta Rangwal (geetarangwal91@gmail.com).

\bibliographystyle{mnras}
\bibliography{article_mn}


\label{lastpage}
\end{document}